\documentclass[superscriptaddress,amsmath,amssymb, aps, prl,longbibliography,twocolumn, footinbib]{revtex4-2}
\usepackage{upgreek}
\usepackage{textcomp}

\usepackage[T1]{fontenc}
\usepackage[utf8]{inputenc}

\usepackage{color}
\usepackage[version=4]{mhchem} 
\usepackage{xcolor}
\usepackage{graphicx}
\usepackage{nccmath}
\usepackage{bm}
\usepackage{xr-hyper}
\usepackage{hyperref}
\hypersetup{colorlinks,breaklinks,
            urlcolor=[rgb]{0,0,0.64},
            linkcolor=[rgb]{0,0,0.64},
            citecolor=[rgb]{0,0,0.64},
            filecolor=[rgb]{0,0,0.64}}
\usepackage[mathlines]{lineno}
\usepackage{dcolumn}

\makeatletter
\newcommand*{\addFileDependency}[1]{
  \typeout{(#1)}
  \@addtofilelist{#1}
  \IfFileExists{#1}{}{\typeout{No file #1.}}
}
\makeatother
 
\newcommand*{\myexternaldocument}[1]{%
    \externaldocument{#1}%
    \addFileDependency{#1.tex}%
    \addFileDependency{#1.aux}%
}

\definecolor{mygreen}{RGB}{0,128,0}  

\usepackage{mystyle}

\myexternaldocument{supplement}

\begin{document}

\title{Wideband covariance magnetometry below the diffraction limit}

\author{Xuan~Hoang~Le}
\affiliation{\HarvardPhysics}
\affiliation{\HarvardCCB}

\author{Pavel~E.~Dolgirev}
\affiliation{\HarvardPhysics}

\author{Piotr~Put}
\affiliation{\HarvardPhysics}
\affiliation{\HarvardCCB}

\author{Eric~L.~Peterson}
\affiliation{\HarvardPhysics}

\author{Arjun~Pillai}
\affiliation{\HarvardCCB}

\author{Alexander~A.~Zibrov}
\affiliation{\HarvardPhysics}
\affiliation{\HarvardCCB}

\author{Eugene~Demler}
\affiliation{\ETH}

\author{Hongkun~Park}
\affiliation{\HarvardPhysics}
\affiliation{\HarvardCCB}

\author{Mikhail~D.~Lukin}
\thanks{Corresponding author: \href{mailto:lukin@physics.harvard.edu}{lukin@physics.harvard.edu}}
\affiliation{\HarvardPhysics}

\date{April 30, 2025}

\begin{abstract}
We experimentally demonstrate a method for measuring correlations of wideband magnetic signals with spatial resolution below the optical diffraction limit. Our technique employs two nitrogen-vacancy (NV) centers in diamond as nanoscale magnetometers, spectrally resolved by inhomogeneous optical transitions. Using high-fidelity optical readout and long spin coherence time, we probe correlated MHz-range noise with sensitivity of 15\,nT\,Hz$^{-1/4}$. 
In addition, we use this system for correlated $T_1$ relaxometry, enabling correlation
measurements of GHz-range noise. 
Under such externally applied noise, while individual NV centers exhibit featureless relaxation, their correlation displays rich coherent and incoherent dynamics reminiscent of superradiance physics.
This capability to probe high-frequency correlations provides a powerful tool for investigating a variety of condensed-matter phenomena characterized by nonlocal correlations.
\end{abstract}

\maketitle

\begin{figure*}[hbt!]  
    \centering
    \includegraphics[width=1\textwidth]{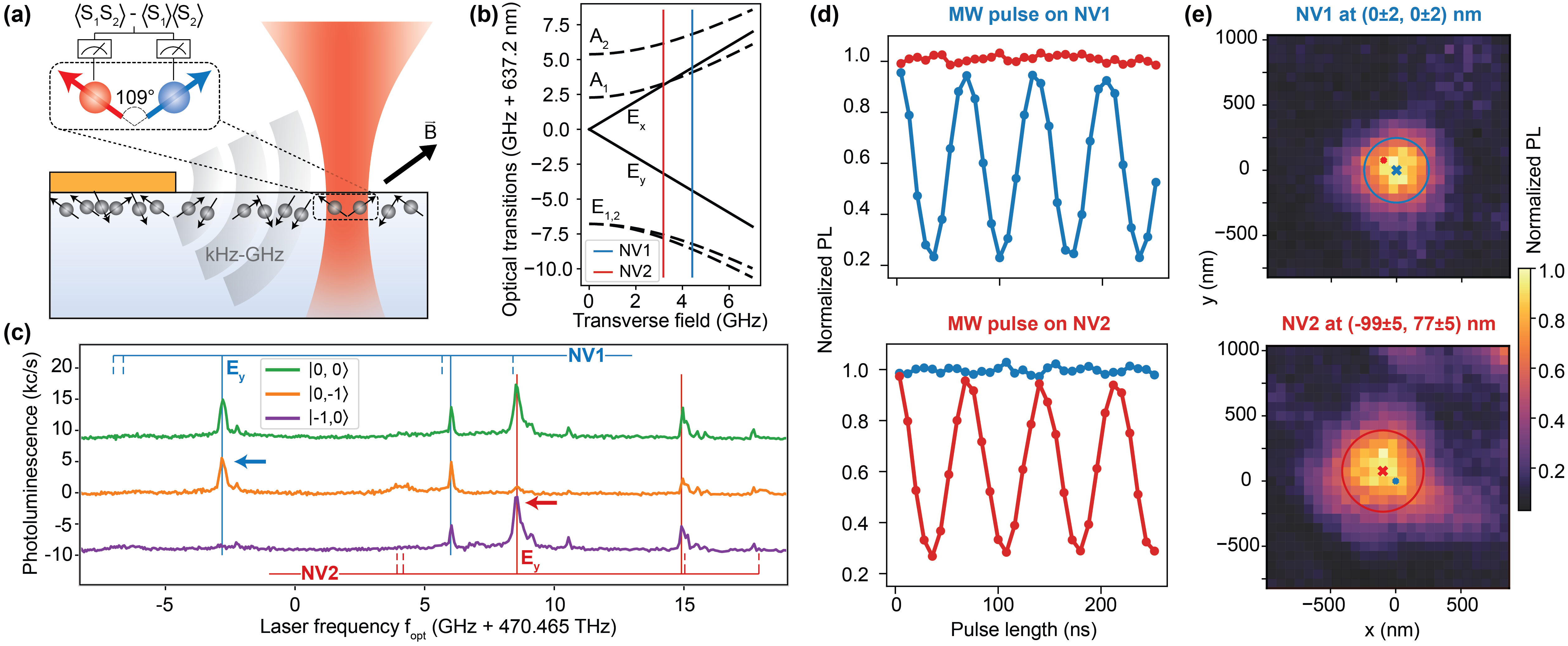} 
    \caption{Super-resolution and independent spin control. (a) Schematics of the experimental system, consisting of two NV centers in a single optical spot experiencing a common magnetic field. The field is probed by measuring spin-spin correlation of the two NVs. NV1 (blue) and NV2 (red---color code used throughout the paper) align differently to a static magnetic field. Both the MW signals driving spin transitions and test signals producing correlated magnetic fields are delivered through a gold stripline (yellow rectangle) fabricated on the diamond.
    (b) Calculated optical transitions of an NV as a function of transverse strain/electric field. The annotations, used as names of each transition, are the destination orbital states in the excited-state manifold. Vertical lines mark each NV's transverse field obtained from (c). Solid (dashed) black curves in (b) and vertical lines in (c) are $m_s=0$ ($\pm1$) transitions.
    (c) PLE spectroscopy, utilizing different initial spin state combinations of two NVs (three spectra, offset for clarity), enables identifying each NV's set of transitions. Arrows indicate transitions used for readout. 
    (d) Independent spin control confirmed by driving Rabi oscillations of one NV and reading both defects.
    (e) Confocal maps taken by resonantly exciting  NV1 (top) and NV2 (bottom). The location of the on-resonant NV (cross) is determined from a fit to a 2D Gaussian (full width at half maximum denoted by a circle). The off-resonant NV is marked by a smaller dot. All experiments in this work are done in a cryostat at 11$\,$K.}
    \label{fig::fig1}
\end{figure*}

Two-point spatio-temporal correlations play an important role in characterizing many physical systems, 
including non-classical electromagnetic field~\cite{scully}, macroscopic quantum coherence in superconductors and Bose-Einstein condensates~\cite{tinkham, andrews1997, altman2004, folling2005}, quantum fluctuations of spin ensembles~\cite{sorensen1998, crooker2004}, and magnetic orderings in quantum simulators~\cite{simon2011, fukuhara2013, parsons2016, manovitz2025, bernaschi2024} and materials~\cite{marshall1968, endoh1988, tianhenghan2012, rybakov2024, 
ishiguro2013, 
callen1965, zapf2008, 
imai1993, carretta2000, kitagawa2018, 
haines1985,  
ron2019, 
back1995, chenhaojin2020,
luyiyang2014,
defilippis2012, sriram2022}. 
In particular, the ability to quantify magnetic correlation could significantly advance understanding of nonlocal collective phenomena in two-dimensional (2D) systems, such as novel magnetic order~\cite{qinghuawang2022, ciorciaro2023, jihosung2025}, hydrodynamic transport~\cite{lucas2018,ruolanxue2024}, and superconductivity~\cite{saito2016}. However, accessing the properties of collective magnetic excitations in realistic materials typically requires probing spectrally broadband correlations at nanometer length scales, which is challenging.   
Examples include kHz- to MHz-range longitudinal and GHz-range transverse spin fluctuations in 2D magnets~\cite{mclaughlin2022, ruolanxue2024}, DC- to GHz-range noise from stimulated phonon emission in graphene~\cite{andersen2019}, current fluctuations~\cite{dolgirev2022, chatterjee2022,zhongyuanliu2025} and Berezinskii-Kosterlitz-Thouless vortex unbinding in 2D superconductors~\cite{curtis2024}.  

Several approaches have been explored  
to probe spin-spin correlations with nanoscale resolution. For example,  
neutron~\cite{marshall1968, endoh1988, tianhenghan2012, rybakov2024} and X-ray scattering~\cite{ishiguro2013}, magnetostriction measurements~\cite{callen1965, zapf2008}, and nuclear magnetic/quadrupole resonance~\cite{imai1993, carretta2000, kitagawa2018} can be used, albeit only for macroscopic, homogeneous samples. Optical techniques~\cite{haines1985, ron2019, back1995, chenhaojin2020, luyiyang2014, defilippis2012, sriram2022} are suitable for studying two dimensional and thin-film materials, but diffraction typically limits the resolution to a fraction of optical wavelength. Moreover, with the exception of optical pump-probe spectroscopy~\cite{defilippis2012, sriram2022}, none of the techniques above are suitable to probe broadband (DC to GHz) dynamics due to various technical limitations~\cite{boothroydChap10, imai1993, zapf2008, ron2019, chenhaojin2020}.

Atom-like systems such as  nitrogen-vacancy (NV) centers in diamond offer a promising approach to this challenging problem.  Such solid-state spin defect magnetometers 
have been used 
in ``single-point'' modality (i.e., probing magnetic field at each point by a single defect or averaging over multiple defects in one optical spot) to study a wide range of static and dynamical phenomena in materials~\cite{casola2018, rovny2024, thiel2019, bhattacharyya2024, ku2020, kolkowitz2015, chunhuidu2017, ruolanxue2024}. 
Recently, Rovny et al.~\cite{rovny2022} demonstrated that this system can be used to  
measure two-point spatio-temporal magnetic correlations. This approach has been extended to parallel measurement of multiple single NVs in different optical spots~\cite{kaihungcheng2024, cambria2024}, or two NVs in the same spot~\cite{huxter2024, joliffe2024}. These methods have been used to probe the magnetic noise up to MHz frequency range, but they can not be directly extended to  
GHz noise, which ubiquitously appears in various condensed-matter systems~\cite{mclaughlin2022, carmiggelt2023, wolfe2014, chunhuidu2017, ericleewong2020, mccullian2020, geraldyan2022, machado2023, ruolanxue2024, ziffer2024, yuxinli2024, zhongyuanliu2025, andersen2019, kolkowitz2015, ariyaratne2018, chatterjee2022, dolgirev2022, zhongyuanliu2025}.

In this Letter, we extend covariance magnetometry to both new spatial and frequency scales by measuring magnetic field correlations up to GHz frequencies below the diffraction limit. Utilizing optical super-resolution techniques to address individual NV centers based on inhomogeneous optical transitions~\cite{bersin2019, monge2023, wentaoji2024, delord2024}, together with resonantly-assisted spin-to-charge conversion~\cite{irber2021}, we achieve low readout noise, resulting in large detectable correlations. We use this system to probe a correlated MHz-range magnetic noise by a Ramsey-based protocol. 
In addition, we demonstrate that co‐relaxation of two NVs enables detailed studies of GHz-range signals.
In particular, by externally applying such magnetic noise with correlated amplitudes, we observe superradiant-like dynamics -- both coherent and incoherent -- arising from local detuning and dephasing. 
This technique can be integrated with other state-of-the-art NV sensing platforms, such as scanning tips and nanopillar arrays~\cite{orphalkobin2023}, while scaling favorably to few-NV clusters. It is therefore particularly suited for probing two-point and higher-order magnetic correlations in condensed matter at different lengthscales (from nm to \textmu m) and across a broad spectrum (from DC to GHz).

\textit{Experimental system.}---Our experiments make use of two shallowly implanted NV centers in the same optical spot addressed by a single-path scanning confocal microscope (Fig.~\ref{fig::fig1}(a)) (for sample details see Supplemental Material~\cite{supp}). We choose a pair of NVs with different crystallographic orientations, resulting in distinct microwave (MW) transitions that enable independent spin control. Although the spin state of each NV and their correlation can be obtained from global readout~\cite{huxter2024, joliffe2024}, such techniques require more repetitions and may suffer from spurious correlations. Instead, we independently read each NV by a defect-selective, spin-selective optical excitation~\cite{bersin2019, monge2023, delord2024, wentaoji2024}. This is possible because at cryogenic temperatures, the spin-dependent ground-to-excited state transitions are spectrally resolvable, and the inhomogeneous local strain/electric field separates each NV's set of optical transitions~\cite{batalov2009} (Fig.~\ref{fig::fig1}(b)). By preparing the NVs in different spin state combinations and performing photoluminescence excitation (PLE) spectroscopy, we identify the set of transitions for each NV (Fig.~\ref{fig::fig1}(c)). We use the bright, high spin contrast $m_s=0$, $E_y$ transition of both NVs for readout in all following experiments.

To confirm independent spin control, we drive Rabi oscillations of one NV and read both defects sequentially in the same repetition. Fig.~\ref{fig::fig1}(d) shows high spin contrast (75-80\%) for the driven NV and virtually no change in the undriven one. Super-resolution imaging of NVs is also straightforward in this modality and is accomplished by performing two confocal scans, each with the laser on-resonant with one NV (Fig.~\ref{fig::fig1}(e)). The measured 125\,nm lateral distance and estimated $\sim$ 20\,nm $z$-distance~\cite{supp} imply negligible magnetic dipolar interactions between the two NVs.

\begin{figure}[b!]
\centering
\includegraphics[width=1\linewidth]{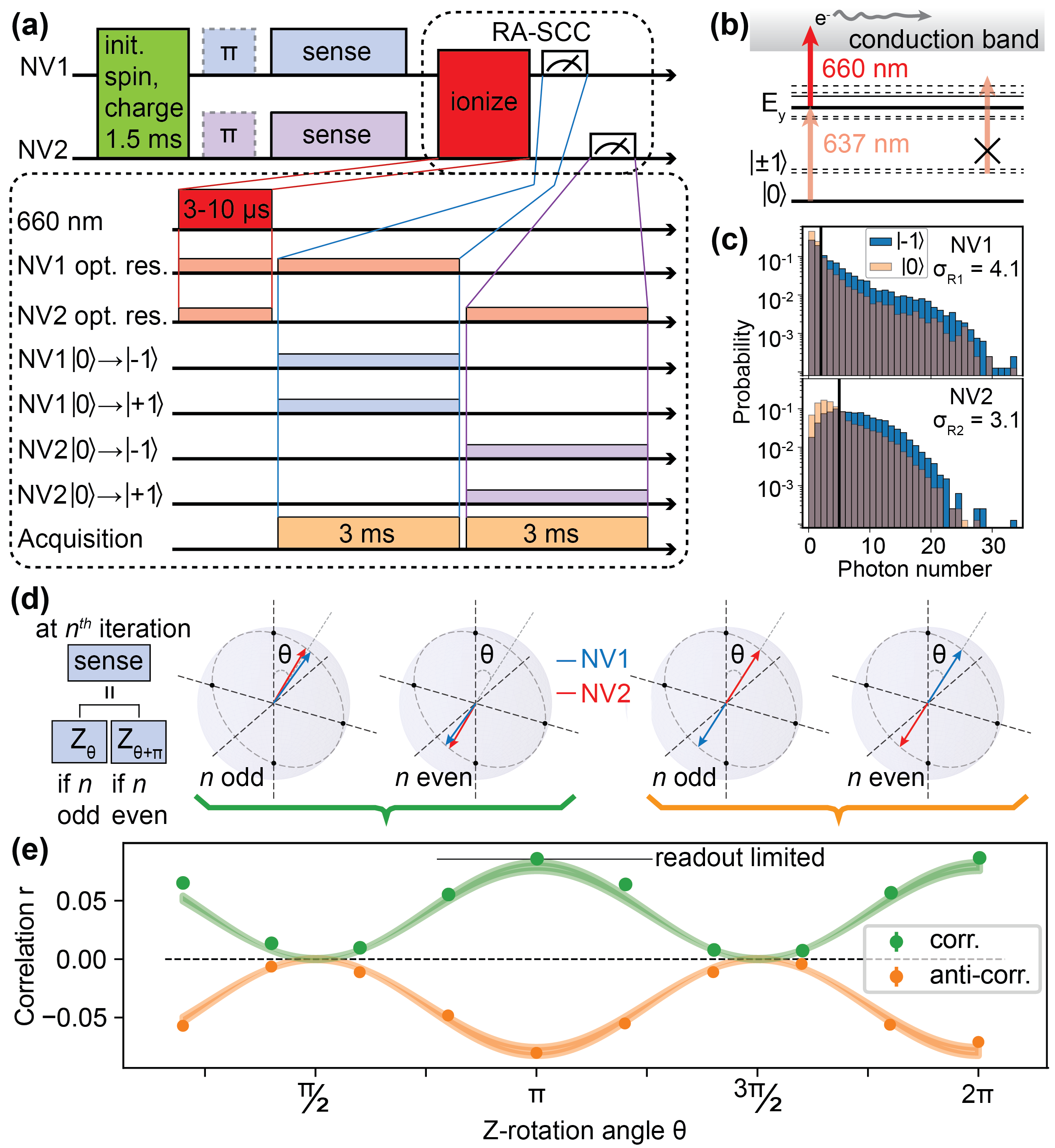} 
\caption{High-fidelity readout and maximum detectable correlation. (a) General protocol for correlation measurements: after initialization to the (NV$^-$, $\ket{0}$) state by 532$\,$nm light, an optional $\pi$-pulse prepares the spin state for sensing. High-fidelity readout is achieved by RA-SCC (dashed box), consisting of spin-dependent ionization by a two-photon process shown in (b), followed by sequential readout of each NV charge state by resonant excitation. During this spin-agnostic readout, low-power MW continuously mixes the spin states to counteract pumping into $\ket{\pm1}$~\cite{irber2021}.
(b) The resonant 637$\,$nm laser selectively excites the $\ket{0}$ state, and a high-power 660$\,$nm laser ionizes this population. 
(c) Photon statistics after RA-SCC for each NV, prepared in $\ket{0}$ or $\ket{-1}$. Readout with lower (higher) photon counts than the threshold (bold vertical lines) are assigned to $\ket{0}$ ($\ket{-1}$).
(d) Demonstration of correlation measurement by direct NV spin driving, alternating between polar angle $\theta$ or $\theta+\pi$. 
(e) Pearson correlation coefficient $r$ for both correlation (same initial states, green) and anti-correlation (opposite initial states, orange) configurations. Filled curves are theory predictions; the curve width denotes uncertainty of measured $\sigma_{\rm R}$.} 
\label{fig::fig2}
\end{figure}

\begin{figure}[b!]
\centering
\includegraphics[width=1\linewidth]{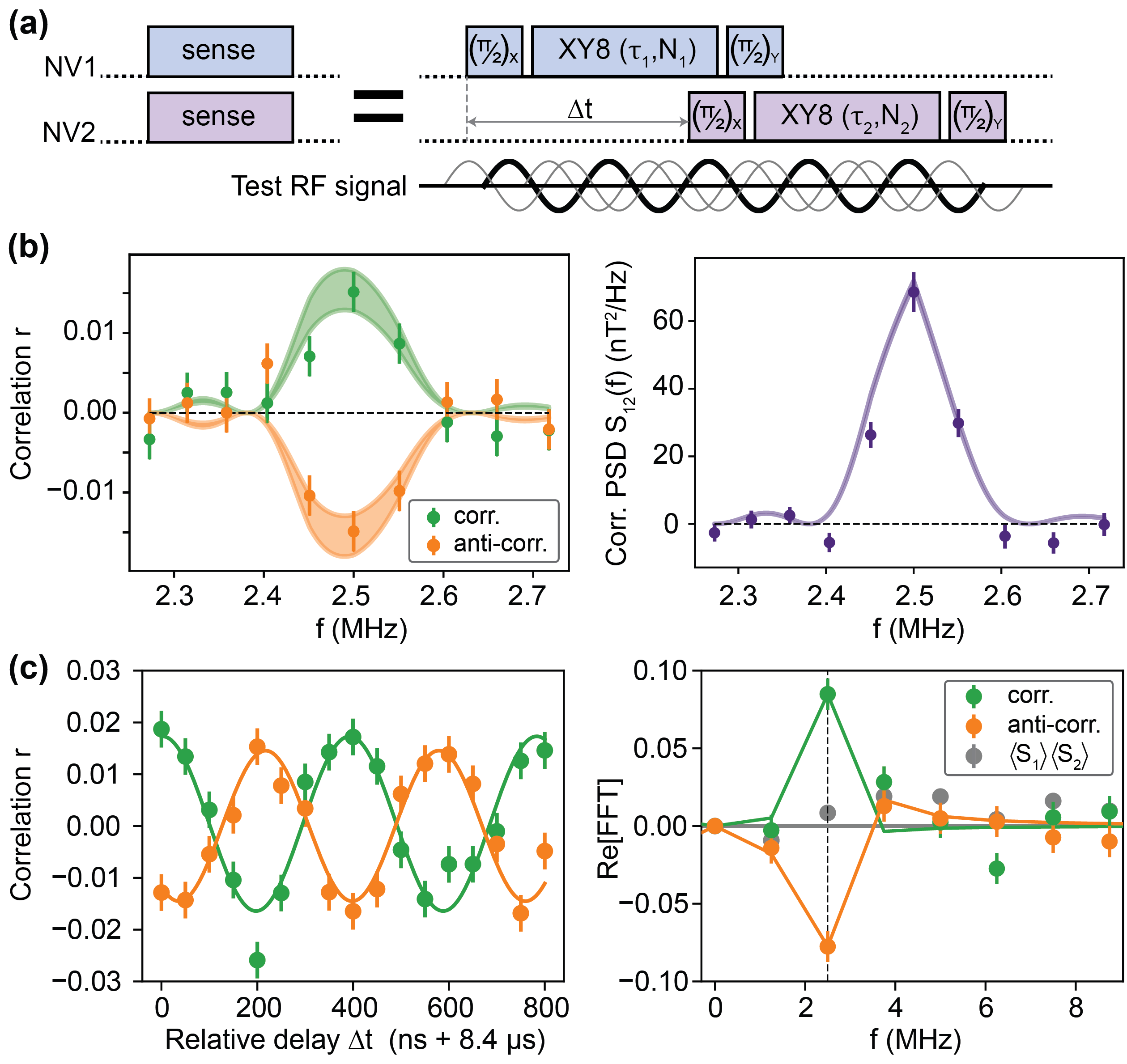} 
\caption{Correlated $T_2$ spectroscopy. (a) The sensing sequence with phase-noisy AC test signal. (b) Equal-time ($\Delta t=0$) spectroscopy, sweeping $\tau_1=\tau_2=\tau$ to probe $f=1/(2\tau)$. PSD of the correlated noise (right) as calculated from the correlations (left). Filled curves represent theory fits, with a global scaling as the sole fit parameter; the curve widths indicate the uncertainty in the measured values of \(\sigma_{\rm R},\, B_{1},\) and \(B_{2}\)~\cite{supp}.
(c) Two-time spectroscopy, sweeping $\Delta t$ while keeping $\tau_1=\tau_2=1/(2f_0)$, showing correlations with sinusoidal fits (left) and the real part of the corresponding Fourier transforms (right). Grey datapoints denote $\langle S_1\rangle$$\langle S_2\rangle$; vertical dashed line marks $f_0$.}
\label{fig::fig3}
\end{figure}

\begin{figure*}[hbt!]  
    \centering
    \includegraphics[width=1\textwidth]{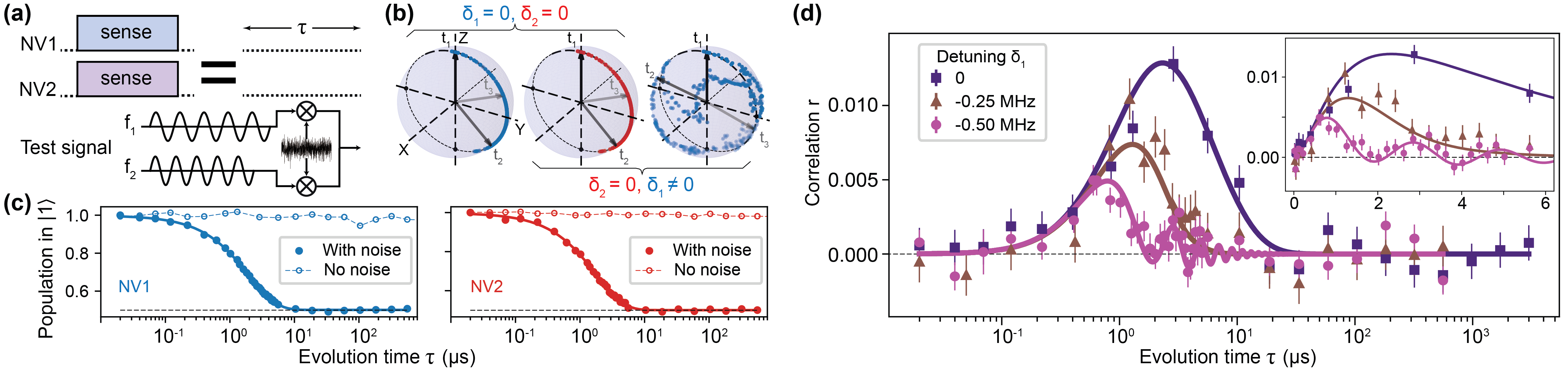} 
    \caption{Correlated $T_1$ spectroscopy. 
    (a) The test signal is generated by combining two MW signals, each resonant with one NV center and independently tuned to have similar powers. Their amplitudes are modulated by the same 10$\,$MHz-bandwidth Gaussian noise. 
    (b) Illustration of dynamics: With no detunings for both MW signals, the two NVs exhibit stochastic yet identical dynamics (first two panels), resulting in high correlation. Three Bloch vectors represent the state of the NVs at arbitrary times \(t_1 < t_2 < t_3\). When the test signal for NV1 is detuned (third panel), despite the noise profile being identical to that of NV2 (second panel), NV1's trajectory includes precession around the $z$-axis. As such, the correlation diminishes and oscillates at the precession rate. The fading color indicates later times. 
    (c) Spin polarization relaxation of each NV under the test noise overlaid with a simple exponential decay, where $T_{\rm 1, NV1}=1.96(2)\,$\textmu s and $T_{\rm 1, NV2}=1.77(3)\,$\textmu s.
    (d) Pearson correlation $r$ for different detunings $\delta_1$ (with $\delta_2=0$ fixed), showing correlation build-up despite featureless single-NV relaxation observed in (c). The inset zooms into the first 6\,\textmu s, highlighting oscillations of correlation at the largest detuning \(\delta_1=-0.5\)\,MHz. Solid curves are global fits of the entire dataset using a model based on the quantum master equation (Eq.~\eqref{eqn::QME}).}
    \label{fig::fig4}
\end{figure*}

\textit{High-fidelity readout and correlation measurements.}---Our goal is to measure the shot-to-shot spin-spin correlation of two NVs, quantified by the Pearson correlation coefficient~\cite{rovny2022}
\begin{align}
    r = \frac{\langle S_1(r_1, t_1)S_2(r_2, t_2) \rangle - \langle S_1(r_1, t_1) \rangle \langle S_2(r_2, t_2) \rangle}{(\sigma_{\rm p1}\sigma_{\rm p2})(\sigma_{\rm R1}\sigma_{\rm R2})},
\label{eqn::eq1}
\end{align}
where $i$ is the NV index, $S_i(r_i,t_i) \in \{0,1\}$ represents the spin state at time $t_i$, $\sigma_{\mathrm{p}i}$ is the quantum projection noise, and $\sigma_{\mathrm{R}i}$ is the single-shot readout noise. $S=0$ ($1$) corresponds to the $\ket{m_s=0}$ ($\ket{-1}$) state; $\langle\, \dots \,\rangle$ denotes averaging over all experimental repetitions. Since $r$ depends on $1/\sigma^2_{\mathrm{R}}$, it is imperative to have low $\sigma_{\rm R}$. Despite high spin contrast, the resonant readout here only achieves $\sigma_{\rm R}=15$ because spin mixing within the excited states limits the readout duration to a few microseconds~\cite{robledo2011, supp}. To collect more photons per shot, we convert the NV spin state to charge state (NV$^-$/NV$^0$), which can be distinguished by the same resonant excitation as before, but for millisecond periods~\cite{supp}. To achieve good spin selectivity for this conversion, we again utilize the resonant excitation~\cite{irber2021}: when the NV is simultaneously illuminated with this laser and a high-power 660$\,$nm light, only the $\ket{0}$ population undergoes two-photon ionization to NV$^0$ and yields low photon counts during later readout, while $\ket{\pm1}$ remain in NV$^-$, thus is bright (Fig.~\ref{fig::fig2}(b), (c)). With this resonantly-assisted spin-to-charge conversion (RA-SCC), we achieve $\sigma_{\mathrm{R}}=3 \text{–} 4$ for both defects.

The overall protocol for all correlation measurements is as follows (Fig.~\ref{fig::fig2}(a)): initialize each NV charge and spin state to (NV$^-$, $\ket{0}$) by 532$\,$nm light, prepare them in $\ket{0}$ or $\ket{-1}$ depending on the specific experiment, apply a MW sensing sequence to each NV independently, and do RA-SCC with sequential charge readout of each NV.

We benchmark our protocol by measuring correlations arising from direct NV spin driving (Fig.~\ref{fig::fig2}(d),~\cite{kaihungcheng2024, cambria2024}). In the correlation configuration (green), both defects are prepared in the same spin state, alternating between either polar angle $\theta$ or $\theta+\pi$ on the Bloch sphere in each iteration. The measured correlation in Fig.~\ref{fig::fig2}(e) agrees well with the theory prediction $r=\cos^2(\theta)/(\sigma_{\rm R1}\sigma_{\rm R2})$~\cite{supp}. Intuitively, although alternating between $\theta$ and $\theta+\pi$ cancels the single-NV quantum projection noise (QPN) and makes $\langle S_i \rangle=\frac{1}{2}$ regardless of $\theta$, it does not cancel the QPN of the product $S_1S_2$, thus $\langle S_1S_2 \rangle$ remains $\theta$-dependent~\cite{supp}. Therefore, $r$ is maximized when $\theta=n\pi$ for integer $n$ due to no QPN, and $r\approx0$ for $\theta=(n+\frac{1}{2})\pi$ because the maximum QPN destroys any existing correlation. We also measure in the anti-correlation configuration (orange), in which both defects are prepared in opposite states, and observe the same results but with opposite sign of $r$, as expected. All reported correlation values hereafter are obtained after subtracting the small background correlation common to both correlation and anti-correlation configurations. Because the NVs undergo no free evolution in this experiment, the effects of spin decoherence are minimized, which enables quantifying the maximum detectable correlation $r_{\rm max}=1/(\sigma_{\rm R1}\sigma_{\rm R2})$. The measured $r_{\rm max}=0.083(3)$ agrees well with $r_{\rm max}^{\rm calc}=0.079(4)$ calculated from the measured $\sigma_{\rm R1}$, $\sigma_{\rm R2}$ in Fig.~\ref{fig::fig2}(c).

\textit{Correlated $T_2$ spectroscopy.}---Next, we demonstrate sensing correlation of a $f_0=2.5$$\,$MHz magnetic field phase-modulated by a Gaussian noise with 25\,kHz bandwidth. 
The XY8-$N$ sensing sequence of each NV can be tuned independently with interpulse spacing $\tau_i$ to probe frequency $1/(2\tau_i)$, number of repetitions $N_{i} \in \{5,8 \}$, and relative delay $\Delta t$ between each NV's sequence (Fig.~\ref{fig::fig3}(a)). We first detect the signal spectroscopically by equal-time correlation measurement ($\Delta t=0$), sweeping $\tau_1=\tau_2$. 
Figure~\ref{fig::fig3}(b) (left) shows agreement between experimental data and calculated correlation lineshape based on the measured $\sigma_{\rm R1}$, $\sigma_{\rm R2}$, and AC field amplitudes at each NV $B_{1}=1.944(11)$$\,$\textmu T and $B_2=0.56B_1$ (due to different orientations), as obtained from single-NV magnetometry~\cite{supp}.
The power spectral density (PSD) of the correlated noise in Fig.~\ref{fig::fig3}(b) (right) shows that despite sensing for only 8\,\textmu s in each shot, much shorter than $T_2>1$\,ms, and 40 minutes in total for each datapoint, we can detect noise strength of $\sim70\,$nT$^2$/Hz with uncertainty of $4\,$nT$^2$/Hz, corresponding to magnetic sensitivity of 15\,nT\,Hz$^{-1/4}$~\cite{supp}.  
Extending sensing to 1\,ms, as allowed by the long $T_2$, with negligible cost to total experiment time (due to long readout) could enable detection of correlated noise strength $< 1\,$nT$^2$/Hz~\cite{supp}. To measure two-time correlation, we sweep the relative delay $\Delta t$ at fixed $\tau_1=\tau_2=1/(2f_0)$, observing the oscillation of $r$ at $f_0$ as expected, while $\langle S_1\rangle$$\langle S_2\rangle$ remains featureless because the noise has zero mean (Fig.~\ref{fig::fig3}(c)).


\textit{Correlated $T_1$ spectroscopy.}---We now extend the technique to probing correlated few-GHz noise, which is beyond the scope of $T_2$ spectroscopy due to the impractically high Rabi frequencies required.
To this end, we apply a test noise such that each NV is subject to a magnetic field $\bm B_i(t) \propto G(t) \cos(\omega_i t)\hat{x} $, where $\omega_i$ is close to the $i$-th NV's $\ket{0}$$\rightarrow$$\ket{-1}$ transition $\Delta_i$ with controllable detuning $\delta_i \equiv \Delta_i - \omega_i$, and $G(t)$ describes a Gaussian noise with zero mean and bandwidth $\nu_c \simeq 10\,$MHz. 
Such noise, common to both sensors and mimicking GHz magnetic field correlations, represents amplitude-modulated but phase-locked fluctuations (Fig.~\ref{fig::fig4}(a)).

While this noise causes the mean spin polarizations to experience simple exponential decays (Fig.~\ref{fig::fig4}(c)), the correlation $r$ undergoes richer dynamics (Fig.~\ref{fig::fig4}(d)): initially, it develops as the two NVs co-relax; next, application of a finite local detuning $\delta_1 \neq 0$ on NV1 results in oscillations of $r$ at $\delta_1$; finally, $r$ decays to zero at longer times. Notably, the observed build-up, oscillations, and long-time decay of spin-spin correlation strikingly resemble the dynamics of dipole-dipole correlations in atomic ensembles undergoing superradiance~\cite{gross1982, yelin2005, rubiesbigorda2022}.

We explain these experimental observations by the following theoretical model. 
The effective Hamiltonian in co-rotating frames of each NV reads~\cite{supp}:
\begin{align}
    H = \sum_{i = 1,2}  \delta_i S_i^z + 2{\cal A}_i G(t) S_i^x ,
    \label{eqn::H}
\end{align}
where $S_i^{\alpha = x,y,z}$ are the spin-1/2 operators and ${\cal A}_i$ encodes the test signal amplitude. Provided that the bandwidth $\nu_c$ is much larger than other relevant dynamical scales, the noise is effectively Markovian and the time-evolution of the two-NV density matrix $\rho(t)$ is governed by the quantum master equation (QME):
\begin{align}
    \partial_t \rho = i \sum_i \delta_i \big[S_i^z, \rho \big] + 2 \big[  {\cal L} ,\big[  \rho,  {\cal L} \big] \big], 
    \label{eqn::QME}
\end{align}
where $\big[ A,B\big] \equiv AB-BA$ denotes the commutator and ${\cal L} \equiv  (2\pi)^{1/4} \sqrt{1/\nu_c} ({\cal A}_1 S_1^x + {\cal A}_2 S_2^x)$ is the jump operator. 

Equation~\eqref{eqn::QME} shows that \(\langle S_i \rangle = 1/2 - \langle S_i^z \rangle\) relaxes as \(\exp(-t/T_{{\rm 1,NV}i})\), in agreement with the data in Fig.~\ref{fig::fig4}(c).
The extracted relaxation times $T_{{\rm 1,NV}i}$ correspond to field amplitudes $B_{1} = 5.72(3)\,$\textmu T and $B_{2} = 6.02(5)\,$\textmu T.

To explain the dynamics of the correlation $r$, we note that 
when each MW signal is resonant with the respective NV, i.e., $\delta_1 = \delta_2 = 0$, as follows from Eq.~\eqref{eqn::H}, both NVs are stochastically driven around the $x$-axis, albeit with identical trajectories that remain in the $yz$-plane (Fig.~\ref{fig::fig4}(b), first two panels). 
The correlation is thus expected to develop over time and remain high.
When the test signal is detuned for the first NV, i.e., $\delta_1 \neq 0$ and $\delta_2 = 0$, the correlation dynamics become more complex as NV1 undergoes both stochastic rotations around the $x$-axis and precession around the $z$-axis at $\delta_1$ (Fig.~\ref{fig::fig4}(b), last two panels). 
In this case, Eq.~\eqref{eqn::QME} predicts oscillations of the correlation at $\delta_1$. 
The observed decay at longer times originates from local and correlated dephasing effects, which we introduce phenomenologically~\cite{supp}.  
With five fitting parameters for the entire dataset, this QME model closely matches the experimental data (Fig.~\ref{fig::fig4}(d)).

The analogy to superradiance can now be understood by looking at the structure of the jump operator ${\cal L}$ in Eq.~\eqref{eqn::QME}, which is a sum of two terms that describe a common decay channel.
In atomic superradiance, such a common decay channel originates from different atoms coupling to the same photonic modes, whose wavelengths significantly exceed the atom-atom spacings. In our system, this correlated decay is deliberately engineered via the common noise and only evident in the rotating frames of each NV center.

\textit{Conclusion and outlook.}---We have demonstrated covariance magnetometry with spatial resolution below the diffraction limit and probed correlated signals in both the MHz and GHz frequency bands.
The combined effect of high-fidelity readout and long coherence time enables detecting correlated noise with magnetic sensitivity of 15\,nT\,Hz$^{-1/4}$, corresponding to detectable noise strength of 4\,nT$^2$/Hz in 40 minutes of total experimental time. 
This noise strength is well below that of critical fluctuations observed in several 2D magnets and superconductors~\cite{ruolanxue2024, yuxinli2024, ziffer2024, zhongyuanliu2025}, and is comparable to magnon density fluctuations in \ce{CrCl3}~\cite{ruolanxue2024}. 
With correlated $T_1$ spectroscopy, we observe correlation of two detuned GHz signals modulated by a common noise profile. Notably, the engineered magnetic field amplitudes are comparable to those reported for NV centers proximal to a magnon bath~\cite{chunhuidu2017, fukami2024}. The observed superradiant-type dynamics of this correlation capture the coherent effect of local detuning and incoherent effect of dephasing.
Combining our correlated $T_1$ protocol with dynamical decoupling schemes could be an interesting avenue for future work.

Our technique can be naturally extended in both the spatial and spectral domains, such as applying to few-NV clusters~\cite{monge2023, delord2024} in scanning-NV platforms~\cite{maletinsky2012, huxter2024}, or incorporating techniques well-developed for ``single-point'' NV sensing such as frequency conversion~\cite{guoqingwang2022} and high-resolution spectroscopy~\cite{laraoui2013, glenn2018}. Sensitivity could also be improved with various methods, from increasing photon collection efficiency with diamond patterning~\cite{robledo2011, maletinsky2012} to utilizing dipolar interactions in NV ensembles~\cite{koyluoglu2025}. This work demonstrates that NV centers constitute a powerful nanoscale wideband probe of two-point and higher-order magnetic correlations, particularly suited to studying condensed-matter systems with intriguing correlated dynamics~\cite{Hosseinabadi2025}, such as superconductors~\cite{chatterjee2022, dolgirev2022, curtis2024, zhongyuanliu2025}, critical fluctuations in 2D magnets~\cite{machado2023, ruolanxue2024, ziffer2024, yuxinli2024}, non-equilibrium~\cite{andersen2019, yifanzhang2024} and correlated transport phenomena~\cite{agarwal2017, ku2020}, and nonlinear magnonic processes~\cite{chunhuidu2017, wolfe2014, carmiggelt2023, ericleewong2020, mccullian2020, geraldyan2022, demokritov2006, mclaughlin2022, ruolanxue2024}.


We gratefully acknowledge discussions with N. P. de Leon, J. Rovny, A. Yacoby, N. Maksimovic, R. Xue, C. Meriles, T. Delord, R. Monge, J. Marino, H. Hosseinabadi, S. Gopalakrishnan, N. U. K{\"o}yl{\"u}o{\u{g}}lu, N. Leitao, H. Gao, M. Abobeih, and experimental support from B. Dwyer, P-J. C. Stas, Y. Q. Huan, J. D. Schaefer, and T. Madhavan. A.P. acknowledges support from DoD NDSEG Graduate Research Fellowship. This work was supported by Gordon and Betty Moore Foundation Grant No. 7797-01, National Science Foundation (grant number PHY-2012023), and the Center for Ultracold Atoms (an NSF Physics Frontiers Center). All fabrication was performed at the Harvard Center for Nanoscale Systems (CNS), a member of the National Nanotechnology Coordinated Infrastructure Network (NNCI).

{\it Note added:} During the preparation of our manuscript we became aware of related work demonstrating super-resolution covariance magnetometry~\cite{rovny2025}, which reports 
correlation measurements of signals with frequency up to a few MHz in two NVs within a single optical spot by complementary detection methods, including utilizing an entangled pair of NV centers.

\bibliography{Bib_main}

\begin{thebibliography}{86}%
\makeatletter
\providecommand \@ifxundefined [1]{%
 \@ifx{#1\undefined}
}%
\providecommand \@ifnum [1]{%
 \ifnum #1\expandafter \@firstoftwo
 \else \expandafter \@secondoftwo
 \fi
}%
\providecommand \@ifx [1]{%
 \ifx #1\expandafter \@firstoftwo
 \else \expandafter \@secondoftwo
 \fi
}%
\providecommand \natexlab [1]{#1}%
\providecommand \enquote  [1]{``#1''}%
\providecommand \bibnamefont  [1]{#1}%
\providecommand \bibfnamefont [1]{#1}%
\providecommand \citenamefont [1]{#1}%
\providecommand \href@noop [0]{\@secondoftwo}%
\providecommand \href [0]{\begingroup \@sanitize@url \@href}%
\providecommand \@href[1]{\@@startlink{#1}\@@href}%
\providecommand \@@href[1]{\endgroup#1\@@endlink}%
\providecommand \@sanitize@url [0]{\catcode `\\12\catcode `\$12\catcode
  `\&12\catcode `\#12\catcode `\^12\catcode `\_12\catcode `\%12\relax}%
\providecommand \@@startlink[1]{}%
\providecommand \@@endlink[0]{}%
\providecommand \url  [0]{\begingroup\@sanitize@url \@url }%
\providecommand \@url [1]{\endgroup\@href {#1}{\urlprefix }}%
\providecommand \urlprefix  [0]{URL }%
\providecommand \Eprint [0]{\href }%
\providecommand \doibase [0]{https://doi.org/}%
\providecommand \selectlanguage [0]{\@gobble}%
\providecommand \bibinfo  [0]{\@secondoftwo}%
\providecommand \bibfield  [0]{\@secondoftwo}%
\providecommand \translation [1]{[#1]}%
\providecommand \BibitemOpen [0]{}%
\providecommand \bibitemStop [0]{}%
\providecommand \bibitemNoStop [0]{.\EOS\space}%
\providecommand \EOS [0]{\spacefactor3000\relax}%
\providecommand \BibitemShut  [1]{\csname bibitem#1\endcsname}%
\let\auto@bib@innerbib\@empty
\bibitem [{\citenamefont {Scully}\ and\ \citenamefont
  {Zubairy}(1997)}]{scully}%
  \BibitemOpen
  \bibfield  {author} {\bibinfo {author} {\bibfnamefont {M.~O.}\ \bibnamefont
  {Scully}}\ and\ \bibinfo {author} {\bibfnamefont {M.~S.}\ \bibnamefont
  {Zubairy}},\ }\href@noop {} {\emph {\bibinfo {title} {{Quantum Optics}}}}\
  (\bibinfo  {publisher} {Cambridge University Press},\ \bibinfo {address}
  {Cambridge, UK},\ \bibinfo {year} {1997})\BibitemShut {NoStop}%
\bibitem [{\citenamefont {Tinkham}(2004)}]{tinkham}%
  \BibitemOpen
  \bibfield  {author} {\bibinfo {author} {\bibfnamefont {M.}~\bibnamefont
  {Tinkham}},\ }\href@noop {} {\emph {\bibinfo {title} {{Introduction to
  Superconductivity}}}},\ \bibinfo {edition} {2nd}\ ed.\ (\bibinfo  {publisher}
  {Dover Publications},\ \bibinfo {address} {Mineola, NY},\ \bibinfo {year}
  {2004})\BibitemShut {NoStop}%
\bibitem [{\citenamefont {Andrews}\ \emph {et~al.}(1997)\citenamefont
  {Andrews}, \citenamefont {Townsend}, \citenamefont {Miesner}, \citenamefont
  {Durfee}, \citenamefont {Kurn},\ and\ \citenamefont
  {Ketterle}}]{andrews1997}%
  \BibitemOpen
  \bibfield  {author} {\bibinfo {author} {\bibfnamefont {M.~R.}\ \bibnamefont
  {Andrews}}, \bibinfo {author} {\bibfnamefont {C.~G.}\ \bibnamefont
  {Townsend}}, \bibinfo {author} {\bibfnamefont {H.-J.}\ \bibnamefont
  {Miesner}}, \bibinfo {author} {\bibfnamefont {D.~S.}\ \bibnamefont {Durfee}},
  \bibinfo {author} {\bibfnamefont {D.~M.}\ \bibnamefont {Kurn}},\ and\
  \bibinfo {author} {\bibfnamefont {W.}~\bibnamefont {Ketterle}},\ }\bibfield
  {title} {\bibinfo {title} {{Observation of Interference Between Two Bose
  Condensates}},\ }\href {https://doi.org/10.1126/science.275.5300.637}
  {\bibfield  {journal} {\bibinfo  {journal} {Science}\ }\textbf {\bibinfo
  {volume} {275}},\ \bibinfo {pages} {637} (\bibinfo {year}
  {1997})}\BibitemShut {NoStop}%
\bibitem [{\citenamefont {Altman}\ \emph {et~al.}(2004)\citenamefont {Altman},
  \citenamefont {Demler},\ and\ \citenamefont {Lukin}}]{altman2004}%
  \BibitemOpen
  \bibfield  {author} {\bibinfo {author} {\bibfnamefont {E.}~\bibnamefont
  {Altman}}, \bibinfo {author} {\bibfnamefont {E.}~\bibnamefont {Demler}},\
  and\ \bibinfo {author} {\bibfnamefont {M.~D.}\ \bibnamefont {Lukin}},\
  }\bibfield  {title} {\bibinfo {title} {{Probing many-body states of ultracold
  atoms via noise correlations}},\ }\href
  {https://doi.org/10.1103/PhysRevA.70.013603} {\bibfield  {journal} {\bibinfo
  {journal} {Physical Review A}\ }\textbf {\bibinfo {volume} {70}},\ \bibinfo
  {pages} {013603} (\bibinfo {year} {2004})}\BibitemShut {NoStop}%
\bibitem [{\citenamefont {Fölling}\ \emph {et~al.}(2005)\citenamefont
  {Fölling}, \citenamefont {Gerbier}, \citenamefont {Widera}, \citenamefont
  {Mandel}, \citenamefont {Gericke},\ and\ \citenamefont
  {Bloch}}]{folling2005}%
  \BibitemOpen
  \bibfield  {author} {\bibinfo {author} {\bibfnamefont {S.}~\bibnamefont
  {Fölling}}, \bibinfo {author} {\bibfnamefont {F.}~\bibnamefont {Gerbier}},
  \bibinfo {author} {\bibfnamefont {A.}~\bibnamefont {Widera}}, \bibinfo
  {author} {\bibfnamefont {O.}~\bibnamefont {Mandel}}, \bibinfo {author}
  {\bibfnamefont {T.}~\bibnamefont {Gericke}},\ and\ \bibinfo {author}
  {\bibfnamefont {I.}~\bibnamefont {Bloch}},\ }\bibfield  {title} {\bibinfo
  {title} {{Spatial quantum noise interferometry in expanding ultracold atom
  clouds}},\ }\href {https://doi.org/10.1038/nature03500} {\bibfield  {journal}
  {\bibinfo  {journal} {Nature}\ }\textbf {\bibinfo {volume} {434}},\ \bibinfo
  {pages} {481} (\bibinfo {year} {2005})}\BibitemShut {NoStop}%
\bibitem [{\citenamefont {Sørensen}\ \emph {et~al.}(1998)\citenamefont
  {Sørensen}, \citenamefont {Hald},\ and\ \citenamefont
  {Polzik}}]{sorensen1998}%
  \BibitemOpen
  \bibfield  {author} {\bibinfo {author} {\bibfnamefont {J.~L.}\ \bibnamefont
  {Sørensen}}, \bibinfo {author} {\bibfnamefont {J.}~\bibnamefont {Hald}},\
  and\ \bibinfo {author} {\bibfnamefont {E.~S.}\ \bibnamefont {Polzik}},\
  }\bibfield  {title} {\bibinfo {title} {{Quantum Noise of an Atomic Spin
  Polarization Measurement}},\ }\href
  {https://doi.org/10.1103/PhysRevLett.80.3487} {\bibfield  {journal} {\bibinfo
   {journal} {Physical Review Letters}\ }\textbf {\bibinfo {volume} {80}},\
  \bibinfo {pages} {3487} (\bibinfo {year} {1998})}\BibitemShut {NoStop}%
\bibitem [{\citenamefont {Crooker}\ \emph {et~al.}(2004)\citenamefont
  {Crooker}, \citenamefont {Rickel}, \citenamefont {Balatsky},\ and\
  \citenamefont {Smith}}]{crooker2004}%
  \BibitemOpen
  \bibfield  {author} {\bibinfo {author} {\bibfnamefont {S.~A.}\ \bibnamefont
  {Crooker}}, \bibinfo {author} {\bibfnamefont {D.~G.}\ \bibnamefont {Rickel}},
  \bibinfo {author} {\bibfnamefont {A.~V.}\ \bibnamefont {Balatsky}},\ and\
  \bibinfo {author} {\bibfnamefont {D.~L.}\ \bibnamefont {Smith}},\ }\bibfield
  {title} {\bibinfo {title} {{Spectroscopy of spontaneous spin noise as a probe
  of spin dynamics and magnetic resonance}},\ }\href
  {https://doi.org/10.1038/nature02804} {\bibfield  {journal} {\bibinfo
  {journal} {Nature}\ }\textbf {\bibinfo {volume} {431}},\ \bibinfo {pages}
  {49} (\bibinfo {year} {2004})}\BibitemShut {NoStop}%
\bibitem [{\citenamefont {Simon}\ \emph {et~al.}(2011)\citenamefont {Simon},
  \citenamefont {Bakr}, \citenamefont {Ma}, \citenamefont {Tai}, \citenamefont
  {Preiss},\ and\ \citenamefont {Greiner}}]{simon2011}%
  \BibitemOpen
  \bibfield  {author} {\bibinfo {author} {\bibfnamefont {J.}~\bibnamefont
  {Simon}}, \bibinfo {author} {\bibfnamefont {W.~S.}\ \bibnamefont {Bakr}},
  \bibinfo {author} {\bibfnamefont {R.}~\bibnamefont {Ma}}, \bibinfo {author}
  {\bibfnamefont {M.~E.}\ \bibnamefont {Tai}}, \bibinfo {author} {\bibfnamefont
  {P.~M.}\ \bibnamefont {Preiss}},\ and\ \bibinfo {author} {\bibfnamefont
  {M.}~\bibnamefont {Greiner}},\ }\bibfield  {title} {\bibinfo {title}
  {{Quantum Simulation of Antiferromagnetic Spin Chains in an Optical
  Lattice}},\ }\href {https://doi.org/10.1038/nature09994} {\bibfield
  {journal} {\bibinfo  {journal} {Nature}\ }\textbf {\bibinfo {volume} {472}},\
  \bibinfo {pages} {307} (\bibinfo {year} {2011})}\BibitemShut {NoStop}%
\bibitem [{\citenamefont {Fukuhara}\ \emph {et~al.}(2013)\citenamefont
  {Fukuhara}, \citenamefont {Schau\ss}, \citenamefont {Endres}, \citenamefont
  {Hild}, \citenamefont {Cheneau}, \citenamefont {Bloch},\ and\ \citenamefont
  {Gross}}]{fukuhara2013}%
  \BibitemOpen
  \bibfield  {author} {\bibinfo {author} {\bibfnamefont {T.}~\bibnamefont
  {Fukuhara}}, \bibinfo {author} {\bibfnamefont {P.}~\bibnamefont {Schau\ss}},
  \bibinfo {author} {\bibfnamefont {M.}~\bibnamefont {Endres}}, \bibinfo
  {author} {\bibfnamefont {S.}~\bibnamefont {Hild}}, \bibinfo {author}
  {\bibfnamefont {M.}~\bibnamefont {Cheneau}}, \bibinfo {author} {\bibfnamefont
  {I.}~\bibnamefont {Bloch}},\ and\ \bibinfo {author} {\bibfnamefont
  {C.}~\bibnamefont {Gross}},\ }\bibfield  {title} {\bibinfo {title}
  {{Microscopic Observation of Magnon Bound States and Their Dynamics}},\
  }\href {https://doi.org/10.1038/nature12541} {\bibfield  {journal} {\bibinfo
  {journal} {Nature}\ }\textbf {\bibinfo {volume} {502}},\ \bibinfo {pages}
  {76} (\bibinfo {year} {2013})}\BibitemShut {NoStop}%
\bibitem [{\citenamefont {Parsons}\ \emph {et~al.}(2016)\citenamefont
  {Parsons}, \citenamefont {Mazurenko}, \citenamefont {Chiu}, \citenamefont
  {Ji}, \citenamefont {Greif},\ and\ \citenamefont {Greiner}}]{parsons2016}%
  \BibitemOpen
  \bibfield  {author} {\bibinfo {author} {\bibfnamefont {M.~F.}\ \bibnamefont
  {Parsons}}, \bibinfo {author} {\bibfnamefont {A.}~\bibnamefont {Mazurenko}},
  \bibinfo {author} {\bibfnamefont {C.~S.}\ \bibnamefont {Chiu}}, \bibinfo
  {author} {\bibfnamefont {G.}~\bibnamefont {Ji}}, \bibinfo {author}
  {\bibfnamefont {D.}~\bibnamefont {Greif}},\ and\ \bibinfo {author}
  {\bibfnamefont {M.}~\bibnamefont {Greiner}},\ }\bibfield  {title} {\bibinfo
  {title} {{Site-resolved measurement of the spin-correlation function in the
  Fermi-Hubbard model}},\ }\href {https://doi.org/10.1126/science.aag1430}
  {\bibfield  {journal} {\bibinfo  {journal} {Science}\ }\textbf {\bibinfo
  {volume} {353}},\ \bibinfo {pages} {1253} (\bibinfo {year}
  {2016})}\BibitemShut {NoStop}%
\bibitem [{\citenamefont {Manovitz}\ \emph {et~al.}(2025)\citenamefont
  {Manovitz}, \citenamefont {Li}, \citenamefont {Ebadi}, \citenamefont
  {Samajdar}, \citenamefont {Geim}, \citenamefont {Evered}, \citenamefont
  {Bluvstein}, \citenamefont {Zhou}, \citenamefont {K{\"o}yl{\"u}o{\u{g}}lu},
  \citenamefont {Feldmeier}, \citenamefont {Dolgirev}, \citenamefont {Maskara},
  \citenamefont {Kalinowski}, \citenamefont {Sachdev}, \citenamefont {Huse},
  \citenamefont {Greiner}, \citenamefont {Vuletić},\ and\ \citenamefont
  {Lukin}}]{manovitz2025}%
  \BibitemOpen
  \bibfield  {author} {\bibinfo {author} {\bibfnamefont {T.}~\bibnamefont
  {Manovitz}}, \bibinfo {author} {\bibfnamefont {S.~H.}\ \bibnamefont {Li}},
  \bibinfo {author} {\bibfnamefont {S.}~\bibnamefont {Ebadi}}, \bibinfo
  {author} {\bibfnamefont {R.}~\bibnamefont {Samajdar}}, \bibinfo {author}
  {\bibfnamefont {A.~A.}\ \bibnamefont {Geim}}, \bibinfo {author}
  {\bibfnamefont {S.~J.}\ \bibnamefont {Evered}}, \bibinfo {author}
  {\bibfnamefont {D.}~\bibnamefont {Bluvstein}}, \bibinfo {author}
  {\bibfnamefont {H.}~\bibnamefont {Zhou}}, \bibinfo {author} {\bibfnamefont
  {N.~U.}\ \bibnamefont {K{\"o}yl{\"u}o{\u{g}}lu}}, \bibinfo {author}
  {\bibfnamefont {J.}~\bibnamefont {Feldmeier}}, \bibinfo {author}
  {\bibfnamefont {P.~E.}\ \bibnamefont {Dolgirev}}, \bibinfo {author}
  {\bibfnamefont {N.}~\bibnamefont {Maskara}}, \bibinfo {author} {\bibfnamefont
  {M.}~\bibnamefont {Kalinowski}}, \bibinfo {author} {\bibfnamefont
  {S.}~\bibnamefont {Sachdev}}, \bibinfo {author} {\bibfnamefont {D.~A.}\
  \bibnamefont {Huse}}, \bibinfo {author} {\bibfnamefont {M.}~\bibnamefont
  {Greiner}}, \bibinfo {author} {\bibfnamefont {V.}~\bibnamefont {Vuletić}},\
  and\ \bibinfo {author} {\bibfnamefont {M.~D.}\ \bibnamefont {Lukin}},\
  }\bibfield  {title} {\bibinfo {title} {{Quantum coarsening and collective
  dynamics on a programmable simulator}},\ }\href
  {https://doi.org/10.1038/s41586-024-08353-5} {\bibfield  {journal} {\bibinfo
  {journal} {Nature}\ }\textbf {\bibinfo {volume} {638}},\ \bibinfo {pages}
  {86} (\bibinfo {year} {2025})}\BibitemShut {NoStop}%
\bibitem [{\citenamefont {Bernaschi}\ \emph {et~al.}(2024)\citenamefont
  {Bernaschi}, \citenamefont {González-Adalid~Pemartín}, \citenamefont
  {Martín-Mayor},\ and\ \citenamefont {Parisi}}]{bernaschi2024}%
  \BibitemOpen
  \bibfield  {author} {\bibinfo {author} {\bibfnamefont {M.}~\bibnamefont
  {Bernaschi}}, \bibinfo {author} {\bibfnamefont {I.}~\bibnamefont
  {González-Adalid~Pemartín}}, \bibinfo {author} {\bibfnamefont
  {V.}~\bibnamefont {Martín-Mayor}},\ and\ \bibinfo {author} {\bibfnamefont
  {G.}~\bibnamefont {Parisi}},\ }\bibfield  {title} {\bibinfo {title} {{The
  quantum transition of the two-dimensional Ising spin glass}},\ }\href
  {https://doi.org/10.1038/s41586-024-07647-y} {\bibfield  {journal} {\bibinfo
  {journal} {Nature}\ }\textbf {\bibinfo {volume} {631}},\ \bibinfo {pages}
  {749} (\bibinfo {year} {2024})}\BibitemShut {NoStop}%
\bibitem [{\citenamefont {Marshall}\ and\ \citenamefont
  {Lowde}(1968)}]{marshall1968}%
  \BibitemOpen
  \bibfield  {author} {\bibinfo {author} {\bibfnamefont {W.}~\bibnamefont
  {Marshall}}\ and\ \bibinfo {author} {\bibfnamefont {R.~D.}\ \bibnamefont
  {Lowde}},\ }\bibfield  {title} {\bibinfo {title} {{Magnetic Correlations and
  Neutron Scattering}},\ }\href {https://doi.org/10.1088/0034-4885/31/2/305}
  {\bibfield  {journal} {\bibinfo  {journal} {Reports on Progress in Physics}\
  }\textbf {\bibinfo {volume} {31}},\ \bibinfo {pages} {705} (\bibinfo {year}
  {1968})}\BibitemShut {NoStop}%
\bibitem [{\citenamefont {Endoh}\ \emph {et~al.}(1988)\citenamefont {Endoh},
  \citenamefont {Yamada}, \citenamefont {Birgeneau}, \citenamefont {Gabbe},
  \citenamefont {Jenssen}, \citenamefont {Kastner}, \citenamefont {Peters},
  \citenamefont {Picone}, \citenamefont {Thurston}, \citenamefont {Tranquada},
  \citenamefont {Shirane}, \citenamefont {Hidaka}, \citenamefont {Oda},
  \citenamefont {Enomoto}, \citenamefont {Suzuki},\ and\ \citenamefont
  {Murakami}}]{endoh1988}%
  \BibitemOpen
  \bibfield  {author} {\bibinfo {author} {\bibfnamefont {Y.}~\bibnamefont
  {Endoh}}, \bibinfo {author} {\bibfnamefont {K.}~\bibnamefont {Yamada}},
  \bibinfo {author} {\bibfnamefont {R.~J.}\ \bibnamefont {Birgeneau}}, \bibinfo
  {author} {\bibfnamefont {D.~R.}\ \bibnamefont {Gabbe}}, \bibinfo {author}
  {\bibfnamefont {H.~P.}\ \bibnamefont {Jenssen}}, \bibinfo {author}
  {\bibfnamefont {M.~A.}\ \bibnamefont {Kastner}}, \bibinfo {author}
  {\bibfnamefont {C.~J.}\ \bibnamefont {Peters}}, \bibinfo {author}
  {\bibfnamefont {P.~J.}\ \bibnamefont {Picone}}, \bibinfo {author}
  {\bibfnamefont {T.~R.}\ \bibnamefont {Thurston}}, \bibinfo {author}
  {\bibfnamefont {J.~M.}\ \bibnamefont {Tranquada}}, \bibinfo {author}
  {\bibfnamefont {G.}~\bibnamefont {Shirane}}, \bibinfo {author} {\bibfnamefont
  {Y.}~\bibnamefont {Hidaka}}, \bibinfo {author} {\bibfnamefont
  {M.}~\bibnamefont {Oda}}, \bibinfo {author} {\bibfnamefont {Y.}~\bibnamefont
  {Enomoto}}, \bibinfo {author} {\bibfnamefont {M.}~\bibnamefont {Suzuki}},\
  and\ \bibinfo {author} {\bibfnamefont {T.}~\bibnamefont {Murakami}},\
  }\bibfield  {title} {\bibinfo {title} {{Static and dynamic spin correlations
  in pure and doped La\textsubscript{2}CuO\textsubscript{4}}},\ }\href
  {https://doi.org/10.1103/PhysRevB.37.7443} {\bibfield  {journal} {\bibinfo
  {journal} {Physical Review B}\ }\textbf {\bibinfo {volume} {37}},\ \bibinfo
  {pages} {7443} (\bibinfo {year} {1988})}\BibitemShut {NoStop}%
\bibitem [{\citenamefont {Han}\ \emph {et~al.}(2012)\citenamefont {Han},
  \citenamefont {Helton}, \citenamefont {Chu}, \citenamefont {Nocera},
  \citenamefont {Rodriguez-Rivera}, \citenamefont {Broholm},\ and\
  \citenamefont {Lee}}]{tianhenghan2012}%
  \BibitemOpen
  \bibfield  {author} {\bibinfo {author} {\bibfnamefont {T.-H.}\ \bibnamefont
  {Han}}, \bibinfo {author} {\bibfnamefont {J.~S.}\ \bibnamefont {Helton}},
  \bibinfo {author} {\bibfnamefont {S.}~\bibnamefont {Chu}}, \bibinfo {author}
  {\bibfnamefont {D.~G.}\ \bibnamefont {Nocera}}, \bibinfo {author}
  {\bibfnamefont {J.~A.}\ \bibnamefont {Rodriguez-Rivera}}, \bibinfo {author}
  {\bibfnamefont {C.}~\bibnamefont {Broholm}},\ and\ \bibinfo {author}
  {\bibfnamefont {Y.~S.}\ \bibnamefont {Lee}},\ }\bibfield  {title} {\bibinfo
  {title} {{Fractionalized excitations in the spin-liquid state of a
  kagome-lattice antiferromagnet}},\ }\href
  {https://doi.org/10.1038/nature11659} {\bibfield  {journal} {\bibinfo
  {journal} {Nature}\ }\textbf {\bibinfo {volume} {492}},\ \bibinfo {pages}
  {406} (\bibinfo {year} {2012})}\BibitemShut {NoStop}%
\bibitem [{\citenamefont {Rybakov}\ \emph {et~al.}(2024)\citenamefont
  {Rybakov}, \citenamefont {Boix-Constant}, \citenamefont {Alba~Venero},
  \citenamefont {van~der Zant}, \citenamefont {Mañas-Valero},\ and\
  \citenamefont {Coronado}}]{rybakov2024}%
  \BibitemOpen
  \bibfield  {author} {\bibinfo {author} {\bibfnamefont {A.}~\bibnamefont
  {Rybakov}}, \bibinfo {author} {\bibfnamefont {C.}~\bibnamefont
  {Boix-Constant}}, \bibinfo {author} {\bibfnamefont {D.}~\bibnamefont
  {Alba~Venero}}, \bibinfo {author} {\bibfnamefont {H.~S.~J.}\ \bibnamefont
  {van~der Zant}}, \bibinfo {author} {\bibfnamefont {S.}~\bibnamefont
  {Mañas-Valero}},\ and\ \bibinfo {author} {\bibfnamefont {E.}~\bibnamefont
  {Coronado}},\ }\bibfield  {title} {\bibinfo {title} {{Probing Short-Range
  Correlations in the van der Waals Magnet CrSBr by Small-Angle Neutron
  Scattering}},\ }\href {https://doi.org/10.1002/smsc.202400244} {\bibfield
  {journal} {\bibinfo  {journal} {Small Science}\ }\textbf {\bibinfo {volume}
  {4}},\ \bibinfo {pages} {2400244} (\bibinfo {year} {2024})}\BibitemShut
  {NoStop}%
\bibitem [{\citenamefont {Ishiguro}\ \emph {et~al.}(2013)\citenamefont
  {Ishiguro}, \citenamefont {Kimura}, \citenamefont {Nakatsuji}, \citenamefont
  {Tsutsui}, \citenamefont {Baron}, \citenamefont {Kimura},\ and\ \citenamefont
  {Wakabayashi}}]{ishiguro2013}%
  \BibitemOpen
  \bibfield  {author} {\bibinfo {author} {\bibfnamefont {Y.}~\bibnamefont
  {Ishiguro}}, \bibinfo {author} {\bibfnamefont {K.}~\bibnamefont {Kimura}},
  \bibinfo {author} {\bibfnamefont {S.}~\bibnamefont {Nakatsuji}}, \bibinfo
  {author} {\bibfnamefont {S.}~\bibnamefont {Tsutsui}}, \bibinfo {author}
  {\bibfnamefont {A.~Q.~R.}\ \bibnamefont {Baron}}, \bibinfo {author}
  {\bibfnamefont {T.}~\bibnamefont {Kimura}},\ and\ \bibinfo {author}
  {\bibfnamefont {Y.}~\bibnamefont {Wakabayashi}},\ }\bibfield  {title}
  {\bibinfo {title} {{Dynamical spin–orbital correlation in the frustrated
  magnet Ba\textsubscript{3}CuSb\textsubscript{2}O\textsubscript{9}}},\ }\href
  {https://doi.org/10.1038/ncomms3022} {\bibfield  {journal} {\bibinfo
  {journal} {Nature Communications}\ }\textbf {\bibinfo {volume} {4}},\
  \bibinfo {pages} {2022} (\bibinfo {year} {2013})}\BibitemShut {NoStop}%
\bibitem [{\citenamefont {Callen}\ and\ \citenamefont
  {Callen}(1965)}]{callen1965}%
  \BibitemOpen
  \bibfield  {author} {\bibinfo {author} {\bibfnamefont {E.}~\bibnamefont
  {Callen}}\ and\ \bibinfo {author} {\bibfnamefont {H.~B.}\ \bibnamefont
  {Callen}},\ }\bibfield  {title} {\bibinfo {title} {{Magnetostriction, Forced
  Magnetostriction, and Anomalous Thermal Expansion in Ferromagnets}},\ }\href
  {https://doi.org/10.1103/PhysRev.139.A455} {\bibfield  {journal} {\bibinfo
  {journal} {Physical Review}\ }\textbf {\bibinfo {volume} {139}},\ \bibinfo
  {pages} {A455} (\bibinfo {year} {1965})}\BibitemShut {NoStop}%
\bibitem [{\citenamefont {Zapf}\ \emph {et~al.}(2008)\citenamefont {Zapf},
  \citenamefont {Correa}, \citenamefont {Sengupta}, \citenamefont {Batista},
  \citenamefont {Tsukamoto}, \citenamefont {Kawashima}, \citenamefont {Egan},
  \citenamefont {Pantea}, \citenamefont {Migliori}, \citenamefont {Betts},
  \citenamefont {Jaime},\ and\ \citenamefont {Paduan-Filho}}]{zapf2008}%
  \BibitemOpen
  \bibfield  {author} {\bibinfo {author} {\bibfnamefont {V.~S.}\ \bibnamefont
  {Zapf}}, \bibinfo {author} {\bibfnamefont {V.~F.}\ \bibnamefont {Correa}},
  \bibinfo {author} {\bibfnamefont {P.}~\bibnamefont {Sengupta}}, \bibinfo
  {author} {\bibfnamefont {C.~D.}\ \bibnamefont {Batista}}, \bibinfo {author}
  {\bibfnamefont {M.}~\bibnamefont {Tsukamoto}}, \bibinfo {author}
  {\bibfnamefont {N.}~\bibnamefont {Kawashima}}, \bibinfo {author}
  {\bibfnamefont {P.}~\bibnamefont {Egan}}, \bibinfo {author} {\bibfnamefont
  {C.}~\bibnamefont {Pantea}}, \bibinfo {author} {\bibfnamefont
  {A.}~\bibnamefont {Migliori}}, \bibinfo {author} {\bibfnamefont {J.~B.}\
  \bibnamefont {Betts}}, \bibinfo {author} {\bibfnamefont {M.}~\bibnamefont
  {Jaime}},\ and\ \bibinfo {author} {\bibfnamefont {A.}~\bibnamefont
  {Paduan-Filho}},\ }\bibfield  {title} {\bibinfo {title} {{Direct measurement
  of spin correlations using magnetostriction}},\ }\href
  {https://doi.org/10.1103/PhysRevB.77.020404} {\bibfield  {journal} {\bibinfo
  {journal} {Physical Review B}\ }\textbf {\bibinfo {volume} {77}},\ \bibinfo
  {pages} {020404} (\bibinfo {year} {2008})}\BibitemShut {NoStop}%
\bibitem [{\citenamefont {Imai}\ \emph {et~al.}(1993)\citenamefont {Imai},
  \citenamefont {Slichter}, \citenamefont {Yoshimura},\ and\ \citenamefont
  {Kosuge}}]{imai1993}%
  \BibitemOpen
  \bibfield  {author} {\bibinfo {author} {\bibfnamefont {T.}~\bibnamefont
  {Imai}}, \bibinfo {author} {\bibfnamefont {C.~P.}\ \bibnamefont {Slichter}},
  \bibinfo {author} {\bibfnamefont {K.}~\bibnamefont {Yoshimura}},\ and\
  \bibinfo {author} {\bibfnamefont {K.}~\bibnamefont {Kosuge}},\ }\bibfield
  {title} {\bibinfo {title} {{Low frequency spin dynamics in undoped and
  Sr-doped La\textsubscript{2}CuO\textsubscript{4}}},\ }\href
  {https://doi.org/10.1103/PhysRevLett.70.1002} {\bibfield  {journal} {\bibinfo
   {journal} {Physical Review Letters}\ }\textbf {\bibinfo {volume} {70}},\
  \bibinfo {pages} {1002} (\bibinfo {year} {1993})}\BibitemShut {NoStop}%
\bibitem [{\citenamefont {Carretta}\ \emph {et~al.}(2000)\citenamefont
  {Carretta}, \citenamefont {Ciabattoni}, \citenamefont {Cuccoli},
  \citenamefont {Mognaschi}, \citenamefont {Rigamonti}, \citenamefont
  {Tognetti},\ and\ \citenamefont {Verrucchi}}]{carretta2000}%
  \BibitemOpen
  \bibfield  {author} {\bibinfo {author} {\bibfnamefont {P.}~\bibnamefont
  {Carretta}}, \bibinfo {author} {\bibfnamefont {T.}~\bibnamefont
  {Ciabattoni}}, \bibinfo {author} {\bibfnamefont {A.}~\bibnamefont {Cuccoli}},
  \bibinfo {author} {\bibfnamefont {E.}~\bibnamefont {Mognaschi}}, \bibinfo
  {author} {\bibfnamefont {A.}~\bibnamefont {Rigamonti}}, \bibinfo {author}
  {\bibfnamefont {V.}~\bibnamefont {Tognetti}},\ and\ \bibinfo {author}
  {\bibfnamefont {P.}~\bibnamefont {Verrucchi}},\ }\bibfield  {title} {\bibinfo
  {title} {{Spin Dynamics and Magnetic Correlation Length in Two-Dimensional
  Quantum Heisenberg Antiferromagnets}},\ }\href
  {https://doi.org/10.1103/PhysRevLett.84.366} {\bibfield  {journal} {\bibinfo
  {journal} {Physical Review Letters}\ }\textbf {\bibinfo {volume} {84}},\
  \bibinfo {pages} {366} (\bibinfo {year} {2000})}\BibitemShut {NoStop}%
\bibitem [{\citenamefont {Kitagawa}\ \emph {et~al.}(2018)\citenamefont
  {Kitagawa}, \citenamefont {Takayama}, \citenamefont {Matsumoto},
  \citenamefont {Kato}, \citenamefont {Takano}, \citenamefont {Kishimoto},
  \citenamefont {Bette}, \citenamefont {Dinnebier}, \citenamefont {Jackeli},\
  and\ \citenamefont {Takagi}}]{kitagawa2018}%
  \BibitemOpen
  \bibfield  {author} {\bibinfo {author} {\bibfnamefont {K.}~\bibnamefont
  {Kitagawa}}, \bibinfo {author} {\bibfnamefont {T.}~\bibnamefont {Takayama}},
  \bibinfo {author} {\bibfnamefont {Y.}~\bibnamefont {Matsumoto}}, \bibinfo
  {author} {\bibfnamefont {A.}~\bibnamefont {Kato}}, \bibinfo {author}
  {\bibfnamefont {R.}~\bibnamefont {Takano}}, \bibinfo {author} {\bibfnamefont
  {Y.}~\bibnamefont {Kishimoto}}, \bibinfo {author} {\bibfnamefont
  {S.}~\bibnamefont {Bette}}, \bibinfo {author} {\bibfnamefont
  {R.}~\bibnamefont {Dinnebier}}, \bibinfo {author} {\bibfnamefont
  {G.}~\bibnamefont {Jackeli}},\ and\ \bibinfo {author} {\bibfnamefont
  {H.}~\bibnamefont {Takagi}},\ }\bibfield  {title} {\bibinfo {title} {{A
  spin--orbital-entangled quantum liquid on a honeycomb lattice}},\ }\href
  {https://doi.org/10.1038/nature25482} {\bibfield  {journal} {\bibinfo
  {journal} {Nature}\ }\textbf {\bibinfo {volume} {554}},\ \bibinfo {pages}
  {341} (\bibinfo {year} {2018})}\BibitemShut {NoStop}%
\bibitem [{\citenamefont {Haines}\ \emph {et~al.}(1985)\citenamefont {Haines},
  \citenamefont {Clauberg},\ and\ \citenamefont {Feder}}]{haines1985}%
  \BibitemOpen
  \bibfield  {author} {\bibinfo {author} {\bibfnamefont {E.~M.}\ \bibnamefont
  {Haines}}, \bibinfo {author} {\bibfnamefont {R.}~\bibnamefont {Clauberg}},\
  and\ \bibinfo {author} {\bibfnamefont {R.}~\bibnamefont {Feder}},\ }\bibfield
   {title} {\bibinfo {title} {{Short-range magnetic order near the Curie
  temperature iron from spin-resolved photoemission}},\ }\href
  {https://doi.org/10.1103/PhysRevLett.54.932} {\bibfield  {journal} {\bibinfo
  {journal} {Physical Review Letters}\ }\textbf {\bibinfo {volume} {54}},\
  \bibinfo {pages} {932} (\bibinfo {year} {1985})}\BibitemShut {NoStop}%
\bibitem [{\citenamefont {Ron}\ \emph {et~al.}(2019)\citenamefont {Ron},
  \citenamefont {Zoghlin}, \citenamefont {Balents}, \citenamefont {Wilson},\
  and\ \citenamefont {Hsieh}}]{ron2019}%
  \BibitemOpen
  \bibfield  {author} {\bibinfo {author} {\bibfnamefont {A.}~\bibnamefont
  {Ron}}, \bibinfo {author} {\bibfnamefont {E.}~\bibnamefont {Zoghlin}},
  \bibinfo {author} {\bibfnamefont {L.}~\bibnamefont {Balents}}, \bibinfo
  {author} {\bibfnamefont {S.~D.}\ \bibnamefont {Wilson}},\ and\ \bibinfo
  {author} {\bibfnamefont {D.}~\bibnamefont {Hsieh}},\ }\bibfield  {title}
  {\bibinfo {title} {{Dimensional crossover in a layered ferromagnet detected
  by spin correlation driven distortions}},\ }\href
  {https://doi.org/10.1038/s41467-019-09663-3} {\bibfield  {journal} {\bibinfo
  {journal} {Nature Communications}\ }\textbf {\bibinfo {volume} {10}},\
  \bibinfo {pages} {1654} (\bibinfo {year} {2019})}\BibitemShut {NoStop}%
\bibitem [{\citenamefont {Back}\ \emph {et~al.}(1995)\citenamefont {Back},
  \citenamefont {Würsch}, \citenamefont {Vaterlaus}, \citenamefont
  {Ramsperger}, \citenamefont {Maier},\ and\ \citenamefont
  {Pescia}}]{back1995}%
  \BibitemOpen
  \bibfield  {author} {\bibinfo {author} {\bibfnamefont {C.~H.}\ \bibnamefont
  {Back}}, \bibinfo {author} {\bibfnamefont {C.}~\bibnamefont {Würsch}},
  \bibinfo {author} {\bibfnamefont {A.}~\bibnamefont {Vaterlaus}}, \bibinfo
  {author} {\bibfnamefont {U.}~\bibnamefont {Ramsperger}}, \bibinfo {author}
  {\bibfnamefont {U.}~\bibnamefont {Maier}},\ and\ \bibinfo {author}
  {\bibfnamefont {D.}~\bibnamefont {Pescia}},\ }\bibfield  {title} {\bibinfo
  {title} {{Experimental confirmation of universality for a phase transition in
  two dimensions}},\ }\href {https://doi.org/10.1038/378597a0} {\bibfield
  {journal} {\bibinfo  {journal} {Nature}\ }\textbf {\bibinfo {volume} {378}},\
  \bibinfo {pages} {597} (\bibinfo {year} {1995})}\BibitemShut {NoStop}%
\bibitem [{\citenamefont {Jin}\ \emph {et~al.}(2020)\citenamefont {Jin},
  \citenamefont {Tao}, \citenamefont {Kang}, \citenamefont {Watanabe},
  \citenamefont {Taniguchi}, \citenamefont {Mak},\ and\ \citenamefont
  {Shan}}]{chenhaojin2020}%
  \BibitemOpen
  \bibfield  {author} {\bibinfo {author} {\bibfnamefont {C.}~\bibnamefont
  {Jin}}, \bibinfo {author} {\bibfnamefont {Z.}~\bibnamefont {Tao}}, \bibinfo
  {author} {\bibfnamefont {K.}~\bibnamefont {Kang}}, \bibinfo {author}
  {\bibfnamefont {K.}~\bibnamefont {Watanabe}}, \bibinfo {author}
  {\bibfnamefont {T.}~\bibnamefont {Taniguchi}}, \bibinfo {author}
  {\bibfnamefont {K.~F.}\ \bibnamefont {Mak}},\ and\ \bibinfo {author}
  {\bibfnamefont {J.}~\bibnamefont {Shan}},\ }\bibfield  {title} {\bibinfo
  {title} {{Imaging and control of critical fluctuations in two-dimensional
  magnets}},\ }\href {https://doi.org/10.1038/s41563-020-0706-8} {\bibfield
  {journal} {\bibinfo  {journal} {Nature Materials}\ }\textbf {\bibinfo
  {volume} {19}},\ \bibinfo {pages} {1290} (\bibinfo {year}
  {2020})}\BibitemShut {NoStop}%
\bibitem [{\citenamefont {Yang}\ \emph {et~al.}(2014)\citenamefont {Yang},
  \citenamefont {Glasenapp}, \citenamefont {Greilich}, \citenamefont {Reuter},
  \citenamefont {Wieck}, \citenamefont {Yakovlev}, \citenamefont {Bayer},\ and\
  \citenamefont {Crooker}}]{luyiyang2014}%
  \BibitemOpen
  \bibfield  {author} {\bibinfo {author} {\bibfnamefont {L.}~\bibnamefont
  {Yang}}, \bibinfo {author} {\bibfnamefont {P.}~\bibnamefont {Glasenapp}},
  \bibinfo {author} {\bibfnamefont {A.}~\bibnamefont {Greilich}}, \bibinfo
  {author} {\bibfnamefont {D.}~\bibnamefont {Reuter}}, \bibinfo {author}
  {\bibfnamefont {A.~D.}\ \bibnamefont {Wieck}}, \bibinfo {author}
  {\bibfnamefont {D.~R.}\ \bibnamefont {Yakovlev}}, \bibinfo {author}
  {\bibfnamefont {M.}~\bibnamefont {Bayer}},\ and\ \bibinfo {author}
  {\bibfnamefont {S.~A.}\ \bibnamefont {Crooker}},\ }\bibfield  {title}
  {\bibinfo {title} {{Two-colour spin noise spectroscopy and fluctuation
  correlations reveal homogeneous linewidths within quantum-dot ensembles}},\
  }\href {https://doi.org/10.1038/ncomms5949} {\bibfield  {journal} {\bibinfo
  {journal} {Nature Communications}\ }\textbf {\bibinfo {volume} {5}},\
  \bibinfo {pages} {4949} (\bibinfo {year} {2014})}\BibitemShut {NoStop}%
\bibitem [{\citenamefont {De~Filippis}\ \emph {et~al.}(2012)\citenamefont
  {De~Filippis}, \citenamefont {Cataudella}, \citenamefont {Nowadnick},
  \citenamefont {Devereaux}, \citenamefont {Mishchenko},\ and\ \citenamefont
  {Nagaosa}}]{defilippis2012}%
  \BibitemOpen
  \bibfield  {author} {\bibinfo {author} {\bibfnamefont {G.}~\bibnamefont
  {De~Filippis}}, \bibinfo {author} {\bibfnamefont {V.}~\bibnamefont
  {Cataudella}}, \bibinfo {author} {\bibfnamefont {E.~A.}\ \bibnamefont
  {Nowadnick}}, \bibinfo {author} {\bibfnamefont {T.~P.}\ \bibnamefont
  {Devereaux}}, \bibinfo {author} {\bibfnamefont {A.~S.}\ \bibnamefont
  {Mishchenko}},\ and\ \bibinfo {author} {\bibfnamefont {N.}~\bibnamefont
  {Nagaosa}},\ }\bibfield  {title} {\bibinfo {title} {{Quantum Dynamics of the
  Hubbard-Holstein Model in Equilibrium and Nonequilibrium: Application to
  Pump-Probe Phenomena}},\ }\href
  {https://doi.org/10.1103/PhysRevLett.109.176402} {\bibfield  {journal}
  {\bibinfo  {journal} {Physical Review Letters}\ }\textbf {\bibinfo {volume}
  {109}},\ \bibinfo {pages} {176402} (\bibinfo {year} {2012})}\BibitemShut
  {NoStop}%
\bibitem [{\citenamefont {Sriram}\ and\ \citenamefont
  {Claassen}(2022)}]{sriram2022}%
  \BibitemOpen
  \bibfield  {author} {\bibinfo {author} {\bibfnamefont {A.}~\bibnamefont
  {Sriram}}\ and\ \bibinfo {author} {\bibfnamefont {M.}~\bibnamefont
  {Claassen}},\ }\bibfield  {title} {\bibinfo {title} {{Light-induced control
  of magnetic phases in Kitaev quantum magnets}},\ }\href
  {https://doi.org/10.1103/PhysRevResearch.4.L032036} {\bibfield  {journal}
  {\bibinfo  {journal} {Physical Review Research}\ }\textbf {\bibinfo {volume}
  {4}},\ \bibinfo {pages} {L032036} (\bibinfo {year} {2022})}\BibitemShut
  {NoStop}%
\bibitem [{\citenamefont {Wang}\ \emph
  {et~al.}(2022{\natexlab{a}})\citenamefont {Wang}, \citenamefont
  {Bedoya-Pinto}, \citenamefont {Blei}, \citenamefont {Dismukes}, \citenamefont
  {Hamo}, \citenamefont {Jenkins}, \citenamefont {Koperski}, \citenamefont
  {Liu}, \citenamefont {Sun}, \citenamefont {Telford}, \citenamefont {Kim},
  \citenamefont {Augustin}, \citenamefont {Vool}, \citenamefont {Yin},
  \citenamefont {Li}, \citenamefont {Falin}, \citenamefont {Dean},
  \citenamefont {Casanova}, \citenamefont {Evans}, \citenamefont {Chshiev},
  \citenamefont {Mishchenko}, \citenamefont {Petrovic}, \citenamefont {He},
  \citenamefont {Zhao}, \citenamefont {Tsen}, \citenamefont {Gerardot},
  \citenamefont {Brotons-Gisbert}, \citenamefont {Guguchia}, \citenamefont
  {Roy}, \citenamefont {Tongay}, \citenamefont {Wang}, \citenamefont {Hasan},
  \citenamefont {Wrachtrup}, \citenamefont {Yacoby}, \citenamefont {Fert},
  \citenamefont {Parkin}, \citenamefont {Novoselov}, \citenamefont {Dai},
  \citenamefont {Balicas},\ and\ \citenamefont {Santos}}]{qinghuawang2022}%
  \BibitemOpen
  \bibfield  {author} {\bibinfo {author} {\bibfnamefont {Q.~H.}\ \bibnamefont
  {Wang}}, \bibinfo {author} {\bibfnamefont {A.}~\bibnamefont {Bedoya-Pinto}},
  \bibinfo {author} {\bibfnamefont {M.}~\bibnamefont {Blei}}, \bibinfo {author}
  {\bibfnamefont {A.~H.}\ \bibnamefont {Dismukes}}, \bibinfo {author}
  {\bibfnamefont {A.}~\bibnamefont {Hamo}}, \bibinfo {author} {\bibfnamefont
  {S.}~\bibnamefont {Jenkins}}, \bibinfo {author} {\bibfnamefont
  {M.}~\bibnamefont {Koperski}}, \bibinfo {author} {\bibfnamefont
  {Y.}~\bibnamefont {Liu}}, \bibinfo {author} {\bibfnamefont {Q.-C.}\
  \bibnamefont {Sun}}, \bibinfo {author} {\bibfnamefont {E.~J.}\ \bibnamefont
  {Telford}}, \bibinfo {author} {\bibfnamefont {H.~H.}\ \bibnamefont {Kim}},
  \bibinfo {author} {\bibfnamefont {M.}~\bibnamefont {Augustin}}, \bibinfo
  {author} {\bibfnamefont {U.}~\bibnamefont {Vool}}, \bibinfo {author}
  {\bibfnamefont {J.-X.}\ \bibnamefont {Yin}}, \bibinfo {author} {\bibfnamefont
  {L.~H.}\ \bibnamefont {Li}}, \bibinfo {author} {\bibfnamefont
  {A.}~\bibnamefont {Falin}}, \bibinfo {author} {\bibfnamefont {C.~R.}\
  \bibnamefont {Dean}}, \bibinfo {author} {\bibfnamefont {F.}~\bibnamefont
  {Casanova}}, \bibinfo {author} {\bibfnamefont {R.~F.~L.}\ \bibnamefont
  {Evans}}, \bibinfo {author} {\bibfnamefont {M.}~\bibnamefont {Chshiev}},
  \bibinfo {author} {\bibfnamefont {A.}~\bibnamefont {Mishchenko}}, \bibinfo
  {author} {\bibfnamefont {C.}~\bibnamefont {Petrovic}}, \bibinfo {author}
  {\bibfnamefont {R.}~\bibnamefont {He}}, \bibinfo {author} {\bibfnamefont
  {L.}~\bibnamefont {Zhao}}, \bibinfo {author} {\bibfnamefont {A.~W.}\
  \bibnamefont {Tsen}}, \bibinfo {author} {\bibfnamefont {B.~D.}\ \bibnamefont
  {Gerardot}}, \bibinfo {author} {\bibfnamefont {M.}~\bibnamefont
  {Brotons-Gisbert}}, \bibinfo {author} {\bibfnamefont {Z.}~\bibnamefont
  {Guguchia}}, \bibinfo {author} {\bibfnamefont {X.}~\bibnamefont {Roy}},
  \bibinfo {author} {\bibfnamefont {S.}~\bibnamefont {Tongay}}, \bibinfo
  {author} {\bibfnamefont {Z.}~\bibnamefont {Wang}}, \bibinfo {author}
  {\bibfnamefont {M.~Z.}\ \bibnamefont {Hasan}}, \bibinfo {author}
  {\bibfnamefont {J.}~\bibnamefont {Wrachtrup}}, \bibinfo {author}
  {\bibfnamefont {A.}~\bibnamefont {Yacoby}}, \bibinfo {author} {\bibfnamefont
  {A.}~\bibnamefont {Fert}}, \bibinfo {author} {\bibfnamefont {S.}~\bibnamefont
  {Parkin}}, \bibinfo {author} {\bibfnamefont {K.~S.}\ \bibnamefont
  {Novoselov}}, \bibinfo {author} {\bibfnamefont {P.}~\bibnamefont {Dai}},
  \bibinfo {author} {\bibfnamefont {L.}~\bibnamefont {Balicas}},\ and\ \bibinfo
  {author} {\bibfnamefont {E.~J.~G.}\ \bibnamefont {Santos}},\ }\bibfield
  {title} {\bibinfo {title} {{The Magnetic Genome of Two-Dimensional van der
  Waals Materials}},\ }\href {https://doi.org/10.1021/acsnano.1c09150}
  {\bibfield  {journal} {\bibinfo  {journal} {ACS Nano}\ }\textbf {\bibinfo
  {volume} {16}},\ \bibinfo {pages} {6960} (\bibinfo {year}
  {2022}{\natexlab{a}})}\BibitemShut {NoStop}%
\bibitem [{\citenamefont {Ciorciaro}\ \emph {et~al.}(2023)\citenamefont
  {Ciorciaro}, \citenamefont {Smole{\'n}ski}, \citenamefont {Morera},
  \citenamefont {Kiper}, \citenamefont {Hiestand}, \citenamefont {Kroner},
  \citenamefont {Zhang}, \citenamefont {Watanabe}, \citenamefont {Taniguchi},
  \citenamefont {Demler} \emph {et~al.}}]{ciorciaro2023}%
  \BibitemOpen
  \bibfield  {author} {\bibinfo {author} {\bibfnamefont {L.}~\bibnamefont
  {Ciorciaro}}, \bibinfo {author} {\bibfnamefont {T.}~\bibnamefont
  {Smole{\'n}ski}}, \bibinfo {author} {\bibfnamefont {I.}~\bibnamefont
  {Morera}}, \bibinfo {author} {\bibfnamefont {N.}~\bibnamefont {Kiper}},
  \bibinfo {author} {\bibfnamefont {S.}~\bibnamefont {Hiestand}}, \bibinfo
  {author} {\bibfnamefont {M.}~\bibnamefont {Kroner}}, \bibinfo {author}
  {\bibfnamefont {Y.}~\bibnamefont {Zhang}}, \bibinfo {author} {\bibfnamefont
  {K.}~\bibnamefont {Watanabe}}, \bibinfo {author} {\bibfnamefont
  {T.}~\bibnamefont {Taniguchi}}, \bibinfo {author} {\bibfnamefont
  {E.}~\bibnamefont {Demler}}, \emph {et~al.},\ }\bibfield  {title} {\bibinfo
  {title} {Kinetic magnetism in triangular moir{\'e} materials},\ }\href
  {https://doi.org/10.1038/s41586-023-06633-0} {\bibfield  {journal} {\bibinfo
  {journal} {Nature}\ }\textbf {\bibinfo {volume} {623}},\ \bibinfo {pages}
  {509} (\bibinfo {year} {2023})}\BibitemShut {NoStop}%
\bibitem [{\citenamefont {Sung}\ \emph {et~al.}(2025)\citenamefont {Sung},
  \citenamefont {Wang}, \citenamefont {Esterlis}, \citenamefont {Volkov},
  \citenamefont {Scuri}, \citenamefont {Zhou}, \citenamefont {Brutschea},
  \citenamefont {Taniguchi}, \citenamefont {Watanabe}, \citenamefont {Yang},
  \citenamefont {Morales}, \citenamefont {Zhang}, \citenamefont {Millis},
  \citenamefont {Lukin}, \citenamefont {Kim}, \citenamefont {Demler},\ and\
  \citenamefont {Park}}]{jihosung2025}%
  \BibitemOpen
  \bibfield  {author} {\bibinfo {author} {\bibfnamefont {J.}~\bibnamefont
  {Sung}}, \bibinfo {author} {\bibfnamefont {J.}~\bibnamefont {Wang}}, \bibinfo
  {author} {\bibfnamefont {I.}~\bibnamefont {Esterlis}}, \bibinfo {author}
  {\bibfnamefont {P.~A.}\ \bibnamefont {Volkov}}, \bibinfo {author}
  {\bibfnamefont {G.}~\bibnamefont {Scuri}}, \bibinfo {author} {\bibfnamefont
  {Y.}~\bibnamefont {Zhou}}, \bibinfo {author} {\bibfnamefont {E.}~\bibnamefont
  {Brutschea}}, \bibinfo {author} {\bibfnamefont {T.}~\bibnamefont
  {Taniguchi}}, \bibinfo {author} {\bibfnamefont {K.}~\bibnamefont {Watanabe}},
  \bibinfo {author} {\bibfnamefont {Y.}~\bibnamefont {Yang}}, \bibinfo {author}
  {\bibfnamefont {M.~A.}\ \bibnamefont {Morales}}, \bibinfo {author}
  {\bibfnamefont {S.}~\bibnamefont {Zhang}}, \bibinfo {author} {\bibfnamefont
  {A.~J.}\ \bibnamefont {Millis}}, \bibinfo {author} {\bibfnamefont {M.~D.}\
  \bibnamefont {Lukin}}, \bibinfo {author} {\bibfnamefont {P.}~\bibnamefont
  {Kim}}, \bibinfo {author} {\bibfnamefont {E.}~\bibnamefont {Demler}},\ and\
  \bibinfo {author} {\bibfnamefont {H.}~\bibnamefont {Park}},\ }\bibfield
  {title} {\bibinfo {title} {An electronic microemulsion phase emerging from a
  quantum crystal-to-liquid transition},\ }\bibfield  {journal} {\bibinfo
  {journal} {Nature Physics}\ }\href
  {https://doi.org/10.1038/s41567-024-02759-8} {10.1038/s41567-024-02759-8}
  (\bibinfo {year} {2025})\BibitemShut {NoStop}%
\bibitem [{\citenamefont {Lucas}\ and\ \citenamefont {Fong}(2018)}]{lucas2018}%
  \BibitemOpen
  \bibfield  {author} {\bibinfo {author} {\bibfnamefont {A.}~\bibnamefont
  {Lucas}}\ and\ \bibinfo {author} {\bibfnamefont {K.~C.}\ \bibnamefont
  {Fong}},\ }\bibfield  {title} {\bibinfo {title} {{Hydrodynamics of electrons
  in graphene}},\ }\href {https://doi.org/10.1088/1361-648X/aaa274} {\bibfield
  {journal} {\bibinfo  {journal} {Journal of Physics: Condensed Matter}\
  }\textbf {\bibinfo {volume} {30}},\ \bibinfo {pages} {053001} (\bibinfo
  {year} {2018})}\BibitemShut {NoStop}%
\bibitem [{\citenamefont {Xue}\ \emph {et~al.}(2024)\citenamefont {Xue},
  \citenamefont {Maksimovic}, \citenamefont {Dolgirev}, \citenamefont {Xia},
  \citenamefont {Kitagawa}, \citenamefont {Müller}, \citenamefont {Machado},
  \citenamefont {Klein}, \citenamefont {MacNeill}, \citenamefont {Watanabe},
  \citenamefont {Taniguchi}, \citenamefont {Jarillo-Herrero}, \citenamefont
  {Lukin}, \citenamefont {Demler},\ and\ \citenamefont
  {Yacoby}}]{ruolanxue2024}%
  \BibitemOpen
  \bibfield  {author} {\bibinfo {author} {\bibfnamefont {R.}~\bibnamefont
  {Xue}}, \bibinfo {author} {\bibfnamefont {N.}~\bibnamefont {Maksimovic}},
  \bibinfo {author} {\bibfnamefont {P.~E.}\ \bibnamefont {Dolgirev}}, \bibinfo
  {author} {\bibfnamefont {L.-Q.}\ \bibnamefont {Xia}}, \bibinfo {author}
  {\bibfnamefont {R.}~\bibnamefont {Kitagawa}}, \bibinfo {author}
  {\bibfnamefont {A.}~\bibnamefont {Müller}}, \bibinfo {author} {\bibfnamefont
  {F.}~\bibnamefont {Machado}}, \bibinfo {author} {\bibfnamefont {D.~R.}\
  \bibnamefont {Klein}}, \bibinfo {author} {\bibfnamefont {D.}~\bibnamefont
  {MacNeill}}, \bibinfo {author} {\bibfnamefont {K.}~\bibnamefont {Watanabe}},
  \bibinfo {author} {\bibfnamefont {T.}~\bibnamefont {Taniguchi}}, \bibinfo
  {author} {\bibfnamefont {P.}~\bibnamefont {Jarillo-Herrero}}, \bibinfo
  {author} {\bibfnamefont {M.~D.}\ \bibnamefont {Lukin}}, \bibinfo {author}
  {\bibfnamefont {E.}~\bibnamefont {Demler}},\ and\ \bibinfo {author}
  {\bibfnamefont {A.}~\bibnamefont {Yacoby}},\ }\bibfield  {title} {\bibinfo
  {title} {{Signatures of magnon hydrodynamics in an atomically-thin
  ferromagnet}},\ }\href {https://arxiv.org/abs/2403.01057} {\bibfield
  {journal} {\bibinfo  {journal} {arXiv preprint arXiv:2403.01057}\ } (\bibinfo
  {year} {2024})}\BibitemShut {NoStop}%
\bibitem [{\citenamefont {Saito}\ \emph {et~al.}(2016)\citenamefont {Saito},
  \citenamefont {Nojima},\ and\ \citenamefont {Iwasa}}]{saito2016}%
  \BibitemOpen
  \bibfield  {author} {\bibinfo {author} {\bibfnamefont {Y.}~\bibnamefont
  {Saito}}, \bibinfo {author} {\bibfnamefont {T.}~\bibnamefont {Nojima}},\ and\
  \bibinfo {author} {\bibfnamefont {Y.}~\bibnamefont {Iwasa}},\ }\bibfield
  {title} {\bibinfo {title} {{Highly Crystalline 2D Superconductors}},\ }\href
  {https://doi.org/10.1038/natrevmats.2016.94} {\bibfield  {journal} {\bibinfo
  {journal} {Nature Reviews Materials}\ }\textbf {\bibinfo {volume} {2}},\
  \bibinfo {pages} {16094} (\bibinfo {year} {2016})}\BibitemShut {NoStop}%
\bibitem [{\citenamefont {McLaughlin}\ \emph {et~al.}(2022)\citenamefont
  {McLaughlin}, \citenamefont {Hu}, \citenamefont {Huang}, \citenamefont
  {Zhang}, \citenamefont {Lu}, \citenamefont {Yan}, \citenamefont {Wang},
  \citenamefont {Tserkovnyak}, \citenamefont {Ni},\ and\ \citenamefont
  {Du}}]{mclaughlin2022}%
  \BibitemOpen
  \bibfield  {author} {\bibinfo {author} {\bibfnamefont {N.~J.}\ \bibnamefont
  {McLaughlin}}, \bibinfo {author} {\bibfnamefont {C.}~\bibnamefont {Hu}},
  \bibinfo {author} {\bibfnamefont {M.}~\bibnamefont {Huang}}, \bibinfo
  {author} {\bibfnamefont {S.}~\bibnamefont {Zhang}}, \bibinfo {author}
  {\bibfnamefont {H.}~\bibnamefont {Lu}}, \bibinfo {author} {\bibfnamefont
  {G.~Q.}\ \bibnamefont {Yan}}, \bibinfo {author} {\bibfnamefont
  {H.}~\bibnamefont {Wang}}, \bibinfo {author} {\bibfnamefont {Y.}~\bibnamefont
  {Tserkovnyak}}, \bibinfo {author} {\bibfnamefont {N.}~\bibnamefont {Ni}},\
  and\ \bibinfo {author} {\bibfnamefont {C.~R.}\ \bibnamefont {Du}},\
  }\bibfield  {title} {\bibinfo {title} {{Quantum Imaging of Magnetic Phase
  Transitions and Spin Fluctuations in Intrinsic Magnetic Topological
  Nanoflakes}},\ }\href {https://doi.org/10.1021/acs.nanolett.2c01390}
  {\bibfield  {journal} {\bibinfo  {journal} {Nano Letters}\ }\textbf {\bibinfo
  {volume} {22}},\ \bibinfo {pages} {5810} (\bibinfo {year}
  {2022})}\BibitemShut {NoStop}%
\bibitem [{\citenamefont {Andersen}\ \emph {et~al.}(2019)\citenamefont
  {Andersen}, \citenamefont {Dwyer}, \citenamefont {Sanchez-Yamagishi},
  \citenamefont {Rodriguez-Nieva}, \citenamefont {Agarwal}, \citenamefont
  {Watanabe}, \citenamefont {Taniguchi}, \citenamefont {Demler}, \citenamefont
  {Kim}, \citenamefont {Park},\ and\ \citenamefont {Lukin}}]{andersen2019}%
  \BibitemOpen
  \bibfield  {author} {\bibinfo {author} {\bibfnamefont {T.~I.}\ \bibnamefont
  {Andersen}}, \bibinfo {author} {\bibfnamefont {B.~L.}\ \bibnamefont {Dwyer}},
  \bibinfo {author} {\bibfnamefont {J.~D.}\ \bibnamefont {Sanchez-Yamagishi}},
  \bibinfo {author} {\bibfnamefont {J.~F.}\ \bibnamefont {Rodriguez-Nieva}},
  \bibinfo {author} {\bibfnamefont {K.}~\bibnamefont {Agarwal}}, \bibinfo
  {author} {\bibfnamefont {K.}~\bibnamefont {Watanabe}}, \bibinfo {author}
  {\bibfnamefont {T.}~\bibnamefont {Taniguchi}}, \bibinfo {author}
  {\bibfnamefont {E.~A.}\ \bibnamefont {Demler}}, \bibinfo {author}
  {\bibfnamefont {P.}~\bibnamefont {Kim}}, \bibinfo {author} {\bibfnamefont
  {H.}~\bibnamefont {Park}},\ and\ \bibinfo {author} {\bibfnamefont {M.~D.}\
  \bibnamefont {Lukin}},\ }\bibfield  {title} {\bibinfo {title}
  {{Electron-phonon instability in graphene revealed by global and local noise
  probes}},\ }\href {https://doi.org/10.1126/science.aaw2104} {\bibfield
  {journal} {\bibinfo  {journal} {Science}\ }\textbf {\bibinfo {volume}
  {364}},\ \bibinfo {pages} {154} (\bibinfo {year} {2019})}\BibitemShut
  {NoStop}%
\bibitem [{\citenamefont {Dolgirev}\ \emph {et~al.}(2022)\citenamefont
  {Dolgirev}, \citenamefont {Chatterjee}, \citenamefont {Esterlis},
  \citenamefont {Zibrov}, \citenamefont {Lukin}, \citenamefont {Yao},\ and\
  \citenamefont {Demler}}]{dolgirev2022}%
  \BibitemOpen
  \bibfield  {author} {\bibinfo {author} {\bibfnamefont {P.~E.}\ \bibnamefont
  {Dolgirev}}, \bibinfo {author} {\bibfnamefont {S.}~\bibnamefont
  {Chatterjee}}, \bibinfo {author} {\bibfnamefont {I.}~\bibnamefont
  {Esterlis}}, \bibinfo {author} {\bibfnamefont {A.~A.}\ \bibnamefont
  {Zibrov}}, \bibinfo {author} {\bibfnamefont {M.~D.}\ \bibnamefont {Lukin}},
  \bibinfo {author} {\bibfnamefont {N.~Y.}\ \bibnamefont {Yao}},\ and\ \bibinfo
  {author} {\bibfnamefont {E.}~\bibnamefont {Demler}},\ }\bibfield  {title}
  {\bibinfo {title} {{Characterizing two-dimensional superconductivity via
  nanoscale noise magnetometry with single-spin qubits}},\ }\href
  {https://doi.org/10.1103/PhysRevB.105.024507} {\bibfield  {journal} {\bibinfo
   {journal} {Physical Review B}\ }\textbf {\bibinfo {volume} {105}},\ \bibinfo
  {pages} {024507} (\bibinfo {year} {2022})}\BibitemShut {NoStop}%
\bibitem [{\citenamefont {Chatterjee}\ \emph {et~al.}(2022)\citenamefont
  {Chatterjee}, \citenamefont {Dolgirev}, \citenamefont {Esterlis},
  \citenamefont {Zibrov}, \citenamefont {Lukin}, \citenamefont {Yao},\ and\
  \citenamefont {Demler}}]{chatterjee2022}%
  \BibitemOpen
  \bibfield  {author} {\bibinfo {author} {\bibfnamefont {S.}~\bibnamefont
  {Chatterjee}}, \bibinfo {author} {\bibfnamefont {P.~E.}\ \bibnamefont
  {Dolgirev}}, \bibinfo {author} {\bibfnamefont {I.}~\bibnamefont {Esterlis}},
  \bibinfo {author} {\bibfnamefont {A.~A.}\ \bibnamefont {Zibrov}}, \bibinfo
  {author} {\bibfnamefont {M.~D.}\ \bibnamefont {Lukin}}, \bibinfo {author}
  {\bibfnamefont {N.~Y.}\ \bibnamefont {Yao}},\ and\ \bibinfo {author}
  {\bibfnamefont {E.}~\bibnamefont {Demler}},\ }\bibfield  {title} {\bibinfo
  {title} {Single-spin qubit magnetic spectroscopy of two-dimensional
  superconductivity},\ }\href@noop {} {\bibfield  {journal} {\bibinfo
  {journal} {Physical Review Research}\ }\textbf {\bibinfo {volume} {4}},\
  \bibinfo {pages} {L012001} (\bibinfo {year} {2022})}\BibitemShut {NoStop}%
\bibitem [{\citenamefont {Liu}\ \emph {et~al.}(2025)\citenamefont {Liu},
  \citenamefont {Gong}, \citenamefont {Kim}, \citenamefont {Diessel},
  \citenamefont {Xu}, \citenamefont {Rehfuss}, \citenamefont {Du},
  \citenamefont {He}, \citenamefont {Singh}, \citenamefont {Eo}, \citenamefont
  {Henriksen}, \citenamefont {Gu}, \citenamefont {Yao}, \citenamefont
  {Machado}, \citenamefont {Ran}, \citenamefont {Chatterjee},\ and\
  \citenamefont {Zu}}]{zhongyuanliu2025}%
  \BibitemOpen
  \bibfield  {author} {\bibinfo {author} {\bibfnamefont {Z.}~\bibnamefont
  {Liu}}, \bibinfo {author} {\bibfnamefont {R.}~\bibnamefont {Gong}}, \bibinfo
  {author} {\bibfnamefont {J.}~\bibnamefont {Kim}}, \bibinfo {author}
  {\bibfnamefont {O.~K.}\ \bibnamefont {Diessel}}, \bibinfo {author}
  {\bibfnamefont {Q.}~\bibnamefont {Xu}}, \bibinfo {author} {\bibfnamefont
  {Z.}~\bibnamefont {Rehfuss}}, \bibinfo {author} {\bibfnamefont
  {X.}~\bibnamefont {Du}}, \bibinfo {author} {\bibfnamefont {G.}~\bibnamefont
  {He}}, \bibinfo {author} {\bibfnamefont {A.}~\bibnamefont {Singh}}, \bibinfo
  {author} {\bibfnamefont {Y.~S.}\ \bibnamefont {Eo}}, \bibinfo {author}
  {\bibfnamefont {E.~A.}\ \bibnamefont {Henriksen}}, \bibinfo {author}
  {\bibfnamefont {G.~D.}\ \bibnamefont {Gu}}, \bibinfo {author} {\bibfnamefont
  {N.~Y.}\ \bibnamefont {Yao}}, \bibinfo {author} {\bibfnamefont
  {F.}~\bibnamefont {Machado}}, \bibinfo {author} {\bibfnamefont
  {S.}~\bibnamefont {Ran}}, \bibinfo {author} {\bibfnamefont {S.}~\bibnamefont
  {Chatterjee}},\ and\ \bibinfo {author} {\bibfnamefont {C.}~\bibnamefont
  {Zu}},\ }\bibfield  {title} {\bibinfo {title} {{Quantum noise spectroscopy of
  superconducting dynamics in thin film Bi$_2$Sr$_2$CaCu$_2$O$_{8+\delta}$}},\
  }\href {https://arxiv.org/abs/2502.04439} {\bibfield  {journal} {\bibinfo
  {journal} {arXiv preprint arXiv:2502.04439}\ } (\bibinfo {year}
  {2025})}\BibitemShut {NoStop}%
\bibitem [{\citenamefont {Curtis}\ \emph {et~al.}(2024)\citenamefont {Curtis},
  \citenamefont {Maksimovic}, \citenamefont {Poniatowski}, \citenamefont
  {Yacoby}, \citenamefont {Halperin}, \citenamefont {Narang},\ and\
  \citenamefont {Demler}}]{curtis2024}%
  \BibitemOpen
  \bibfield  {author} {\bibinfo {author} {\bibfnamefont {J.~B.}\ \bibnamefont
  {Curtis}}, \bibinfo {author} {\bibfnamefont {N.}~\bibnamefont {Maksimovic}},
  \bibinfo {author} {\bibfnamefont {N.~R.}\ \bibnamefont {Poniatowski}},
  \bibinfo {author} {\bibfnamefont {A.}~\bibnamefont {Yacoby}}, \bibinfo
  {author} {\bibfnamefont {B.}~\bibnamefont {Halperin}}, \bibinfo {author}
  {\bibfnamefont {P.}~\bibnamefont {Narang}},\ and\ \bibinfo {author}
  {\bibfnamefont {E.}~\bibnamefont {Demler}},\ }\bibfield  {title} {\bibinfo
  {title} {{Probing the Berezinskii-Kosterlitz-Thouless vortex unbinding
  transition in two-dimensional superconductors using local noise
  magnetometry}},\ }\href {https://doi.org/10.1103/PhysRevB.110.144518}
  {\bibfield  {journal} {\bibinfo  {journal} {Physical Review B}\ }\textbf
  {\bibinfo {volume} {110}},\ \bibinfo {pages} {144518} (\bibinfo {year}
  {2024})}\BibitemShut {NoStop}%
\bibitem [{\citenamefont {Boothroyd}(2020)}]{boothroydChap10}%
  \BibitemOpen
  \bibfield  {author} {\bibinfo {author} {\bibfnamefont {A.~T.}\ \bibnamefont
  {Boothroyd}},\ }\bibfield  {title} {\bibinfo {title} {{Practical Aspects of
  Neutron Scattering}},\ }in\ \href
  {https://doi.org/10.1093/oso/9780198862314.003.0010} {\emph {\bibinfo
  {booktitle} {Principles of Neutron Scattering from Condensed Matter}}}\
  (\bibinfo  {publisher} {Oxford University Press},\ \bibinfo {year} {2020})\
  pp.\ \bibinfo {pages} {343--404}\BibitemShut {NoStop}%
\bibitem [{\citenamefont {Casola}\ \emph {et~al.}(2018)\citenamefont {Casola},
  \citenamefont {van~der Sar},\ and\ \citenamefont {Yacoby}}]{casola2018}%
  \BibitemOpen
  \bibfield  {author} {\bibinfo {author} {\bibfnamefont {F.}~\bibnamefont
  {Casola}}, \bibinfo {author} {\bibfnamefont {T.}~\bibnamefont {van~der
  Sar}},\ and\ \bibinfo {author} {\bibfnamefont {A.}~\bibnamefont {Yacoby}},\
  }\bibfield  {title} {\bibinfo {title} {{Probing condensed matter physics with
  magnetometry based on nitrogen-vacancy centres in diamond}},\ }\href
  {https://doi.org/10.1038/natrevmats.2017.88} {\bibfield  {journal} {\bibinfo
  {journal} {Nature Reviews Materials}\ }\textbf {\bibinfo {volume} {3}},\
  \bibinfo {pages} {17088} (\bibinfo {year} {2018})}\BibitemShut {NoStop}%
\bibitem [{\citenamefont {Rovny}\ \emph {et~al.}(2024)\citenamefont {Rovny},
  \citenamefont {Gopalakrishnan}, \citenamefont {Jayich}, \citenamefont
  {Maletinsky}, \citenamefont {Demler},\ and\ \citenamefont
  {de~Leon}}]{rovny2024}%
  \BibitemOpen
  \bibfield  {author} {\bibinfo {author} {\bibfnamefont {J.}~\bibnamefont
  {Rovny}}, \bibinfo {author} {\bibfnamefont {S.}~\bibnamefont
  {Gopalakrishnan}}, \bibinfo {author} {\bibfnamefont {A.~C.~B.}\ \bibnamefont
  {Jayich}}, \bibinfo {author} {\bibfnamefont {P.}~\bibnamefont {Maletinsky}},
  \bibinfo {author} {\bibfnamefont {E.}~\bibnamefont {Demler}},\ and\ \bibinfo
  {author} {\bibfnamefont {N.~P.}\ \bibnamefont {de~Leon}},\ }\bibfield
  {title} {\bibinfo {title} {{Nanoscale diamond quantum sensors for many-body
  physics}},\ }\href {https://doi.org/10.1038/s42254-024-00775-4} {\bibfield
  {journal} {\bibinfo  {journal} {Nature Reviews Physics}\ }\textbf {\bibinfo
  {volume} {6}},\ \bibinfo {pages} {753} (\bibinfo {year} {2024})}\BibitemShut
  {NoStop}%
\bibitem [{\citenamefont {Thiel}\ \emph {et~al.}(2019)\citenamefont {Thiel},
  \citenamefont {Wang}, \citenamefont {Tschudin}, \citenamefont {Rohner},
  \citenamefont {Gutiérrez-Lezama}, \citenamefont {Ubrig}, \citenamefont
  {Gibertini}, \citenamefont {Giannini}, \citenamefont {Morpurgo},\ and\
  \citenamefont {Maletinsky}}]{thiel2019}%
  \BibitemOpen
  \bibfield  {author} {\bibinfo {author} {\bibfnamefont {L.}~\bibnamefont
  {Thiel}}, \bibinfo {author} {\bibfnamefont {Z.}~\bibnamefont {Wang}},
  \bibinfo {author} {\bibfnamefont {M.~A.}\ \bibnamefont {Tschudin}}, \bibinfo
  {author} {\bibfnamefont {D.}~\bibnamefont {Rohner}}, \bibinfo {author}
  {\bibfnamefont {I.}~\bibnamefont {Gutiérrez-Lezama}}, \bibinfo {author}
  {\bibfnamefont {N.}~\bibnamefont {Ubrig}}, \bibinfo {author} {\bibfnamefont
  {M.}~\bibnamefont {Gibertini}}, \bibinfo {author} {\bibfnamefont
  {E.}~\bibnamefont {Giannini}}, \bibinfo {author} {\bibfnamefont {A.~F.}\
  \bibnamefont {Morpurgo}},\ and\ \bibinfo {author} {\bibfnamefont
  {P.}~\bibnamefont {Maletinsky}},\ }\bibfield  {title} {\bibinfo {title}
  {{Probing magnetism in 2D materials at the nanoscale with single-spin
  microscopy}},\ }\href {https://doi.org/10.1126/science.aav6926} {\bibfield
  {journal} {\bibinfo  {journal} {Science}\ }\textbf {\bibinfo {volume}
  {364}},\ \bibinfo {pages} {973} (\bibinfo {year} {2019})}\BibitemShut
  {NoStop}%
\bibitem [{\citenamefont {Bhattacharyya}\ \emph {et~al.}(2024)\citenamefont
  {Bhattacharyya}, \citenamefont {Chen}, \citenamefont {Huang}, \citenamefont
  {Chatterjee}, \citenamefont {Huang}, \citenamefont {Kobrin}, \citenamefont
  {Lyu}, \citenamefont {Smart}, \citenamefont {Block}, \citenamefont {Wang},
  \citenamefont {Wang}, \citenamefont {Wu}, \citenamefont {Hsieh},
  \citenamefont {Ma}, \citenamefont {Mandyam}, \citenamefont {Chen},
  \citenamefont {Davis}, \citenamefont {Geballe}, \citenamefont {Zu},
  \citenamefont {Struzhkin}, \citenamefont {Jeanloz}, \citenamefont {Moore},
  \citenamefont {Cui}, \citenamefont {Galli}, \citenamefont {Halperin},
  \citenamefont {Laumann},\ and\ \citenamefont {Yao}}]{bhattacharyya2024}%
  \BibitemOpen
  \bibfield  {author} {\bibinfo {author} {\bibfnamefont {P.}~\bibnamefont
  {Bhattacharyya}}, \bibinfo {author} {\bibfnamefont {W.}~\bibnamefont {Chen}},
  \bibinfo {author} {\bibfnamefont {X.}~\bibnamefont {Huang}}, \bibinfo
  {author} {\bibfnamefont {S.}~\bibnamefont {Chatterjee}}, \bibinfo {author}
  {\bibfnamefont {B.}~\bibnamefont {Huang}}, \bibinfo {author} {\bibfnamefont
  {B.}~\bibnamefont {Kobrin}}, \bibinfo {author} {\bibfnamefont
  {Y.}~\bibnamefont {Lyu}}, \bibinfo {author} {\bibfnamefont {T.~J.}\
  \bibnamefont {Smart}}, \bibinfo {author} {\bibfnamefont {M.}~\bibnamefont
  {Block}}, \bibinfo {author} {\bibfnamefont {E.}~\bibnamefont {Wang}},
  \bibinfo {author} {\bibfnamefont {Z.}~\bibnamefont {Wang}}, \bibinfo {author}
  {\bibfnamefont {W.}~\bibnamefont {Wu}}, \bibinfo {author} {\bibfnamefont
  {S.}~\bibnamefont {Hsieh}}, \bibinfo {author} {\bibfnamefont
  {H.}~\bibnamefont {Ma}}, \bibinfo {author} {\bibfnamefont {S.}~\bibnamefont
  {Mandyam}}, \bibinfo {author} {\bibfnamefont {B.}~\bibnamefont {Chen}},
  \bibinfo {author} {\bibfnamefont {E.}~\bibnamefont {Davis}}, \bibinfo
  {author} {\bibfnamefont {Z.~M.}\ \bibnamefont {Geballe}}, \bibinfo {author}
  {\bibfnamefont {C.}~\bibnamefont {Zu}}, \bibinfo {author} {\bibfnamefont
  {V.}~\bibnamefont {Struzhkin}}, \bibinfo {author} {\bibfnamefont
  {R.}~\bibnamefont {Jeanloz}}, \bibinfo {author} {\bibfnamefont {J.~E.}\
  \bibnamefont {Moore}}, \bibinfo {author} {\bibfnamefont {T.}~\bibnamefont
  {Cui}}, \bibinfo {author} {\bibfnamefont {G.}~\bibnamefont {Galli}}, \bibinfo
  {author} {\bibfnamefont {B.~I.}\ \bibnamefont {Halperin}}, \bibinfo {author}
  {\bibfnamefont {C.~R.}\ \bibnamefont {Laumann}},\ and\ \bibinfo {author}
  {\bibfnamefont {N.~Y.}\ \bibnamefont {Yao}},\ }\bibfield  {title} {\bibinfo
  {title} {Imaging the meissner effect in hydride superconductors using quantum
  sensors},\ }\href {https://doi.org/10.1038/s41586-024-07026-7} {\bibfield
  {journal} {\bibinfo  {journal} {Nature}\ }\textbf {\bibinfo {volume} {627}},\
  \bibinfo {pages} {73} (\bibinfo {year} {2024})}\BibitemShut {NoStop}%
\bibitem [{\citenamefont {Ku}\ \emph {et~al.}(2020)\citenamefont {Ku},
  \citenamefont {Zhou}, \citenamefont {Li}, \citenamefont {Shin}, \citenamefont
  {Shi}, \citenamefont {Burch}, \citenamefont {Anderson}, \citenamefont
  {Pierce}, \citenamefont {Xie}, \citenamefont {Hamo}, \citenamefont {Vool},
  \citenamefont {Zhang}, \citenamefont {Casola}, \citenamefont {Taniguchi},
  \citenamefont {Watanabe}, \citenamefont {Fogler}, \citenamefont {Kim},
  \citenamefont {Yacoby},\ and\ \citenamefont {Walsworth}}]{ku2020}%
  \BibitemOpen
  \bibfield  {author} {\bibinfo {author} {\bibfnamefont {M.~J.~H.}\
  \bibnamefont {Ku}}, \bibinfo {author} {\bibfnamefont {T.~X.}\ \bibnamefont
  {Zhou}}, \bibinfo {author} {\bibfnamefont {Q.}~\bibnamefont {Li}}, \bibinfo
  {author} {\bibfnamefont {Y.~J.}\ \bibnamefont {Shin}}, \bibinfo {author}
  {\bibfnamefont {J.~K.}\ \bibnamefont {Shi}}, \bibinfo {author} {\bibfnamefont
  {C.}~\bibnamefont {Burch}}, \bibinfo {author} {\bibfnamefont {L.~E.}\
  \bibnamefont {Anderson}}, \bibinfo {author} {\bibfnamefont {A.~T.}\
  \bibnamefont {Pierce}}, \bibinfo {author} {\bibfnamefont {Y.}~\bibnamefont
  {Xie}}, \bibinfo {author} {\bibfnamefont {A.}~\bibnamefont {Hamo}}, \bibinfo
  {author} {\bibfnamefont {U.}~\bibnamefont {Vool}}, \bibinfo {author}
  {\bibfnamefont {H.}~\bibnamefont {Zhang}}, \bibinfo {author} {\bibfnamefont
  {F.}~\bibnamefont {Casola}}, \bibinfo {author} {\bibfnamefont
  {T.}~\bibnamefont {Taniguchi}}, \bibinfo {author} {\bibfnamefont
  {K.}~\bibnamefont {Watanabe}}, \bibinfo {author} {\bibfnamefont {M.~M.}\
  \bibnamefont {Fogler}}, \bibinfo {author} {\bibfnamefont {P.}~\bibnamefont
  {Kim}}, \bibinfo {author} {\bibfnamefont {A.}~\bibnamefont {Yacoby}},\ and\
  \bibinfo {author} {\bibfnamefont {R.~L.}\ \bibnamefont {Walsworth}},\
  }\bibfield  {title} {\bibinfo {title} {{Imaging viscous flow of the Dirac
  fluid in graphene}},\ }\href {https://doi.org/10.1038/s41586-020-2507-2}
  {\bibfield  {journal} {\bibinfo  {journal} {Nature}\ }\textbf {\bibinfo
  {volume} {583}},\ \bibinfo {pages} {537} (\bibinfo {year}
  {2020})}\BibitemShut {NoStop}%
\bibitem [{\citenamefont {Kolkowitz}\ \emph {et~al.}(2015)\citenamefont
  {Kolkowitz}, \citenamefont {Safira}, \citenamefont {High}, \citenamefont
  {Devlin}, \citenamefont {Choi}, \citenamefont {Unterreithmeier},
  \citenamefont {Patterson}, \citenamefont {Zibrov}, \citenamefont
  {Manucharyan}, \citenamefont {Park},\ and\ \citenamefont
  {Lukin}}]{kolkowitz2015}%
  \BibitemOpen
  \bibfield  {author} {\bibinfo {author} {\bibfnamefont {S.}~\bibnamefont
  {Kolkowitz}}, \bibinfo {author} {\bibfnamefont {A.}~\bibnamefont {Safira}},
  \bibinfo {author} {\bibfnamefont {A.~A.}\ \bibnamefont {High}}, \bibinfo
  {author} {\bibfnamefont {R.~C.}\ \bibnamefont {Devlin}}, \bibinfo {author}
  {\bibfnamefont {S.}~\bibnamefont {Choi}}, \bibinfo {author} {\bibfnamefont
  {Q.~P.}\ \bibnamefont {Unterreithmeier}}, \bibinfo {author} {\bibfnamefont
  {D.}~\bibnamefont {Patterson}}, \bibinfo {author} {\bibfnamefont {A.~S.}\
  \bibnamefont {Zibrov}}, \bibinfo {author} {\bibfnamefont {V.~E.}\
  \bibnamefont {Manucharyan}}, \bibinfo {author} {\bibfnamefont
  {H.}~\bibnamefont {Park}},\ and\ \bibinfo {author} {\bibfnamefont {M.~D.}\
  \bibnamefont {Lukin}},\ }\bibfield  {title} {\bibinfo {title} {{Probing
  Johnson noise and ballistic transport in normal metals with a single-spin
  qubit}},\ }\href {https://doi.org/10.1126/science.aaa4298} {\bibfield
  {journal} {\bibinfo  {journal} {Science}\ }\textbf {\bibinfo {volume}
  {347}},\ \bibinfo {pages} {1129} (\bibinfo {year} {2015})}\BibitemShut
  {NoStop}%
\bibitem [{\citenamefont {Du}\ \emph {et~al.}(2017)\citenamefont {Du},
  \citenamefont {van~der Sar}, \citenamefont {Zhou}, \citenamefont {Upadhyaya},
  \citenamefont {Casola}, \citenamefont {Zhang}, \citenamefont {Onbasli},
  \citenamefont {Ross}, \citenamefont {Walsworth}, \citenamefont
  {Tserkovnyak},\ and\ \citenamefont {Yacoby}}]{chunhuidu2017}%
  \BibitemOpen
  \bibfield  {author} {\bibinfo {author} {\bibfnamefont {C.}~\bibnamefont
  {Du}}, \bibinfo {author} {\bibfnamefont {T.}~\bibnamefont {van~der Sar}},
  \bibinfo {author} {\bibfnamefont {T.~X.}\ \bibnamefont {Zhou}}, \bibinfo
  {author} {\bibfnamefont {P.}~\bibnamefont {Upadhyaya}}, \bibinfo {author}
  {\bibfnamefont {F.}~\bibnamefont {Casola}}, \bibinfo {author} {\bibfnamefont
  {H.}~\bibnamefont {Zhang}}, \bibinfo {author} {\bibfnamefont {M.~C.}\
  \bibnamefont {Onbasli}}, \bibinfo {author} {\bibfnamefont {C.~A.}\
  \bibnamefont {Ross}}, \bibinfo {author} {\bibfnamefont {R.~L.}\ \bibnamefont
  {Walsworth}}, \bibinfo {author} {\bibfnamefont {Y.}~\bibnamefont
  {Tserkovnyak}},\ and\ \bibinfo {author} {\bibfnamefont {A.}~\bibnamefont
  {Yacoby}},\ }\bibfield  {title} {\bibinfo {title} {{Control and local
  measurement of the spin chemical potential in a magnetic insulator}},\ }\href
  {https://doi.org/10.1126/science.aak9611} {\bibfield  {journal} {\bibinfo
  {journal} {Science}\ }\textbf {\bibinfo {volume} {357}},\ \bibinfo {pages}
  {195} (\bibinfo {year} {2017})}\BibitemShut {NoStop}%
\bibitem [{\citenamefont {Rovny}\ \emph {et~al.}(2022)\citenamefont {Rovny},
  \citenamefont {Yuan}, \citenamefont {Fitzpatrick}, \citenamefont {Abdalla},
  \citenamefont {Futamura}, \citenamefont {Fox}, \citenamefont {Cambria},
  \citenamefont {Kolkowitz},\ and\ \citenamefont {de~Leon}}]{rovny2022}%
  \BibitemOpen
  \bibfield  {author} {\bibinfo {author} {\bibfnamefont {J.}~\bibnamefont
  {Rovny}}, \bibinfo {author} {\bibfnamefont {Z.}~\bibnamefont {Yuan}},
  \bibinfo {author} {\bibfnamefont {M.}~\bibnamefont {Fitzpatrick}}, \bibinfo
  {author} {\bibfnamefont {A.~I.}\ \bibnamefont {Abdalla}}, \bibinfo {author}
  {\bibfnamefont {L.}~\bibnamefont {Futamura}}, \bibinfo {author}
  {\bibfnamefont {C.}~\bibnamefont {Fox}}, \bibinfo {author} {\bibfnamefont
  {M.~C.}\ \bibnamefont {Cambria}}, \bibinfo {author} {\bibfnamefont
  {S.}~\bibnamefont {Kolkowitz}},\ and\ \bibinfo {author} {\bibfnamefont
  {N.~P.}\ \bibnamefont {de~Leon}},\ }\bibfield  {title} {\bibinfo {title}
  {{Nanoscale covariance magnetometry with diamond quantum sensors}},\ }\href
  {https://doi.org/10.1126/science.ade9858} {\bibfield  {journal} {\bibinfo
  {journal} {Science}\ }\textbf {\bibinfo {volume} {378}},\ \bibinfo {pages}
  {1301} (\bibinfo {year} {2022})}\BibitemShut {NoStop}%
\bibitem [{\citenamefont {Cheng}\ \emph {et~al.}(2024)\citenamefont {Cheng},
  \citenamefont {Kazi}, \citenamefont {Rovny}, \citenamefont {Zhang},
  \citenamefont {Nassar}, \citenamefont {Thompson},\ and\ \citenamefont
  {de~Leon}}]{kaihungcheng2024}%
  \BibitemOpen
  \bibfield  {author} {\bibinfo {author} {\bibfnamefont {K.-H.}\ \bibnamefont
  {Cheng}}, \bibinfo {author} {\bibfnamefont {Z.}~\bibnamefont {Kazi}},
  \bibinfo {author} {\bibfnamefont {J.}~\bibnamefont {Rovny}}, \bibinfo
  {author} {\bibfnamefont {B.}~\bibnamefont {Zhang}}, \bibinfo {author}
  {\bibfnamefont {L.}~\bibnamefont {Nassar}}, \bibinfo {author} {\bibfnamefont
  {J.~D.}\ \bibnamefont {Thompson}},\ and\ \bibinfo {author} {\bibfnamefont
  {N.~P.}\ \bibnamefont {de~Leon}},\ }\bibfield  {title} {\bibinfo {title}
  {{Massively multiplexed nanoscale magnetometry with diamond quantum
  sensors}},\ }\href {https://arxiv.org/abs/2408.11666} {\bibfield  {journal}
  {\bibinfo  {journal} {arXiv preprint arXiv:2408.11666}\ } (\bibinfo {year}
  {2024})}\BibitemShut {NoStop}%
\bibitem [{\citenamefont {Cambria}\ \emph {et~al.}(2024)\citenamefont
  {Cambria}, \citenamefont {Chand},\ and\ \citenamefont
  {Kolkowitz}}]{cambria2024}%
  \BibitemOpen
  \bibfield  {author} {\bibinfo {author} {\bibfnamefont {M.}~\bibnamefont
  {Cambria}}, \bibinfo {author} {\bibfnamefont {S.}~\bibnamefont {Chand}},\
  and\ \bibinfo {author} {\bibfnamefont {S.}~\bibnamefont {Kolkowitz}},\
  }\bibfield  {title} {\bibinfo {title} {{Scalable parallel measurement of
  individual nitrogen-vacancy centers}},\ }\href
  {https://arxiv.org/abs/2408.11715} {\bibfield  {journal} {\bibinfo  {journal}
  {arXiv preprint arXiv:2408.11715}\ } (\bibinfo {year} {2024})}\BibitemShut
  {NoStop}%
\bibitem [{\citenamefont {Huxter}\ \emph {et~al.}(2024)\citenamefont {Huxter},
  \citenamefont {Dalmagioni},\ and\ \citenamefont {Degen}}]{huxter2024}%
  \BibitemOpen
  \bibfield  {author} {\bibinfo {author} {\bibfnamefont {W.~S.}\ \bibnamefont
  {Huxter}}, \bibinfo {author} {\bibfnamefont {F.}~\bibnamefont {Dalmagioni}},\
  and\ \bibinfo {author} {\bibfnamefont {C.~L.}\ \bibnamefont {Degen}},\
  }\bibfield  {title} {\bibinfo {title} {{Multiplexed scanning microscopy with
  dual-qubit spin sensors}},\ }\href {https://arxiv.org/abs/2407.19576}
  {\bibfield  {journal} {\bibinfo  {journal} {arXiv preprint arXiv:2407.19576}\
  } (\bibinfo {year} {2024})}\BibitemShut {NoStop}%
\bibitem [{\citenamefont {Joliffe}\ \emph {et~al.}(2024)\citenamefont
  {Joliffe}, \citenamefont {Vorobyov},\ and\ \citenamefont
  {Wrachtrup}}]{joliffe2024}%
  \BibitemOpen
  \bibfield  {author} {\bibinfo {author} {\bibfnamefont {M.}~\bibnamefont
  {Joliffe}}, \bibinfo {author} {\bibfnamefont {V.}~\bibnamefont {Vorobyov}},\
  and\ \bibinfo {author} {\bibfnamefont {J.}~\bibnamefont {Wrachtrup}},\
  }\bibfield  {title} {\bibinfo {title} {{Readout of strongly coupled NV
  center-pair spin states with deep neural networks}},\ }\href
  {https://arxiv.org/abs/2412.19581} {\bibfield  {journal} {\bibinfo  {journal}
  {arXiv preprint arXiv:2412.19581}\ } (\bibinfo {year} {2024})}\BibitemShut
  {NoStop}%
\bibitem [{\citenamefont {Carmiggelt}\ \emph {et~al.}(2023)\citenamefont
  {Carmiggelt}, \citenamefont {Bertelli}, \citenamefont {Mulder}, \citenamefont
  {Teepe}, \citenamefont {Elyasi}, \citenamefont {Simon}, \citenamefont
  {Bauer}, \citenamefont {Blanter},\ and\ \citenamefont {van~der
  Sar}}]{carmiggelt2023}%
  \BibitemOpen
  \bibfield  {author} {\bibinfo {author} {\bibfnamefont {J.~J.}\ \bibnamefont
  {Carmiggelt}}, \bibinfo {author} {\bibfnamefont {I.}~\bibnamefont
  {Bertelli}}, \bibinfo {author} {\bibfnamefont {R.~W.}\ \bibnamefont
  {Mulder}}, \bibinfo {author} {\bibfnamefont {A.}~\bibnamefont {Teepe}},
  \bibinfo {author} {\bibfnamefont {M.}~\bibnamefont {Elyasi}}, \bibinfo
  {author} {\bibfnamefont {B.~G.}\ \bibnamefont {Simon}}, \bibinfo {author}
  {\bibfnamefont {G.~E.~W.}\ \bibnamefont {Bauer}}, \bibinfo {author}
  {\bibfnamefont {Y.~M.}\ \bibnamefont {Blanter}},\ and\ \bibinfo {author}
  {\bibfnamefont {T.}~\bibnamefont {van~der Sar}},\ }\bibfield  {title}
  {\bibinfo {title} {{Broadband microwave detection using electron spins in a
  hybrid diamond-magnet sensor chip}},\ }\href
  {https://doi.org/10.1038/s41467-023-36146-3} {\bibfield  {journal} {\bibinfo
  {journal} {Nature Communications}\ }\textbf {\bibinfo {volume} {14}},\
  \bibinfo {pages} {490} (\bibinfo {year} {2023})}\BibitemShut {NoStop}%
\bibitem [{\citenamefont {Wolfe}\ \emph {et~al.}(2014)\citenamefont {Wolfe},
  \citenamefont {Bhallamudi}, \citenamefont {Wang}, \citenamefont {Du},
  \citenamefont {Manuilov}, \citenamefont {Berger}, \citenamefont {Adur},
  \citenamefont {Yang},\ and\ \citenamefont {Hammel}}]{wolfe2014}%
  \BibitemOpen
  \bibfield  {author} {\bibinfo {author} {\bibfnamefont {C.~S.}\ \bibnamefont
  {Wolfe}}, \bibinfo {author} {\bibfnamefont {V.~P.}\ \bibnamefont
  {Bhallamudi}}, \bibinfo {author} {\bibfnamefont {H.~L.}\ \bibnamefont
  {Wang}}, \bibinfo {author} {\bibfnamefont {C.~H.}\ \bibnamefont {Du}},
  \bibinfo {author} {\bibfnamefont {S.}~\bibnamefont {Manuilov}}, \bibinfo
  {author} {\bibfnamefont {A.~J.}\ \bibnamefont {Berger}}, \bibinfo {author}
  {\bibfnamefont {R.}~\bibnamefont {Adur}}, \bibinfo {author} {\bibfnamefont
  {F.~Y.}\ \bibnamefont {Yang}},\ and\ \bibinfo {author} {\bibfnamefont
  {P.~C.}\ \bibnamefont {Hammel}},\ }\bibfield  {title} {\bibinfo {title}
  {{Off-resonant manipulation of spins in diamond via precessing magnetization
  of a proximal ferromagnet}},\ }\href
  {https://doi.org/10.1103/PhysRevB.89.180406} {\bibfield  {journal} {\bibinfo
  {journal} {Physical Review B}\ }\textbf {\bibinfo {volume} {89}},\ \bibinfo
  {pages} {180406} (\bibinfo {year} {2014})}\BibitemShut {NoStop}%
\bibitem [{\citenamefont {Lee-Wong}\ \emph {et~al.}(2020)\citenamefont
  {Lee-Wong}, \citenamefont {Xue}, \citenamefont {Ye}, \citenamefont {Kreisel},
  \citenamefont {van~der Sar}, \citenamefont {Yacoby},\ and\ \citenamefont
  {Du}}]{ericleewong2020}%
  \BibitemOpen
  \bibfield  {author} {\bibinfo {author} {\bibfnamefont {E.}~\bibnamefont
  {Lee-Wong}}, \bibinfo {author} {\bibfnamefont {R.}~\bibnamefont {Xue}},
  \bibinfo {author} {\bibfnamefont {F.}~\bibnamefont {Ye}}, \bibinfo {author}
  {\bibfnamefont {A.}~\bibnamefont {Kreisel}}, \bibinfo {author} {\bibfnamefont
  {T.}~\bibnamefont {van~der Sar}}, \bibinfo {author} {\bibfnamefont
  {A.}~\bibnamefont {Yacoby}},\ and\ \bibinfo {author} {\bibfnamefont {C.~R.}\
  \bibnamefont {Du}},\ }\bibfield  {title} {\bibinfo {title} {{Nanoscale
  Detection of Magnon Excitations with Variable Wavevectors Through a Quantum
  Spin Sensor}},\ }\href {https://doi.org/10.1021/acs.nanolett.0c00085}
  {\bibfield  {journal} {\bibinfo  {journal} {Nano Letters}\ }\textbf {\bibinfo
  {volume} {20}},\ \bibinfo {pages} {3284} (\bibinfo {year}
  {2020})}\BibitemShut {NoStop}%
\bibitem [{\citenamefont {McCullian}\ \emph {et~al.}(2020)\citenamefont
  {McCullian}, \citenamefont {Thabt}, \citenamefont {Gray}, \citenamefont
  {Melendez}, \citenamefont {Wolf}, \citenamefont {Safonov}, \citenamefont
  {Pelekhov}, \citenamefont {Bhallamudi}, \citenamefont {Page},\ and\
  \citenamefont {Hammel}}]{mccullian2020}%
  \BibitemOpen
  \bibfield  {author} {\bibinfo {author} {\bibfnamefont {B.~A.}\ \bibnamefont
  {McCullian}}, \bibinfo {author} {\bibfnamefont {A.~M.}\ \bibnamefont
  {Thabt}}, \bibinfo {author} {\bibfnamefont {B.~A.}\ \bibnamefont {Gray}},
  \bibinfo {author} {\bibfnamefont {A.~L.}\ \bibnamefont {Melendez}}, \bibinfo
  {author} {\bibfnamefont {M.~S.}\ \bibnamefont {Wolf}}, \bibinfo {author}
  {\bibfnamefont {V.~L.}\ \bibnamefont {Safonov}}, \bibinfo {author}
  {\bibfnamefont {D.~V.}\ \bibnamefont {Pelekhov}}, \bibinfo {author}
  {\bibfnamefont {V.~P.}\ \bibnamefont {Bhallamudi}}, \bibinfo {author}
  {\bibfnamefont {M.~R.}\ \bibnamefont {Page}},\ and\ \bibinfo {author}
  {\bibfnamefont {P.~C.}\ \bibnamefont {Hammel}},\ }\bibfield  {title}
  {\bibinfo {title} {{Broadband multi-magnon relaxometry using a quantum spin
  sensor for high frequency ferromagnetic dynamics sensing}},\ }\href
  {https://doi.org/10.1038/s41467-020-19121-0} {\bibfield  {journal} {\bibinfo
  {journal} {Nature Communications}\ }\textbf {\bibinfo {volume} {11}},\
  \bibinfo {pages} {5229} (\bibinfo {year} {2020})}\BibitemShut {NoStop}%
\bibitem [{\citenamefont {Yan}\ \emph {et~al.}(2022)\citenamefont {Yan},
  \citenamefont {Li}, \citenamefont {Lu}, \citenamefont {Huang}, \citenamefont
  {Xiao}, \citenamefont {Wernert}, \citenamefont {Brock}, \citenamefont
  {Fullerton}, \citenamefont {Chen}, \citenamefont {Wang},\ and\ \citenamefont
  {Du}}]{geraldyan2022}%
  \BibitemOpen
  \bibfield  {author} {\bibinfo {author} {\bibfnamefont {G.~Q.}\ \bibnamefont
  {Yan}}, \bibinfo {author} {\bibfnamefont {S.}~\bibnamefont {Li}}, \bibinfo
  {author} {\bibfnamefont {H.}~\bibnamefont {Lu}}, \bibinfo {author}
  {\bibfnamefont {M.}~\bibnamefont {Huang}}, \bibinfo {author} {\bibfnamefont
  {Y.}~\bibnamefont {Xiao}}, \bibinfo {author} {\bibfnamefont {L.}~\bibnamefont
  {Wernert}}, \bibinfo {author} {\bibfnamefont {J.~A.}\ \bibnamefont {Brock}},
  \bibinfo {author} {\bibfnamefont {E.~E.}\ \bibnamefont {Fullerton}}, \bibinfo
  {author} {\bibfnamefont {H.}~\bibnamefont {Chen}}, \bibinfo {author}
  {\bibfnamefont {H.}~\bibnamefont {Wang}},\ and\ \bibinfo {author}
  {\bibfnamefont {C.~R.}\ \bibnamefont {Du}},\ }\bibfield  {title} {\bibinfo
  {title} {{Quantum Sensing and Imaging of Spin–Orbit-Torque-Driven Spin
  Dynamics in the Non-Collinear Antiferromagnet Mn$_3$Sn}},\ }\href
  {https://doi.org/10.1002/adma.202200327} {\bibfield  {journal} {\bibinfo
  {journal} {Advanced Materials}\ }\textbf {\bibinfo {volume} {34}},\ \bibinfo
  {pages} {2200327} (\bibinfo {year} {2022})}\BibitemShut {NoStop}%
\bibitem [{\citenamefont {Machado}\ \emph {et~al.}(2023)\citenamefont
  {Machado}, \citenamefont {Demler}, \citenamefont {Yao},\ and\ \citenamefont
  {Chatterjee}}]{machado2023}%
  \BibitemOpen
  \bibfield  {author} {\bibinfo {author} {\bibfnamefont {F.}~\bibnamefont
  {Machado}}, \bibinfo {author} {\bibfnamefont {E.~A.}\ \bibnamefont {Demler}},
  \bibinfo {author} {\bibfnamefont {N.~Y.}\ \bibnamefont {Yao}},\ and\ \bibinfo
  {author} {\bibfnamefont {S.}~\bibnamefont {Chatterjee}},\ }\bibfield  {title}
  {\bibinfo {title} {{Quantum Noise Spectroscopy of Dynamical Critical
  Phenomena}},\ }\href {https://doi.org/10.1103/PhysRevLett.131.070801}
  {\bibfield  {journal} {\bibinfo  {journal} {Physical Review Letters}\
  }\textbf {\bibinfo {volume} {131}},\ \bibinfo {pages} {070801} (\bibinfo
  {year} {2023})}\BibitemShut {NoStop}%
\bibitem [{\citenamefont {Ziffer}\ \emph {et~al.}(2024)\citenamefont {Ziffer},
  \citenamefont {Machado}, \citenamefont {Ursprung}, \citenamefont {Lozovoi},
  \citenamefont {Tazi}, \citenamefont {Yuan}, \citenamefont {Ziebel},
  \citenamefont {Delord}, \citenamefont {Zeng}, \citenamefont {Telford},
  \citenamefont {Chica}, \citenamefont {deQuilettes}, \citenamefont {Zhu},
  \citenamefont {Hone}, \citenamefont {Shepard}, \citenamefont {Roy},
  \citenamefont {de~Leon}, \citenamefont {Davis}, \citenamefont {Chatterjee},
  \citenamefont {Meriles}, \citenamefont {Owen}, \citenamefont {Schuck},\ and\
  \citenamefont {Pasupathy}}]{ziffer2024}%
  \BibitemOpen
  \bibfield  {author} {\bibinfo {author} {\bibfnamefont {M.~E.}\ \bibnamefont
  {Ziffer}}, \bibinfo {author} {\bibfnamefont {F.}~\bibnamefont {Machado}},
  \bibinfo {author} {\bibfnamefont {B.}~\bibnamefont {Ursprung}}, \bibinfo
  {author} {\bibfnamefont {A.}~\bibnamefont {Lozovoi}}, \bibinfo {author}
  {\bibfnamefont {A.~B.}\ \bibnamefont {Tazi}}, \bibinfo {author}
  {\bibfnamefont {Z.}~\bibnamefont {Yuan}}, \bibinfo {author} {\bibfnamefont
  {M.~E.}\ \bibnamefont {Ziebel}}, \bibinfo {author} {\bibfnamefont
  {T.}~\bibnamefont {Delord}}, \bibinfo {author} {\bibfnamefont
  {N.}~\bibnamefont {Zeng}}, \bibinfo {author} {\bibfnamefont {E.}~\bibnamefont
  {Telford}}, \bibinfo {author} {\bibfnamefont {D.~G.}\ \bibnamefont {Chica}},
  \bibinfo {author} {\bibfnamefont {D.~W.}\ \bibnamefont {deQuilettes}},
  \bibinfo {author} {\bibfnamefont {X.}~\bibnamefont {Zhu}}, \bibinfo {author}
  {\bibfnamefont {J.~C.}\ \bibnamefont {Hone}}, \bibinfo {author}
  {\bibfnamefont {K.~L.}\ \bibnamefont {Shepard}}, \bibinfo {author}
  {\bibfnamefont {X.}~\bibnamefont {Roy}}, \bibinfo {author} {\bibfnamefont
  {N.~P.}\ \bibnamefont {de~Leon}}, \bibinfo {author} {\bibfnamefont {E.~J.}\
  \bibnamefont {Davis}}, \bibinfo {author} {\bibfnamefont {S.}~\bibnamefont
  {Chatterjee}}, \bibinfo {author} {\bibfnamefont {C.~A.}\ \bibnamefont
  {Meriles}}, \bibinfo {author} {\bibfnamefont {J.~S.}\ \bibnamefont {Owen}},
  \bibinfo {author} {\bibfnamefont {P.~J.}\ \bibnamefont {Schuck}},\ and\
  \bibinfo {author} {\bibfnamefont {A.~N.}\ \bibnamefont {Pasupathy}},\
  }\bibfield  {title} {\bibinfo {title} {{Quantum Noise Spectroscopy of
  Critical Slowing Down in an Atomically Thin Magnet}},\ }\href
  {https://arxiv.org/abs/2407.05614} {\bibfield  {journal} {\bibinfo  {journal}
  {arXiv preprint arXiv:2407.05614}\ } (\bibinfo {year} {2024})}\BibitemShut
  {NoStop}%
\bibitem [{\citenamefont {Li}\ \emph {et~al.}(2024)\citenamefont {Li},
  \citenamefont {Ding}, \citenamefont {Wang}, \citenamefont {Sun},
  \citenamefont {Chen}, \citenamefont {Wang}, \citenamefont {Wang},
  \citenamefont {Gong}, \citenamefont {Zeng}, \citenamefont {Shi},\ and\
  \citenamefont {Du}}]{yuxinli2024}%
  \BibitemOpen
  \bibfield  {author} {\bibinfo {author} {\bibfnamefont {Y.}~\bibnamefont
  {Li}}, \bibinfo {author} {\bibfnamefont {Z.}~\bibnamefont {Ding}}, \bibinfo
  {author} {\bibfnamefont {C.}~\bibnamefont {Wang}}, \bibinfo {author}
  {\bibfnamefont {H.}~\bibnamefont {Sun}}, \bibinfo {author} {\bibfnamefont
  {Z.}~\bibnamefont {Chen}}, \bibinfo {author} {\bibfnamefont {P.}~\bibnamefont
  {Wang}}, \bibinfo {author} {\bibfnamefont {Y.}~\bibnamefont {Wang}}, \bibinfo
  {author} {\bibfnamefont {M.}~\bibnamefont {Gong}}, \bibinfo {author}
  {\bibfnamefont {H.}~\bibnamefont {Zeng}}, \bibinfo {author} {\bibfnamefont
  {F.}~\bibnamefont {Shi}},\ and\ \bibinfo {author} {\bibfnamefont
  {J.}~\bibnamefont {Du}},\ }\bibfield  {title} {\bibinfo {title} {{Critical
  fluctuation and noise spectra in two-dimensional Fe$_{3}$GeTe$_{2}$
  magnets}},\ }\href {https://arxiv.org/abs/2407.00647} {\bibfield  {journal}
  {\bibinfo  {journal} {arXiv preprint arXiv:2407.00647}\ } (\bibinfo {year}
  {2024})}\BibitemShut {NoStop}%
\bibitem [{\citenamefont {Ariyaratne}\ \emph {et~al.}(2018)\citenamefont
  {Ariyaratne}, \citenamefont {Bluvstein}, \citenamefont {Myers},\ and\
  \citenamefont {Jayich}}]{ariyaratne2018}%
  \BibitemOpen
  \bibfield  {author} {\bibinfo {author} {\bibfnamefont {A.}~\bibnamefont
  {Ariyaratne}}, \bibinfo {author} {\bibfnamefont {D.}~\bibnamefont
  {Bluvstein}}, \bibinfo {author} {\bibfnamefont {B.~A.}\ \bibnamefont
  {Myers}},\ and\ \bibinfo {author} {\bibfnamefont {A.~C.~B.}\ \bibnamefont
  {Jayich}},\ }\bibfield  {title} {\bibinfo {title} {{Nanoscale electrical
  conductivity imaging using a nitrogen-vacancy center in diamond}},\ }\href
  {https://doi.org/10.1038/s41467-018-04798-1} {\bibfield  {journal} {\bibinfo
  {journal} {Nature Communications}\ }\textbf {\bibinfo {volume} {9}},\
  \bibinfo {pages} {2406} (\bibinfo {year} {2018})}\BibitemShut {NoStop}%
\bibitem [{\citenamefont {Bersin}\ \emph {et~al.}(2019)\citenamefont {Bersin},
  \citenamefont {Walsh}, \citenamefont {Mouradian}, \citenamefont {Trusheim},
  \citenamefont {Schröder},\ and\ \citenamefont {Englund}}]{bersin2019}%
  \BibitemOpen
  \bibfield  {author} {\bibinfo {author} {\bibfnamefont {E.}~\bibnamefont
  {Bersin}}, \bibinfo {author} {\bibfnamefont {M.}~\bibnamefont {Walsh}},
  \bibinfo {author} {\bibfnamefont {S.~L.}\ \bibnamefont {Mouradian}}, \bibinfo
  {author} {\bibfnamefont {M.~E.}\ \bibnamefont {Trusheim}}, \bibinfo {author}
  {\bibfnamefont {T.}~\bibnamefont {Schröder}},\ and\ \bibinfo {author}
  {\bibfnamefont {D.}~\bibnamefont {Englund}},\ }\bibfield  {title} {\bibinfo
  {title} {{Individual Control and Readout of Qubits in a Sub-Diffraction
  Volume}},\ }\href {https://doi.org/10.1038/s41534-019-0154-y} {\bibfield
  {journal} {\bibinfo  {journal} {npj Quantum Information}\ }\textbf {\bibinfo
  {volume} {5}},\ \bibinfo {pages} {38} (\bibinfo {year} {2019})}\BibitemShut
  {NoStop}%
\bibitem [{\citenamefont {Monge}\ \emph {et~al.}(2023)\citenamefont {Monge},
  \citenamefont {Delord},\ and\ \citenamefont {Meriles}}]{monge2023}%
  \BibitemOpen
  \bibfield  {author} {\bibinfo {author} {\bibfnamefont {R.}~\bibnamefont
  {Monge}}, \bibinfo {author} {\bibfnamefont {T.}~\bibnamefont {Delord}},\ and\
  \bibinfo {author} {\bibfnamefont {C.~A.}\ \bibnamefont {Meriles}},\
  }\bibfield  {title} {\bibinfo {title} {{Reversible optical data storage below
  the diffraction limit}},\ }\href {https://doi.org/10.1038/s41565-023-01542-9}
  {\bibfield  {journal} {\bibinfo  {journal} {Nature Nanotechnology}\ }\textbf
  {\bibinfo {volume} {19}},\ \bibinfo {pages} {202} (\bibinfo {year}
  {2023})}\BibitemShut {NoStop}%
\bibitem [{\citenamefont {Ji}\ \emph {et~al.}(2024)\citenamefont {Ji},
  \citenamefont {Liu}, \citenamefont {Guo}, \citenamefont {Hu}, \citenamefont
  {Zhou}, \citenamefont {Dai}, \citenamefont {Chen}, \citenamefont {Yu},
  \citenamefont {Wang}, \citenamefont {Xia}, \citenamefont {Shi}, \citenamefont
  {Wang},\ and\ \citenamefont {Du}}]{wentaoji2024}%
  \BibitemOpen
  \bibfield  {author} {\bibinfo {author} {\bibfnamefont {W.}~\bibnamefont
  {Ji}}, \bibinfo {author} {\bibfnamefont {Z.}~\bibnamefont {Liu}}, \bibinfo
  {author} {\bibfnamefont {Y.}~\bibnamefont {Guo}}, \bibinfo {author}
  {\bibfnamefont {Z.}~\bibnamefont {Hu}}, \bibinfo {author} {\bibfnamefont
  {J.}~\bibnamefont {Zhou}}, \bibinfo {author} {\bibfnamefont {S.}~\bibnamefont
  {Dai}}, \bibinfo {author} {\bibfnamefont {Y.}~\bibnamefont {Chen}}, \bibinfo
  {author} {\bibfnamefont {P.}~\bibnamefont {Yu}}, \bibinfo {author}
  {\bibfnamefont {M.}~\bibnamefont {Wang}}, \bibinfo {author} {\bibfnamefont
  {K.}~\bibnamefont {Xia}}, \bibinfo {author} {\bibfnamefont {F.}~\bibnamefont
  {Shi}}, \bibinfo {author} {\bibfnamefont {Y.}~\bibnamefont {Wang}},\ and\
  \bibinfo {author} {\bibfnamefont {J.}~\bibnamefont {Du}},\ }\bibfield
  {title} {\bibinfo {title} {{Correlated sensing with a solid-state quantum
  multisensor system for atomic-scale structural analysis}},\ }\href
  {https://doi.org/10.1038/s41566-023-01352-4} {\bibfield  {journal} {\bibinfo
  {journal} {Nature Photonics}\ }\textbf {\bibinfo {volume} {18}},\ \bibinfo
  {pages} {43} (\bibinfo {year} {2024})}\BibitemShut {NoStop}%
\bibitem [{\citenamefont {Delord}\ \emph {et~al.}(2024)\citenamefont {Delord},
  \citenamefont {Monge},\ and\ \citenamefont {Meriles}}]{delord2024}%
  \BibitemOpen
  \bibfield  {author} {\bibinfo {author} {\bibfnamefont {T.}~\bibnamefont
  {Delord}}, \bibinfo {author} {\bibfnamefont {R.}~\bibnamefont {Monge}},\ and\
  \bibinfo {author} {\bibfnamefont {C.~A.}\ \bibnamefont {Meriles}},\
  }\bibfield  {title} {\bibinfo {title} {{Correlated spectroscopy of electric
  noise with color center clusters}},\ }\href
  {https://doi.org/10.1021/acs.nanolett.4c00222} {\bibfield  {journal}
  {\bibinfo  {journal} {Nano Letters}\ }\textbf {\bibinfo {volume} {24}},\
  \bibinfo {pages} {6474} (\bibinfo {year} {2024})}\BibitemShut {NoStop}%
\bibitem [{\citenamefont {Irber}\ \emph {et~al.}(2021)\citenamefont {Irber},
  \citenamefont {Poggiali}, \citenamefont {Kong}, \citenamefont {Kieschnick},
  \citenamefont {Lühmann}, \citenamefont {Kwiatkowski}, \citenamefont
  {Meijer}, \citenamefont {Du}, \citenamefont {Shi},\ and\ \citenamefont
  {Reinhard}}]{irber2021}%
  \BibitemOpen
  \bibfield  {author} {\bibinfo {author} {\bibfnamefont {D.~M.}\ \bibnamefont
  {Irber}}, \bibinfo {author} {\bibfnamefont {F.}~\bibnamefont {Poggiali}},
  \bibinfo {author} {\bibfnamefont {F.}~\bibnamefont {Kong}}, \bibinfo {author}
  {\bibfnamefont {M.}~\bibnamefont {Kieschnick}}, \bibinfo {author}
  {\bibfnamefont {T.}~\bibnamefont {Lühmann}}, \bibinfo {author}
  {\bibfnamefont {D.}~\bibnamefont {Kwiatkowski}}, \bibinfo {author}
  {\bibfnamefont {J.}~\bibnamefont {Meijer}}, \bibinfo {author} {\bibfnamefont
  {J.}~\bibnamefont {Du}}, \bibinfo {author} {\bibfnamefont {F.}~\bibnamefont
  {Shi}},\ and\ \bibinfo {author} {\bibfnamefont {F.}~\bibnamefont
  {Reinhard}},\ }\bibfield  {title} {\bibinfo {title} {{Robust all-optical
  single-shot readout of nitrogen-vacancy centers in diamond}},\ }\href
  {https://doi.org/10.1038/s41467-020-20755-3} {\bibfield  {journal} {\bibinfo
  {journal} {Nature Communications}\ }\textbf {\bibinfo {volume} {12}},\
  \bibinfo {pages} {532} (\bibinfo {year} {2021})}\BibitemShut {NoStop}%
\bibitem [{\citenamefont {Orphal-Kobin}\ \emph {et~al.}(2023)\citenamefont
  {Orphal-Kobin}, \citenamefont {Unterguggenberger}, \citenamefont
  {Pregnolato}, \citenamefont {Kemf}, \citenamefont {Matalla}, \citenamefont
  {Unger}, \citenamefont {Ostermay}, \citenamefont {Pieplow},\ and\
  \citenamefont {Schröder}}]{orphalkobin2023}%
  \BibitemOpen
  \bibfield  {author} {\bibinfo {author} {\bibfnamefont {L.}~\bibnamefont
  {Orphal-Kobin}}, \bibinfo {author} {\bibfnamefont {K.}~\bibnamefont
  {Unterguggenberger}}, \bibinfo {author} {\bibfnamefont {T.}~\bibnamefont
  {Pregnolato}}, \bibinfo {author} {\bibfnamefont {N.}~\bibnamefont {Kemf}},
  \bibinfo {author} {\bibfnamefont {M.}~\bibnamefont {Matalla}}, \bibinfo
  {author} {\bibfnamefont {R.-S.}\ \bibnamefont {Unger}}, \bibinfo {author}
  {\bibfnamefont {I.}~\bibnamefont {Ostermay}}, \bibinfo {author}
  {\bibfnamefont {G.}~\bibnamefont {Pieplow}},\ and\ \bibinfo {author}
  {\bibfnamefont {T.}~\bibnamefont {Schröder}},\ }\bibfield  {title} {\bibinfo
  {title} {{Optically Coherent Nitrogen-Vacancy Defect Centers in Diamond
  Nanostructures}},\ }\href {https://doi.org/10.1103/PhysRevX.13.011042}
  {\bibfield  {journal} {\bibinfo  {journal} {Physical Review X}\ }\textbf
  {\bibinfo {volume} {13}},\ \bibinfo {pages} {011042} (\bibinfo {year}
  {2023})}\BibitemShut {NoStop}%
\bibitem [{sup()}]{supp}%
  \BibitemOpen
  \href@noop {} {}\bibinfo {note} {See Supplemental Material for experimental
  details and theoretical derivations of correlation.}\BibitemShut {Stop}%
\bibitem [{\citenamefont {Batalov}\ \emph {et~al.}(2009)\citenamefont
  {Batalov}, \citenamefont {Jacques}, \citenamefont {Kaiser}, \citenamefont
  {Siyushev}, \citenamefont {Neumann}, \citenamefont {Rogers}, \citenamefont
  {McMurtrie}, \citenamefont {Manson}, \citenamefont {Jelezko},\ and\
  \citenamefont {Wrachtrup}}]{batalov2009}%
  \BibitemOpen
  \bibfield  {author} {\bibinfo {author} {\bibfnamefont {A.}~\bibnamefont
  {Batalov}}, \bibinfo {author} {\bibfnamefont {V.}~\bibnamefont {Jacques}},
  \bibinfo {author} {\bibfnamefont {F.}~\bibnamefont {Kaiser}}, \bibinfo
  {author} {\bibfnamefont {P.}~\bibnamefont {Siyushev}}, \bibinfo {author}
  {\bibfnamefont {P.}~\bibnamefont {Neumann}}, \bibinfo {author} {\bibfnamefont
  {L.~J.}\ \bibnamefont {Rogers}}, \bibinfo {author} {\bibfnamefont {R.~L.}\
  \bibnamefont {McMurtrie}}, \bibinfo {author} {\bibfnamefont {N.~B.}\
  \bibnamefont {Manson}}, \bibinfo {author} {\bibfnamefont {F.}~\bibnamefont
  {Jelezko}},\ and\ \bibinfo {author} {\bibfnamefont {J.}~\bibnamefont
  {Wrachtrup}},\ }\bibfield  {title} {\bibinfo {title} {{Low Temperature
  Studies of the Excited-State Structure of Negatively Charged Nitrogen-Vacancy
  Color Centers in Diamond}},\ }\href
  {https://doi.org/10.1103/PhysRevLett.102.195506} {\bibfield  {journal}
  {\bibinfo  {journal} {Physical Review Letters}\ }\textbf {\bibinfo {volume}
  {102}},\ \bibinfo {pages} {195506} (\bibinfo {year} {2009})}\BibitemShut
  {NoStop}%
\bibitem [{\citenamefont {Robledo}\ \emph {et~al.}(2011)\citenamefont
  {Robledo}, \citenamefont {Childress}, \citenamefont {Bernien}, \citenamefont
  {Hensen}, \citenamefont {Alkemade},\ and\ \citenamefont
  {Hanson}}]{robledo2011}%
  \BibitemOpen
  \bibfield  {author} {\bibinfo {author} {\bibfnamefont {L.}~\bibnamefont
  {Robledo}}, \bibinfo {author} {\bibfnamefont {L.}~\bibnamefont {Childress}},
  \bibinfo {author} {\bibfnamefont {H.}~\bibnamefont {Bernien}}, \bibinfo
  {author} {\bibfnamefont {B.}~\bibnamefont {Hensen}}, \bibinfo {author}
  {\bibfnamefont {P.~F.~A.}\ \bibnamefont {Alkemade}},\ and\ \bibinfo {author}
  {\bibfnamefont {R.}~\bibnamefont {Hanson}},\ }\bibfield  {title} {\bibinfo
  {title} {{High-fidelity projective read-out of a solid-state spin quantum
  register}},\ }\href {https://doi.org/10.1038/nature10401} {\bibfield
  {journal} {\bibinfo  {journal} {Nature}\ }\textbf {\bibinfo {volume} {477}},\
  \bibinfo {pages} {574} (\bibinfo {year} {2011})}\BibitemShut {NoStop}%
\bibitem [{\citenamefont {Gross}\ and\ \citenamefont
  {Haroche}(1982)}]{gross1982}%
  \BibitemOpen
  \bibfield  {author} {\bibinfo {author} {\bibfnamefont {M.}~\bibnamefont
  {Gross}}\ and\ \bibinfo {author} {\bibfnamefont {S.}~\bibnamefont
  {Haroche}},\ }\bibfield  {title} {\bibinfo {title} {{Superradiance: An Essay
  on the Theory of Collective Spontaneous Emission}},\ }\href
  {https://doi.org/10.1016/0370-1573(82)90102-8} {\bibfield  {journal}
  {\bibinfo  {journal} {Physics Reports}\ }\textbf {\bibinfo {volume} {93}},\
  \bibinfo {pages} {301} (\bibinfo {year} {1982})}\BibitemShut {NoStop}%
\bibitem [{\citenamefont {Yelin}\ \emph {et~al.}(2005)\citenamefont {Yelin},
  \citenamefont {Koštrun}, \citenamefont {Wang},\ and\ \citenamefont
  {Fleischhauer}}]{yelin2005}%
  \BibitemOpen
  \bibfield  {author} {\bibinfo {author} {\bibfnamefont {S.~F.}\ \bibnamefont
  {Yelin}}, \bibinfo {author} {\bibfnamefont {M.}~\bibnamefont {Koštrun}},
  \bibinfo {author} {\bibfnamefont {T.}~\bibnamefont {Wang}},\ and\ \bibinfo
  {author} {\bibfnamefont {M.}~\bibnamefont {Fleischhauer}},\ }\href
  {https://arxiv.org/abs/quant-ph/0509184} {\bibinfo {title} {{Correlation in
  superradiance: A closed-form approach to cooperative effects}}} (\bibinfo
  {year} {2005}),\ \Eprint {https://arxiv.org/abs/quant-ph/0509184}
  {arXiv:quant-ph/0509184 [quant-ph]} \BibitemShut {NoStop}%
\bibitem [{\citenamefont {Rubies-Bigorda}\ and\ \citenamefont
  {Yelin}(2022)}]{rubiesbigorda2022}%
  \BibitemOpen
  \bibfield  {author} {\bibinfo {author} {\bibfnamefont {O.}~\bibnamefont
  {Rubies-Bigorda}}\ and\ \bibinfo {author} {\bibfnamefont {S.~F.}\
  \bibnamefont {Yelin}},\ }\bibfield  {title} {\bibinfo {title} {{Superradiance
  and subradiance in inverted atomic arrays}},\ }\href
  {https://doi.org/10.1103/PhysRevA.106.053717} {\bibfield  {journal} {\bibinfo
   {journal} {Physical Review A}\ }\textbf {\bibinfo {volume} {106}},\ \bibinfo
  {pages} {053717} (\bibinfo {year} {2022})}\BibitemShut {NoStop}%
\bibitem [{\citenamefont {Fukami}\ \emph {et~al.}(2024)\citenamefont {Fukami},
  \citenamefont {Marcks}, \citenamefont {Candido}, \citenamefont {Weiss},
  \citenamefont {Soloway}, \citenamefont {Sullivan}, \citenamefont {Delegan},
  \citenamefont {Heremans}, \citenamefont {Flatt\'e},\ and\ \citenamefont
  {Awschalom}}]{fukami2024}%
  \BibitemOpen
  \bibfield  {author} {\bibinfo {author} {\bibfnamefont {M.}~\bibnamefont
  {Fukami}}, \bibinfo {author} {\bibfnamefont {J.~C.}\ \bibnamefont {Marcks}},
  \bibinfo {author} {\bibfnamefont {D.~R.}\ \bibnamefont {Candido}}, \bibinfo
  {author} {\bibfnamefont {L.~R.}\ \bibnamefont {Weiss}}, \bibinfo {author}
  {\bibfnamefont {B.}~\bibnamefont {Soloway}}, \bibinfo {author} {\bibfnamefont
  {S.~E.}\ \bibnamefont {Sullivan}}, \bibinfo {author} {\bibfnamefont
  {N.}~\bibnamefont {Delegan}}, \bibinfo {author} {\bibfnamefont {F.~J.}\
  \bibnamefont {Heremans}}, \bibinfo {author} {\bibfnamefont {M.~E.}\
  \bibnamefont {Flatt\'e}},\ and\ \bibinfo {author} {\bibfnamefont {D.~D.}\
  \bibnamefont {Awschalom}},\ }\bibfield  {title} {\bibinfo {title}
  {{Magnon-mediated qubit coupling determined via dissipation measurements}},\
  }\href {https://doi.org/10.1073/pnas.2313754120} {\bibfield  {journal}
  {\bibinfo  {journal} {Proceedings of the National Academy of Sciences of the
  United States of America}\ }\textbf {\bibinfo {volume} {121}},\ \bibinfo
  {pages} {e2313754120} (\bibinfo {year} {2024})}\BibitemShut {NoStop}%
\bibitem [{\citenamefont {{Maletinsky, Patrick and Hong, Sungkun and Grinolds,
  Michael S. and Hausmann, B. and Lukin, Mikhail D. and Walsworth, Ronald L.
  and Loncar, Marko and Yacoby, Amir}}(2012)}]{maletinsky2012}%
  \BibitemOpen
  \bibfield  {author} {\bibinfo {author} {\bibnamefont {{Maletinsky, Patrick
  and Hong, Sungkun and Grinolds, Michael S. and Hausmann, B. and Lukin,
  Mikhail D. and Walsworth, Ronald L. and Loncar, Marko and Yacoby, Amir}}},\
  }\bibfield  {title} {\bibinfo {title} {{A robust scanning diamond sensor for
  nanoscale imaging with single nitrogen-vacancy centres}},\ }\href
  {https://doi.org/10.1038/nnano.2012.50} {\bibfield  {journal} {\bibinfo
  {journal} {Nature Nanotechnology}\ }\textbf {\bibinfo {volume} {7}},\
  \bibinfo {pages} {320} (\bibinfo {year} {2012})}\BibitemShut {NoStop}%
\bibitem [{\citenamefont {Wang}\ \emph
  {et~al.}(2022{\natexlab{b}})\citenamefont {Wang}, \citenamefont {Liu},
  \citenamefont {Schloss}, \citenamefont {Alsid}, \citenamefont {Braje},\ and\
  \citenamefont {Cappellaro}}]{guoqingwang2022}%
  \BibitemOpen
  \bibfield  {author} {\bibinfo {author} {\bibfnamefont {G.}~\bibnamefont
  {Wang}}, \bibinfo {author} {\bibfnamefont {Y.-X.}\ \bibnamefont {Liu}},
  \bibinfo {author} {\bibfnamefont {J.~M.}\ \bibnamefont {Schloss}}, \bibinfo
  {author} {\bibfnamefont {S.~T.}\ \bibnamefont {Alsid}}, \bibinfo {author}
  {\bibfnamefont {D.~A.}\ \bibnamefont {Braje}},\ and\ \bibinfo {author}
  {\bibfnamefont {P.}~\bibnamefont {Cappellaro}},\ }\bibfield  {title}
  {\bibinfo {title} {{Sensing of Arbitrary-Frequency Fields Using a Quantum
  Mixer}},\ }\href {https://doi.org/10.1103/PhysRevX.12.021061} {\bibfield
  {journal} {\bibinfo  {journal} {Physical Review X}\ }\textbf {\bibinfo
  {volume} {12}},\ \bibinfo {pages} {021061} (\bibinfo {year}
  {2022}{\natexlab{b}})}\BibitemShut {NoStop}%
\bibitem [{\citenamefont {Laraoui}\ \emph {et~al.}(2013)\citenamefont
  {Laraoui}, \citenamefont {Dolde}, \citenamefont {Burk}, \citenamefont
  {Reinhard}, \citenamefont {Wrachtrup},\ and\ \citenamefont
  {Meriles}}]{laraoui2013}%
  \BibitemOpen
  \bibfield  {author} {\bibinfo {author} {\bibfnamefont {A.}~\bibnamefont
  {Laraoui}}, \bibinfo {author} {\bibfnamefont {F.}~\bibnamefont {Dolde}},
  \bibinfo {author} {\bibfnamefont {C.}~\bibnamefont {Burk}}, \bibinfo {author}
  {\bibfnamefont {F.}~\bibnamefont {Reinhard}}, \bibinfo {author}
  {\bibfnamefont {J.}~\bibnamefont {Wrachtrup}},\ and\ \bibinfo {author}
  {\bibfnamefont {C.~A.}\ \bibnamefont {Meriles}},\ }\bibfield  {title}
  {\bibinfo {title} {{High-resolution correlation spectroscopy of $^{13}$C
  spins near a nitrogen-vacancy centre in diamond}},\ }\href
  {https://doi.org/10.1038/ncomms2685} {\bibfield  {journal} {\bibinfo
  {journal} {Nature Communications}\ }\textbf {\bibinfo {volume} {4}},\
  \bibinfo {pages} {1651} (\bibinfo {year} {2013})}\BibitemShut {NoStop}%
\bibitem [{\citenamefont {Glenn}\ \emph {et~al.}(2018)\citenamefont {Glenn},
  \citenamefont {Bucher}, \citenamefont {Lee}, \citenamefont {Lukin},
  \citenamefont {Park},\ and\ \citenamefont {Walsworth}}]{glenn2018}%
  \BibitemOpen
  \bibfield  {author} {\bibinfo {author} {\bibfnamefont {D.~R.}\ \bibnamefont
  {Glenn}}, \bibinfo {author} {\bibfnamefont {D.~B.}\ \bibnamefont {Bucher}},
  \bibinfo {author} {\bibfnamefont {J.}~\bibnamefont {Lee}}, \bibinfo {author}
  {\bibfnamefont {M.~D.}\ \bibnamefont {Lukin}}, \bibinfo {author}
  {\bibfnamefont {H.}~\bibnamefont {Park}},\ and\ \bibinfo {author}
  {\bibfnamefont {R.~L.}\ \bibnamefont {Walsworth}},\ }\bibfield  {title}
  {\bibinfo {title} {{High-resolution magnetic resonance spectroscopy using a
  solid-state spin sensor}},\ }\href {https://doi.org/10.1038/nature25781}
  {\bibfield  {journal} {\bibinfo  {journal} {Nature}\ }\textbf {\bibinfo
  {volume} {555}},\ \bibinfo {pages} {351} (\bibinfo {year}
  {2018})}\BibitemShut {NoStop}%
\bibitem [{\citenamefont {K{\"o}yl{\"u}o{\u{g}}lu}\ \emph
  {et~al.}(2025)\citenamefont {K{\"o}yl{\"u}o{\u{g}}lu} \emph
  {et~al.}}]{koyluoglu2025}%
  \BibitemOpen
  \bibfield  {author} {\bibinfo {author} {\bibfnamefont {N.~U.}\ \bibnamefont
  {K{\"o}yl{\"u}o{\u{g}}lu}} \emph {et~al.},\ }\bibfield  {title} {\bibinfo
  {title} {{Interaction-enhanced many-body covariance magnetometry for strongly
  interacting quantum sensors}}} (\bibinfo {year} {2025}),\ \bibinfo {note} {in
  preparation}\BibitemShut {NoStop}%
\bibitem [{\citenamefont {Hosseinabadi}\ \emph {et~al.}(2025)\citenamefont
  {Hosseinabadi} \emph {et~al.}}]{Hosseinabadi2025}%
  \BibitemOpen
  \bibfield  {author} {\bibinfo {author} {\bibfnamefont {H.}~\bibnamefont
  {Hosseinabadi}} \emph {et~al.},\ }\bibfield  {title} {\bibinfo {title}
  {Theory of correlated spectroscopy with two {NV} centers}} (\bibinfo {year}
  {2025}),\ \bibinfo {note} {in preparation}\BibitemShut {NoStop}%
\bibitem [{\citenamefont {Zhang}\ \emph {et~al.}(2024)\citenamefont {Zhang},
  \citenamefont {Samajdar},\ and\ \citenamefont
  {Gopalakrishnan}}]{yifanzhang2024}%
  \BibitemOpen
  \bibfield  {author} {\bibinfo {author} {\bibfnamefont {Y.}~\bibnamefont
  {Zhang}}, \bibinfo {author} {\bibfnamefont {R.}~\bibnamefont {Samajdar}},\
  and\ \bibinfo {author} {\bibfnamefont {S.}~\bibnamefont {Gopalakrishnan}},\
  }\bibfield  {title} {\bibinfo {title} {{Nanoscale sensing of spatial
  correlations in nonequilibrium current noise}},\ }\href
  {https://arxiv.org/abs/2404.15398} {\bibfield  {journal} {\bibinfo  {journal}
  {arXiv preprint arXiv:2404.15398}\ } (\bibinfo {year} {2024})}\BibitemShut
  {NoStop}%
\bibitem [{\citenamefont {Agarwal}\ \emph {et~al.}(2017)\citenamefont
  {Agarwal}, \citenamefont {Schmidt}, \citenamefont {Halperin}, \citenamefont
  {Oganesyan}, \citenamefont {Zaránd}, \citenamefont {Lukin},\ and\
  \citenamefont {Demler}}]{agarwal2017}%
  \BibitemOpen
  \bibfield  {author} {\bibinfo {author} {\bibfnamefont {K.}~\bibnamefont
  {Agarwal}}, \bibinfo {author} {\bibfnamefont {R.}~\bibnamefont {Schmidt}},
  \bibinfo {author} {\bibfnamefont {B.}~\bibnamefont {Halperin}}, \bibinfo
  {author} {\bibfnamefont {V.}~\bibnamefont {Oganesyan}}, \bibinfo {author}
  {\bibfnamefont {G.}~\bibnamefont {Zaránd}}, \bibinfo {author} {\bibfnamefont
  {M.~D.}\ \bibnamefont {Lukin}},\ and\ \bibinfo {author} {\bibfnamefont
  {E.}~\bibnamefont {Demler}},\ }\bibfield  {title} {\bibinfo {title}
  {{Magnetic noise spectroscopy as a probe of local electronic correlations in
  two-dimensional systems}},\ }\href
  {https://doi.org/10.1103/PhysRevB.95.155107} {\bibfield  {journal} {\bibinfo
  {journal} {Physical Review B}\ }\textbf {\bibinfo {volume} {95}},\ \bibinfo
  {pages} {155107} (\bibinfo {year} {2017})}\BibitemShut {NoStop}%
\bibitem [{\citenamefont {Demokritov}\ \emph {et~al.}(2006)\citenamefont
  {Demokritov}, \citenamefont {Demidov}, \citenamefont {Dzyapko}, \citenamefont
  {Melkov}, \citenamefont {Serga}, \citenamefont {Hillebrands},\ and\
  \citenamefont {Slavin}}]{demokritov2006}%
  \BibitemOpen
  \bibfield  {author} {\bibinfo {author} {\bibfnamefont {S.~O.}\ \bibnamefont
  {Demokritov}}, \bibinfo {author} {\bibfnamefont {V.~E.}\ \bibnamefont
  {Demidov}}, \bibinfo {author} {\bibfnamefont {O.}~\bibnamefont {Dzyapko}},
  \bibinfo {author} {\bibfnamefont {G.~A.}\ \bibnamefont {Melkov}}, \bibinfo
  {author} {\bibfnamefont {A.~A.}\ \bibnamefont {Serga}}, \bibinfo {author}
  {\bibfnamefont {B.}~\bibnamefont {Hillebrands}},\ and\ \bibinfo {author}
  {\bibfnamefont {A.~N.}\ \bibnamefont {Slavin}},\ }\bibfield  {title}
  {\bibinfo {title} {{Bose--Einstein condensation of quasi-equilibrium magnons
  at room temperature under pumping}},\ }\href
  {https://doi.org/10.1038/nature05117} {\bibfield  {journal} {\bibinfo
  {journal} {Nature}\ }\textbf {\bibinfo {volume} {443}},\ \bibinfo {pages}
  {430} (\bibinfo {year} {2006})}\BibitemShut {NoStop}%
\bibitem [{\citenamefont {Rovny}\ \emph {et~al.}(2025)\citenamefont {Rovny},
  \citenamefont {Kolkowitz},\ and\ \citenamefont {de~Leon}}]{rovny2025}%
  \BibitemOpen
  \bibfield  {author} {\bibinfo {author} {\bibfnamefont {J.}~\bibnamefont
  {Rovny}}, \bibinfo {author} {\bibfnamefont {S.}~\bibnamefont {Kolkowitz}},\
  and\ \bibinfo {author} {\bibfnamefont {N.~P.}\ \bibnamefont {de~Leon}},\
  }\bibfield  {title} {\bibinfo {title} {Multi-qubit nanoscale sensing with
  entanglement as a resource},\ }\href {https://arxiv.org/abs/2504.12533}
  {\bibfield  {journal} {\bibinfo  {journal} {arXiv preprint arXiv:2504.12533}\
  } (\bibinfo {year} {2025})}\BibitemShut {NoStop}%
\end{thebibliography}%


\begin{thebibliography}{11}%
\makeatletter
\providecommand \@ifxundefined [1]{%
 \@ifx{#1\undefined}
}%
\providecommand \@ifnum [1]{%
 \ifnum #1\expandafter \@firstoftwo
 \else \expandafter \@secondoftwo
 \fi
}%
\providecommand \@ifx [1]{%
 \ifx #1\expandafter \@firstoftwo
 \else \expandafter \@secondoftwo
 \fi
}%
\providecommand \natexlab [1]{#1}%
\providecommand \enquote  [1]{``#1''}%
\providecommand \bibnamefont  [1]{#1}%
\providecommand \bibfnamefont [1]{#1}%
\providecommand \citenamefont [1]{#1}%
\providecommand \href@noop [0]{\@secondoftwo}%
\providecommand \href [0]{\begingroup \@sanitize@url \@href}%
\providecommand \@href[1]{\@@startlink{#1}\@@href}%
\providecommand \@@href[1]{\endgroup#1\@@endlink}%
\providecommand \@sanitize@url [0]{\catcode `\\12\catcode `\$12\catcode
  `\&12\catcode `\#12\catcode `\^12\catcode `\_12\catcode `\%12\relax}%
\providecommand \@@startlink[1]{}%
\providecommand \@@endlink[0]{}%
\providecommand \url  [0]{\begingroup\@sanitize@url \@url }%
\providecommand \@url [1]{\endgroup\@href {#1}{\urlprefix }}%
\providecommand \urlprefix  [0]{URL }%
\providecommand \Eprint [0]{\href }%
\providecommand \doibase [0]{https://doi.org/}%
\providecommand \selectlanguage [0]{\@gobble}%
\providecommand \bibinfo  [0]{\@secondoftwo}%
\providecommand \bibfield  [0]{\@secondoftwo}%
\providecommand \translation [1]{[#1]}%
\providecommand \BibitemOpen [0]{}%
\providecommand \bibitemStop [0]{}%
\providecommand \bibitemNoStop [0]{.\EOS\space}%
\providecommand \EOS [0]{\spacefactor3000\relax}%
\providecommand \BibitemShut  [1]{\csname bibitem#1\endcsname}%
\let\auto@bib@innerbib\@empty
\bibitem [{\citenamefont {Sangtawesin}\ \emph {et~al.}(2019)\citenamefont
  {Sangtawesin}, \citenamefont {Dwyer}, \citenamefont {Srinivasan},
  \citenamefont {Allred}, \citenamefont {Rodgers}, \citenamefont {De~Greve},
  \citenamefont {Stacey}, \citenamefont {Dontschuk}, \citenamefont {O'Donnell},
  \citenamefont {Hu}, \citenamefont {Evans}, \citenamefont {Jaye},
  \citenamefont {Fischer}, \citenamefont {Markham}, \citenamefont {Twitchen},
  \citenamefont {Park}, \citenamefont {Lukin},\ and\ \citenamefont
  {de~Leon}}]{sangtawesin2019}%
  \BibitemOpen
  \bibfield  {author} {\bibinfo {author} {\bibfnamefont {S.}~\bibnamefont
  {Sangtawesin}}, \bibinfo {author} {\bibfnamefont {B.~L.}\ \bibnamefont
  {Dwyer}}, \bibinfo {author} {\bibfnamefont {S.}~\bibnamefont {Srinivasan}},
  \bibinfo {author} {\bibfnamefont {J.~J.}\ \bibnamefont {Allred}}, \bibinfo
  {author} {\bibfnamefont {L.~V.~H.}\ \bibnamefont {Rodgers}}, \bibinfo
  {author} {\bibfnamefont {K.}~\bibnamefont {De~Greve}}, \bibinfo {author}
  {\bibfnamefont {A.}~\bibnamefont {Stacey}}, \bibinfo {author} {\bibfnamefont
  {N.}~\bibnamefont {Dontschuk}}, \bibinfo {author} {\bibfnamefont {K.~M.}\
  \bibnamefont {O'Donnell}}, \bibinfo {author} {\bibfnamefont {D.}~\bibnamefont
  {Hu}}, \bibinfo {author} {\bibfnamefont {D.~A.}\ \bibnamefont {Evans}},
  \bibinfo {author} {\bibfnamefont {C.}~\bibnamefont {Jaye}}, \bibinfo {author}
  {\bibfnamefont {D.~A.}\ \bibnamefont {Fischer}}, \bibinfo {author}
  {\bibfnamefont {M.~L.}\ \bibnamefont {Markham}}, \bibinfo {author}
  {\bibfnamefont {D.~J.}\ \bibnamefont {Twitchen}}, \bibinfo {author}
  {\bibfnamefont {H.}~\bibnamefont {Park}}, \bibinfo {author} {\bibfnamefont
  {M.~D.}\ \bibnamefont {Lukin}},\ and\ \bibinfo {author} {\bibfnamefont
  {N.~P.}\ \bibnamefont {de~Leon}},\ }\bibfield  {title} {\bibinfo {title}
  {Origins of diamond surface noise probed by correlating single-spin
  measurements with surface spectroscopy},\ }\href
  {https://doi.org/10.1103/PhysRevX.9.031052} {\bibfield  {journal} {\bibinfo
  {journal} {Physical Review X}\ }\textbf {\bibinfo {volume} {9}},\ \bibinfo
  {pages} {031052} (\bibinfo {year} {2019})}\BibitemShut {NoStop}%
\bibitem [{\citenamefont {Pezzagna}\ \emph {et~al.}(2010)\citenamefont
  {Pezzagna}, \citenamefont {Naydenov}, \citenamefont {Jelezko}, \citenamefont
  {Wrachtrup},\ and\ \citenamefont {Meijer}}]{pezzagna2010}%
  \BibitemOpen
  \bibfield  {author} {\bibinfo {author} {\bibfnamefont {S.}~\bibnamefont
  {Pezzagna}}, \bibinfo {author} {\bibfnamefont {B.}~\bibnamefont {Naydenov}},
  \bibinfo {author} {\bibfnamefont {F.}~\bibnamefont {Jelezko}}, \bibinfo
  {author} {\bibfnamefont {J.}~\bibnamefont {Wrachtrup}},\ and\ \bibinfo
  {author} {\bibfnamefont {J.}~\bibnamefont {Meijer}},\ }\bibfield  {title}
  {\bibinfo {title} {Creation efficiency of nitrogen-vacancy centres in
  diamond},\ }\href {https://doi.org/10.1088/1367-2630/12/6/065017} {\bibfield
  {journal} {\bibinfo  {journal} {New Journal of Physics}\ }\textbf {\bibinfo
  {volume} {12}},\ \bibinfo {pages} {065017} (\bibinfo {year}
  {2010})}\BibitemShut {NoStop}%
\bibitem [{\citenamefont {Pham}\ \emph {et~al.}(2016)\citenamefont {Pham},
  \citenamefont {DeVience}, \citenamefont {Casola}, \citenamefont {Lovchinsky},
  \citenamefont {Sushkov}, \citenamefont {Bersin}, \citenamefont {Lee},
  \citenamefont {Urbach}, \citenamefont {Cappellaro}, \citenamefont {Park},
  \citenamefont {Yacoby}, \citenamefont {Lukin},\ and\ \citenamefont
  {Walsworth}}]{linhpham2016}%
  \BibitemOpen
  \bibfield  {author} {\bibinfo {author} {\bibfnamefont {L.~M.}\ \bibnamefont
  {Pham}}, \bibinfo {author} {\bibfnamefont {S.~J.}\ \bibnamefont {DeVience}},
  \bibinfo {author} {\bibfnamefont {F.}~\bibnamefont {Casola}}, \bibinfo
  {author} {\bibfnamefont {I.}~\bibnamefont {Lovchinsky}}, \bibinfo {author}
  {\bibfnamefont {A.~O.}\ \bibnamefont {Sushkov}}, \bibinfo {author}
  {\bibfnamefont {E.}~\bibnamefont {Bersin}}, \bibinfo {author} {\bibfnamefont
  {J.}~\bibnamefont {Lee}}, \bibinfo {author} {\bibfnamefont {E.}~\bibnamefont
  {Urbach}}, \bibinfo {author} {\bibfnamefont {P.}~\bibnamefont {Cappellaro}},
  \bibinfo {author} {\bibfnamefont {H.}~\bibnamefont {Park}}, \bibinfo {author}
  {\bibfnamefont {A.}~\bibnamefont {Yacoby}}, \bibinfo {author} {\bibfnamefont
  {M.}~\bibnamefont {Lukin}},\ and\ \bibinfo {author} {\bibfnamefont {R.~L.}\
  \bibnamefont {Walsworth}},\ }\bibfield  {title} {\bibinfo {title} {{NMR}
  technique for determining the depth of shallow nitrogen-vacancy centers in
  diamond},\ }\href {https://doi.org/10.1103/PhysRevB.93.045425} {\bibfield
  {journal} {\bibinfo  {journal} {Physical Review B}\ }\textbf {\bibinfo
  {volume} {93}},\ \bibinfo {pages} {045425} (\bibinfo {year}
  {2016})}\BibitemShut {NoStop}%
\bibitem [{\citenamefont {Xue}\ \emph {et~al.}(2024)\citenamefont {Xue},
  \citenamefont {Maksimovic}, \citenamefont {Dolgirev}, \citenamefont {Xia},
  \citenamefont {Kitagawa}, \citenamefont {Müller}, \citenamefont {Machado},
  \citenamefont {Klein}, \citenamefont {MacNeill}, \citenamefont {Watanabe},
  \citenamefont {Taniguchi}, \citenamefont {Jarillo-Herrero}, \citenamefont
  {Lukin}, \citenamefont {Demler},\ and\ \citenamefont
  {Yacoby}}]{ruolanxue2024supp}%
  \BibitemOpen
  \bibfield  {author} {\bibinfo {author} {\bibfnamefont {R.}~\bibnamefont
  {Xue}}, \bibinfo {author} {\bibfnamefont {N.}~\bibnamefont {Maksimovic}},
  \bibinfo {author} {\bibfnamefont {P.~E.}\ \bibnamefont {Dolgirev}}, \bibinfo
  {author} {\bibfnamefont {L.-Q.}\ \bibnamefont {Xia}}, \bibinfo {author}
  {\bibfnamefont {R.}~\bibnamefont {Kitagawa}}, \bibinfo {author}
  {\bibfnamefont {A.}~\bibnamefont {Müller}}, \bibinfo {author} {\bibfnamefont
  {F.}~\bibnamefont {Machado}}, \bibinfo {author} {\bibfnamefont {D.~R.}\
  \bibnamefont {Klein}}, \bibinfo {author} {\bibfnamefont {D.}~\bibnamefont
  {MacNeill}}, \bibinfo {author} {\bibfnamefont {K.}~\bibnamefont {Watanabe}},
  \bibinfo {author} {\bibfnamefont {T.}~\bibnamefont {Taniguchi}}, \bibinfo
  {author} {\bibfnamefont {P.}~\bibnamefont {Jarillo-Herrero}}, \bibinfo
  {author} {\bibfnamefont {M.~D.}\ \bibnamefont {Lukin}}, \bibinfo {author}
  {\bibfnamefont {E.}~\bibnamefont {Demler}},\ and\ \bibinfo {author}
  {\bibfnamefont {A.}~\bibnamefont {Yacoby}},\ }\bibfield  {title} {\bibinfo
  {title} {{Signatures of magnon hydrodynamics in an atomically-thin
  ferromagnet}},\ }\href {https://arxiv.org/abs/2403.01057} {\bibfield
  {journal} {\bibinfo  {journal} {arXiv preprint arXiv:2403.01057}\ } (\bibinfo
  {year} {2024})}\BibitemShut {NoStop}%
\bibitem [{\citenamefont {Irber}\ \emph {et~al.}(2021)\citenamefont {Irber},
  \citenamefont {Poggiali}, \citenamefont {Kong}, \citenamefont {Kieschnick},
  \citenamefont {Lühmann}, \citenamefont {Kwiatkowski}, \citenamefont
  {Meijer}, \citenamefont {Du}, \citenamefont {Shi},\ and\ \citenamefont
  {Reinhard}}]{irber2021supp}%
  \BibitemOpen
  \bibfield  {author} {\bibinfo {author} {\bibfnamefont {D.~M.}\ \bibnamefont
  {Irber}}, \bibinfo {author} {\bibfnamefont {F.}~\bibnamefont {Poggiali}},
  \bibinfo {author} {\bibfnamefont {F.}~\bibnamefont {Kong}}, \bibinfo {author}
  {\bibfnamefont {M.}~\bibnamefont {Kieschnick}}, \bibinfo {author}
  {\bibfnamefont {T.}~\bibnamefont {Lühmann}}, \bibinfo {author}
  {\bibfnamefont {D.}~\bibnamefont {Kwiatkowski}}, \bibinfo {author}
  {\bibfnamefont {J.}~\bibnamefont {Meijer}}, \bibinfo {author} {\bibfnamefont
  {J.}~\bibnamefont {Du}}, \bibinfo {author} {\bibfnamefont {F.}~\bibnamefont
  {Shi}},\ and\ \bibinfo {author} {\bibfnamefont {F.}~\bibnamefont
  {Reinhard}},\ }\bibfield  {title} {\bibinfo {title} {{Robust all-optical
  single-shot readout of nitrogen-vacancy centers in diamond}},\ }\href
  {https://doi.org/10.1038/s41467-020-20755-3} {\bibfield  {journal} {\bibinfo
  {journal} {Nature Communications}\ }\textbf {\bibinfo {volume} {12}},\
  \bibinfo {pages} {532} (\bibinfo {year} {2021})}\BibitemShut {NoStop}%
\bibitem [{\citenamefont {Hopper}\ \emph {et~al.}(2018)\citenamefont {Hopper},
  \citenamefont {Shulevitz},\ and\ \citenamefont {Bassett}}]{hopper2018}%
  \BibitemOpen
  \bibfield  {author} {\bibinfo {author} {\bibfnamefont {D.~A.}\ \bibnamefont
  {Hopper}}, \bibinfo {author} {\bibfnamefont {H.~J.}\ \bibnamefont
  {Shulevitz}},\ and\ \bibinfo {author} {\bibfnamefont {L.~C.}\ \bibnamefont
  {Bassett}},\ }\bibfield  {title} {\bibinfo {title} {Spin readout techniques
  of the nitrogen-vacancy center in diamond},\ }\href
  {https://doi.org/10.3390/mi9090437} {\bibfield  {journal} {\bibinfo
  {journal} {Micromachines}\ }\textbf {\bibinfo {volume} {9}},\ \bibinfo
  {pages} {437} (\bibinfo {year} {2018})}\BibitemShut {NoStop}%
\bibitem [{\citenamefont {Rovny}\ \emph {et~al.}(2022)\citenamefont {Rovny},
  \citenamefont {Yuan}, \citenamefont {Fitzpatrick}, \citenamefont {Abdalla},
  \citenamefont {Futamura}, \citenamefont {Fox}, \citenamefont {Cambria},
  \citenamefont {Kolkowitz},\ and\ \citenamefont {de~Leon}}]{rovny2022supp}%
  \BibitemOpen
  \bibfield  {author} {\bibinfo {author} {\bibfnamefont {J.}~\bibnamefont
  {Rovny}}, \bibinfo {author} {\bibfnamefont {Z.}~\bibnamefont {Yuan}},
  \bibinfo {author} {\bibfnamefont {M.}~\bibnamefont {Fitzpatrick}}, \bibinfo
  {author} {\bibfnamefont {A.~I.}\ \bibnamefont {Abdalla}}, \bibinfo {author}
  {\bibfnamefont {L.}~\bibnamefont {Futamura}}, \bibinfo {author}
  {\bibfnamefont {C.}~\bibnamefont {Fox}}, \bibinfo {author} {\bibfnamefont
  {M.~C.}\ \bibnamefont {Cambria}}, \bibinfo {author} {\bibfnamefont
  {S.}~\bibnamefont {Kolkowitz}},\ and\ \bibinfo {author} {\bibfnamefont
  {N.~P.}\ \bibnamefont {de~Leon}},\ }\bibfield  {title} {\bibinfo {title}
  {{Nanoscale covariance magnetometry with diamond quantum sensors}},\ }\href
  {https://doi.org/10.1126/science.ade9858} {\bibfield  {journal} {\bibinfo
  {journal} {Science}\ }\textbf {\bibinfo {volume} {378}},\ \bibinfo {pages}
  {1301} (\bibinfo {year} {2022})}\BibitemShut {NoStop}%
\bibitem [{\citenamefont {Degen}\ \emph {et~al.}(2017)\citenamefont {Degen},
  \citenamefont {Reinhard},\ and\ \citenamefont {Cappellaro}}]{degen2017}%
  \BibitemOpen
  \bibfield  {author} {\bibinfo {author} {\bibfnamefont {C.~L.}\ \bibnamefont
  {Degen}}, \bibinfo {author} {\bibfnamefont {F.}~\bibnamefont {Reinhard}},\
  and\ \bibinfo {author} {\bibfnamefont {P.}~\bibnamefont {Cappellaro}},\
  }\bibfield  {title} {\bibinfo {title} {Quantum sensing},\ }\href
  {https://doi.org/10.1103/RevModPhys.89.035002} {\bibfield  {journal}
  {\bibinfo  {journal} {Reviews of Modern Physics}\ }\textbf {\bibinfo {volume}
  {89}},\ \bibinfo {pages} {035002} (\bibinfo {year} {2017})}\BibitemShut
  {NoStop}%
\bibitem [{\citenamefont {Sza\'nkowski}\ \emph {et~al.}(2016)\citenamefont
  {Sza\'nkowski}, \citenamefont {Trippenbach},\ and\ \citenamefont
  {Cywi\'nski}}]{szankowski2016}%
  \BibitemOpen
  \bibfield  {author} {\bibinfo {author} {\bibfnamefont {P.}~\bibnamefont
  {Sza\'nkowski}}, \bibinfo {author} {\bibfnamefont {M.}~\bibnamefont
  {Trippenbach}},\ and\ \bibinfo {author} {\bibfnamefont {L.}~\bibnamefont
  {Cywi\'nski}},\ }\bibfield  {title} {\bibinfo {title} {Spectroscopy of cross
  correlations of environmental noises with two qubits},\ }\href
  {https://doi.org/10.1103/PhysRevA.94.012109} {\bibfield  {journal} {\bibinfo
  {journal} {Physical Review A}\ }\textbf {\bibinfo {volume} {94}},\ \bibinfo
  {pages} {012109} (\bibinfo {year} {2016})}\BibitemShut {NoStop}%
\bibitem [{\citenamefont {Thibaudeau}\ and\ \citenamefont
  {Beaujouan}(2012)}]{thibaudeau2012thermostatting}%
  \BibitemOpen
  \bibfield  {author} {\bibinfo {author} {\bibfnamefont {P.}~\bibnamefont
  {Thibaudeau}}\ and\ \bibinfo {author} {\bibfnamefont {D.}~\bibnamefont
  {Beaujouan}},\ }\bibfield  {title} {\bibinfo {title} {Thermostatting the
  atomic spin dynamics from controlled demons},\ }\href
  {https://doi.org/10.1016/j.physa.2011.11.030} {\bibfield  {journal} {\bibinfo
   {journal} {Physica A}\ }\textbf {\bibinfo {volume} {391}},\ \bibinfo {pages}
  {1963} (\bibinfo {year} {2012})}\BibitemShut {NoStop}%
\bibitem [{\citenamefont {Childress}\ \emph {et~al.}(2006)\citenamefont
  {Childress}, \citenamefont {Gurudev~Dutt}, \citenamefont {Taylor},
  \citenamefont {Zibrov}, \citenamefont {Jelezko}, \citenamefont {Wrachtrup},
  \citenamefont {Hemmer},\ and\ \citenamefont {Lukin}}]{childress2006}%
  \BibitemOpen
  \bibfield  {author} {\bibinfo {author} {\bibfnamefont {L.}~\bibnamefont
  {Childress}}, \bibinfo {author} {\bibfnamefont {M.~V.}\ \bibnamefont
  {Gurudev~Dutt}}, \bibinfo {author} {\bibfnamefont {J.~M.}\ \bibnamefont
  {Taylor}}, \bibinfo {author} {\bibfnamefont {A.~S.}\ \bibnamefont {Zibrov}},
  \bibinfo {author} {\bibfnamefont {F.}~\bibnamefont {Jelezko}}, \bibinfo
  {author} {\bibfnamefont {J.}~\bibnamefont {Wrachtrup}}, \bibinfo {author}
  {\bibfnamefont {P.~R.}\ \bibnamefont {Hemmer}},\ and\ \bibinfo {author}
  {\bibfnamefont {M.~D.}\ \bibnamefont {Lukin}},\ }\bibfield  {title} {\bibinfo
  {title} {Coherent dynamics of coupled electron and nuclear spin qubits in
  diamond},\ }\href {https://doi.org/10.1126/science.1131871} {\bibfield
  {journal} {\bibinfo  {journal} {Science}\ }\textbf {\bibinfo {volume}
  {314}},\ \bibinfo {pages} {281} (\bibinfo {year} {2006})}\BibitemShut
  {NoStop}%
\end{thebibliography}%

\end{document}


\title{Supplemental Material to \\
``Wideband covariance magnetometry below the diffraction limit''}

\author{Xuan~Hoang~Le}
\affiliation{\HarvardPhysics}
\affiliation{\HarvardCCB}

\author{Pavel~E.~Dolgirev}
\affiliation{\HarvardPhysics}

\author{Piotr~Put}
\affiliation{\HarvardPhysics}
\affiliation{\HarvardCCB}

\author{Eric~L.~Peterson}
\affiliation{\HarvardPhysics}

\author{Arjun~Pillai}
\affiliation{\HarvardCCB}

\author{Alexander~A.~Zibrov}
\affiliation{\HarvardPhysics}
\affiliation{\HarvardCCB}

\author{Eugene~Demler}
\affiliation{\ETH}

\author{Hongkun~Park}
\affiliation{\HarvardPhysics}
\affiliation{\HarvardCCB}

\author{Mikhail~D.~Lukin}
\thanks{Corresponding author: \href{mailto:lukin@physics.harvard.edu}{lukin@physics.harvard.edu}}
\affiliation{\HarvardPhysics}

\date{April 30, 2025}

\maketitle

\beginsupplement

\section{Experimental setup}

\subsection{Diamond sample}
As a platform for correlation sensing, we use an electronic grade CVD-grown diamond (Element Six) with an overgrown diamond layer isotopically enriched in $^{12}$C to minimize nitrogen-vacancy (NV) center coupling to spurious nuclear spins. NV centers are created by $^{14}$N ion implantation at 25\,keV with a dose of $10^{11}$\,cm$^{-2}$, followed by a two-step vacuum annealing process as in Ref.~\cite{sangtawesin2019}: first, at 800\,$^{\circ}\mathrm{C}$ for 8 hours to form NV centers, then at 1200\,$^{\circ}\mathrm{C}$ for 2 hours to remove divacancies and multi-vacancy centers that could result in magnetic noise and inhomogeneous broadening of optical linewidths. The diamond is then cleaned in a refluxing mixture of 1:1:1 sulfuric:nitric:perchloric acid (triacid cleaning) to remove graphitic carbon on the surface. This procedure results in NV centers roughly 50\,nm below the surface with a depth distribution across $\sim$20\,nm, based on Stopping and Range of Ions in Matter (SRIM) simulation with corrections~\cite{pezzagna2010, linhpham2016, ruolanxue2024supp}, with narrow, stable optical transitions (Sec.~\ref{sec:read_char}) and long coherence times (Sec.~\ref{sec:singleNVT2}). The choice of implantation fluence as above yields single resolvable NV centers but also the NV cluster used in this work.

To enable independent driving of the two sensors used for correlation measurements, we select two NV centers oriented along different crystallographic axes and study them throughout this work. 
With the magnetic field aligned along the crystallographic axis of NV1, the resulting optically detected magnetic resonance (ODMR) spectrum, shown in Fig.~\ref{fig::figS1}, exhibits a clear frequency separation between the microwave transitions of NV1 and NV2, allowing independent control (see main text). For all experiments in this work, we use $\ket{m_s = 0}$ and $\ket{m_s = -1}$ as the quantum states for both NVs.

\subsection{Optical setup}

We use a home-built single-path scanning confocal microscope setup integrated with an optical cryostat (Montana Instruments Cryo-Optic X-Plane) that houses the diamond sample on the Agile Temperature Sample Mount (ATSM), temperature-stabilized at $\sim11$\,K throughout the experiment by a LakeShore 335 Temperature Controller. An in-vacuum, room-temperature microscope objective (Zeiss LD EC Epiplan-Neofluar 100x, NA = 0.75, WD = 4\,mm) focuses laser light used for spin and charge initialization, ionization, and readout of the NV centers. The same objective collects photoluminescence (PL), which is separated from all excitation laser beams via a dichroic mirror (Semrock Di03-R660-t1-25x36), coupled to a single-mode fiber, and measured using a single-photon-counting module (Excelitas SPCM-AQRH-14-FC). XY scanning for confocal imaging is implemented using a pair of galvo mirrors (Thorlabs GVS012) combined with Z-focusing achieved by moving the sample mounted on a piezo element (Attocube ANPz101). 

The experiment utilizes 4 lasers. Spin and charge initialization is performed with an optically pumped solid-state green laser (Laserglow Technologies, 532\,nm, 540\,\textmu W) modulated by an acousto-optic modulator (AOM) (Gooch \& Housego AOMO 3080-125). The red resonant excitation, used for both readout and assisting ionization, is provided by a dedicated external cavity diode laser (ECDL) (New Focus TLB-6304H-LN Velocity) for each NV. Each ECDL is modulated by an AOM (Isomet 1250C-974), then coupled to optical fibers and combined by a 50:50 fiber beamsplitter (Thorlabs TW630R5A2). Both ECDLs are actively frequency-stabilized to the NV transitions near 637.22\,nm using a wavemeter (HighFinesse WS7-30) to measure the frequency with 30\,MHz precision and feedback on the piezo controller of the ECDL's cavity. NV ionization is accomplished using a high-power fiber-pigtailed diode laser at 660\,nm (Thorlabs LP660-SF60, 72$\,$mW), pulsed by a current driver board (Highland Technology D200). Since resonant readout only needs low optical power (<~2\,\textmu W, see Sec.~\ref{sec:read_char}), the resonant and 660\,nm excitations are combined by a 10:90 (respectively) fiber optic coupler (Thorlabs TN632R2F1), then launched to free space and combined with the 532\,nm light by a dichroic mirror (Semrock Di02-R532-25x36).

\subsection{Microwave control setup and test signals}

Microwave (MW) drive is delivered to the NV centers via a photolithographically defined stripline (225\,nm layer of Au on 10\,nm adhesive layer of Ti) in the shape of omega-loop (96\,nm inner radius, 50\,nm width) fabricated directly on the diamond and wirebonded to a copper PCB.

The control MW pulses are generated via IQ mixing, using a MW generator (Stanford Research Systems (SRS) SG386) and analog control pulses produced by a high-sampling-rate arbitrary waveform generator (AWG) (Siglent SDG6022X). Each NV is controlled by a dedicated signal generator and AWG combination. The aforementioned generators are referred to as ``main'' generators to distinguish from two additional signal generators (SRS SG384 and Rohde \& Schwarz SMC100A), used to apply MW drive on the $\ket{m_s = 0} \rightarrow\ket{m_s = +1}$ transitions of each NV during charge-state readout (see main text), and controlled by MW switches.

All four MW signals are combined with a power splitter, then split into a \textit{high-power} path used for MW pulses and a \textit{low-power} path used for two purposes: continuous-wave MW drives during charge-state readout, and the test signal for correlated $T_1$ experiments. The \textit{high-power} path has no attenuation, while the \textit{low-power} path has 16\,dB attenuation. Both paths are connected to a MW switch that toggles between them, then the signal is amplified by a medium-gain (Mini-Circuits ZX60-6013E-S+) and a high-gain, high power amplifier (Mini-Circuits ZHL-25W-63+).

The MHz-range test signals for correlated $T_2$ experiments are generated by another AWG with controllable noise bandwidth (Keysight 33522A). These test signals undergo no further amplification before being combined with the post-amplification MW drive signals, then sent to the stripline. In Fig.~3 of the main text, the test signal is a 2.5\,MHz signal, phase-randomized with 25\,kHz bandwidth Gaussian noise. In Fig.~\ref{fig::figS8}, the two test signals from each channel of the AWG are at 2.5\,MHz and 2.0833\,MHz, both phase-randomized with 25\,kHz bandwidth Gaussian noise, then combined before being sent to the stripline.

For correlated $T_1$ experiments, a 10\,MHz bandwidth Gaussian noise from the Keysight AWG amplitude-modulates the two ``main'' generators via their Q-ports, while their I-ports remain controlled by the Siglent AWGs to apply a $\pi-$pulse if needed. This scheme produces the correlated depolarization test signal shown in Fig.~4(a) of the main text. During the sensing time, since no MW pulses are required and also to reduce the noise power, we use the \textit{low-power} MW path.

The single 4-way power splitter in use is a Mini-Circuits ZA4PD-4-S+; all 2-way power splitters are Mini-Circuits ZFRSC-42-S+. Microwave circulators (Pasternack PE8301 and DiTom D312060) are used right after each ``main'' generator's output to avoid spurious effects when driving the 2 NVs at the same time with different-frequency drives. Timing of all optical pulses, MW switch toggling, and triggers for Siglent AWGs and data acquisition, is orchestrated by a high speed programmable pulse generator (SpinCore PulseBlasterESR-PRO-USB-RM2).

\begin{figure*}[hbt!]  
    \centering
    \includegraphics[width=0.5\textwidth]{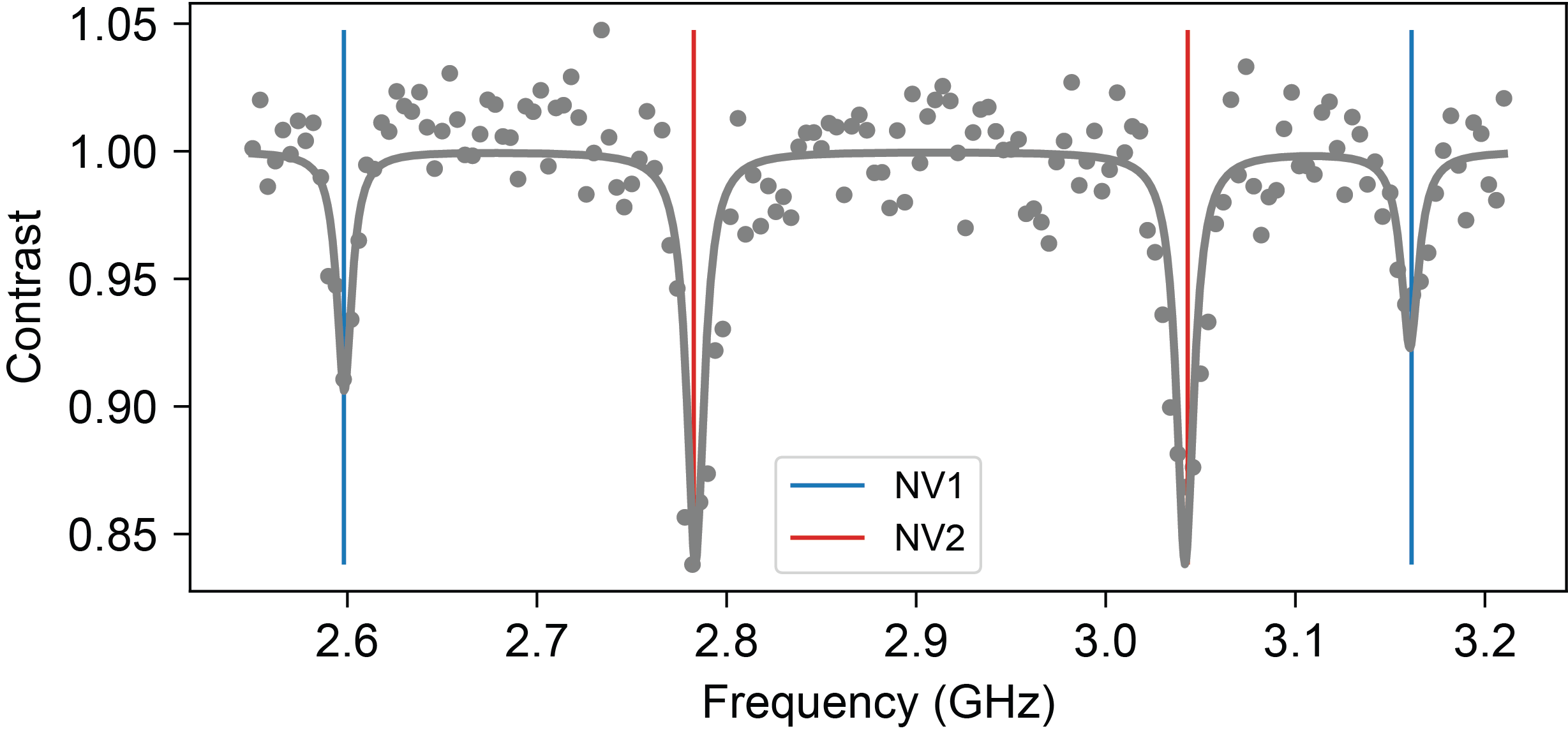}
    \caption{ODMR spectrum taken with conventional green readout, showing distinct MW transitions for each NV. As in the main text, blue (red) denotes NV1 (NV2) throughout the Supplemental Material unless noted otherwise.}
    \label{fig::figS1}
\end{figure*}

\section{Readout characterization}\label{sec:read_char}
Except for the ODMR experiment shown in Fig.~\ref{fig::figS1}, all other measurements in this work utilize resonant laser light for readout, which enables independent readout of NV centers within the same confocal spot (Fig.~1 of the main text). Experiments that do not involve correlation measurements are typically done with resonant readout (RR) of the NV spin state directly, which is faster than with spin-to-charge conversion. On the other hand, for correlation measurements, achieving low-noise spin readout is crucial. To this end, we employ a resonantly-assisted spin-to-charge conversion (RA-SCC) protocol, followed by charge-state readout with the same resonant laser (see main text)~\cite{irber2021supp}.

\begin{figure*}[hbt!]  
    \centering
    \includegraphics[width=0.9\textwidth]{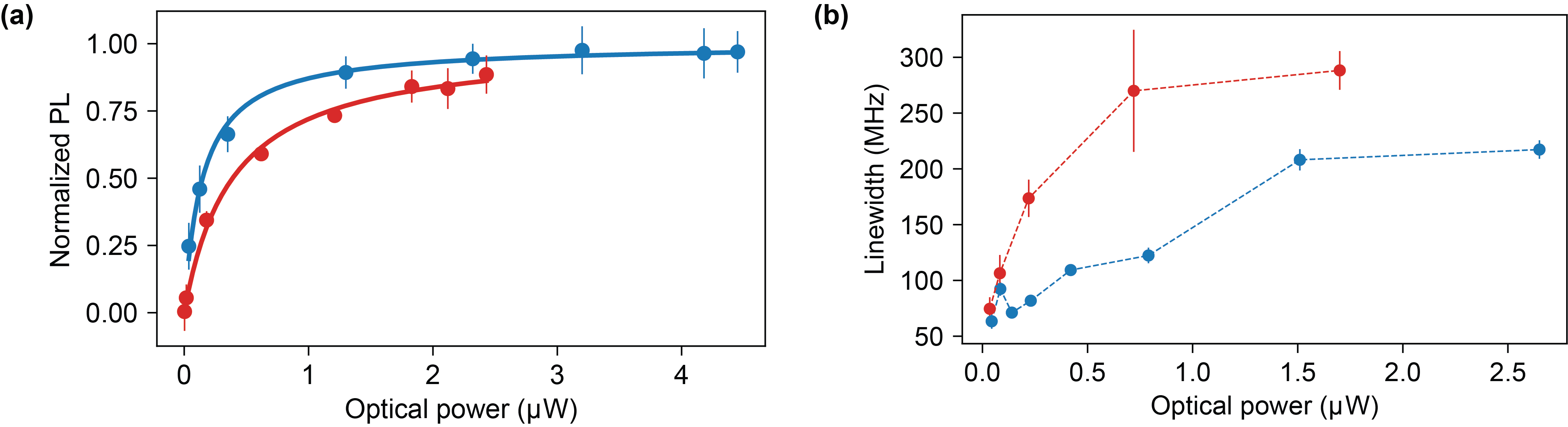}
    \caption{Effects of optical power on resonant readout. (a) Saturation curves with fits (solid lines). (b) Power broadening of the linewidths. Dashed lines are guides to the eye.}
    \label{fig::figS2}
\end{figure*}

The PL count rate for RR as a function of optical power is shown in Fig.~\ref{fig::figS2}(a) for both NV centers. The solid lines represent fits to saturation function $I = I_0\frac{P}{P + P_s}$, yielding saturation powers $P_{S1}=148\pm13$\,nW and  $P_{S2}=389\pm42$\,nW for NV1 and NV2, respectively.
The linewidth of the optical transition used for RR as a function of optical power is shown in Fig.~\ref{fig::figS2}(b), highlighting the linewidth broadening at both high and low optical power levels.

\begin{figure*}[hbt!]  
    \centering
    \includegraphics[width=0.9\textwidth]{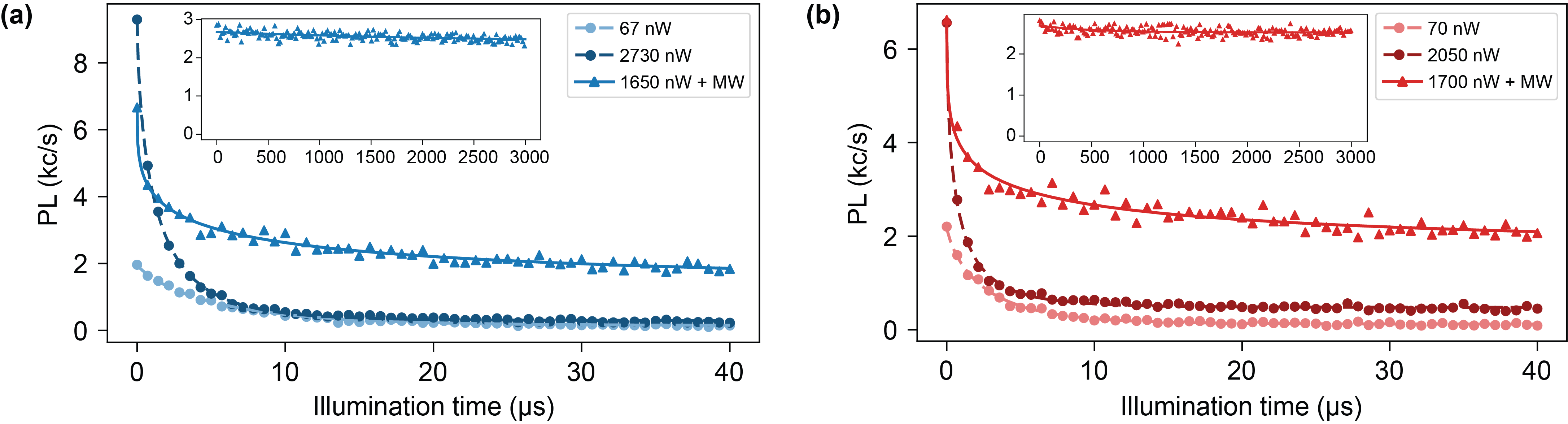}
    \caption{Resonant readout duration for NV1 (a) and NV2 (b). Without MW excitations during readout, the available collection window is very short ($< 5\,$\textmu s) for both high and low optical power (circle datapoints, fitted with dashed curves). With low-power ($f_{\rm Rabi}\approx 2\,$MHz) MW excitations during readout to mix the spin states (triangles, solid curves), the NVs can be read for milliseconds (insets).}
    \label{fig::figS3}
\end{figure*}

The time dynamics of PL under resonant excitation are shown in Fig.~\ref{fig::figS3}, along with the corresponding fit to a stretched exponential function $Ae^{-(t/\tau_{RR})^n} + C$. For both NV centers, we observe that applying a low-power ($f_{\rm Rabi}\approx 2\,$MHz) microwave drive during readout continuously mixes the spin states, counteracting optical pumping into the $\ket{\pm1}$ states due to spin mixing in the excited states manifold~\cite{irber2021supp}. This approach allows collection of more photons over extended time windows (see insets in Fig.~\ref{fig::figS3}). The stable PL counts over 3\,ms windows indicate the charge stability of NV$^{-}$ under continuous resonant laser illumination at optical powers above saturation. 

\begin{figure*}[hbt!]  
    \centering
    \includegraphics[width=0.93\textwidth]{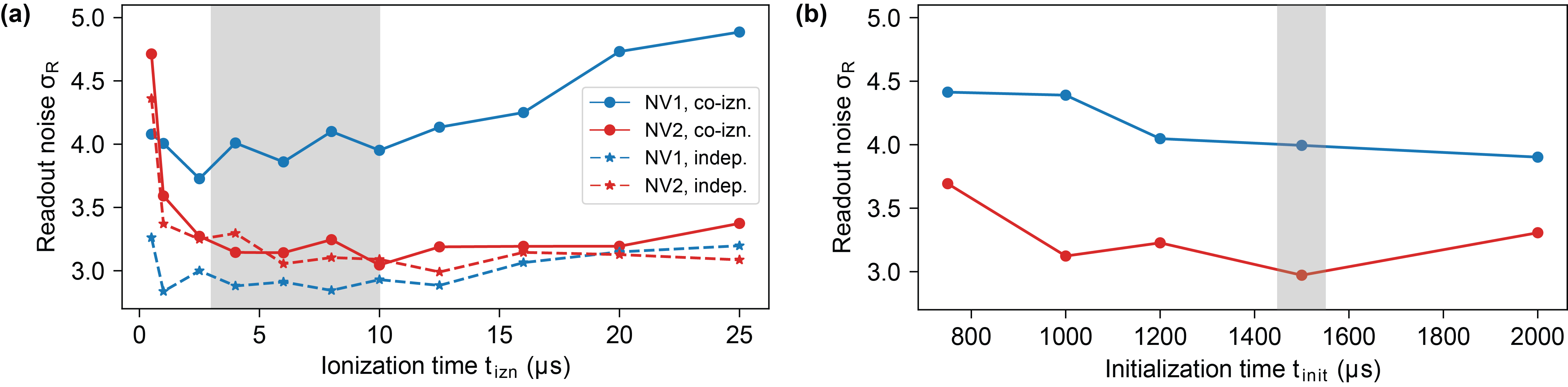}
    \caption{RA-SCC readout noise versus ionization time (a) and initialization time (b). The NVs can be ionized at the same time (``co-izn.'') or independently (``indep.''). All curves are guides to the eye. Grey area denotes the parameter range used for correlation measurements.}
    \label{fig::figS4}
\end{figure*}

To minimize the spin readout noise $\sigma_{r}$ (defined for thresholded readout in~\cite{hopper2018}) in RA-SCC protocol, we optimize the ionization time (Fig.~\ref{fig::figS4}(a)) and the green-laser-based charge and spin initialization time (Fig.~\ref{fig::figS4}(b)). We attribute the increased readout noise of NV1 associated with simultaneous ionization of both NV centers (dashed vs. solid lines in Fig.~\ref{fig::figS4}(a)) to partial overlap of optical transitions of NV1 and NV2 (see Fig.~1(c) of the main text). Specifically, the in-use $m_s=0$, $E_y$ transition of NV2 overlaps with the very dim $m_s=\pm1$, $A_2$ transition of NV1, causing some accidental ionization of $m_s=-1$ population of NV1 when intentionally ionizing $m_s=0$ population of NV2. We note that the typical choice of resonant laser powers, 1.0 and 1.7\,\textmu W for NV1 and NV2, respectively, is to ensure both good readout and good ionization fidelity. 

\begin{figure*}[hbt!]  
    \centering
    \includegraphics[width=0.8\textwidth]{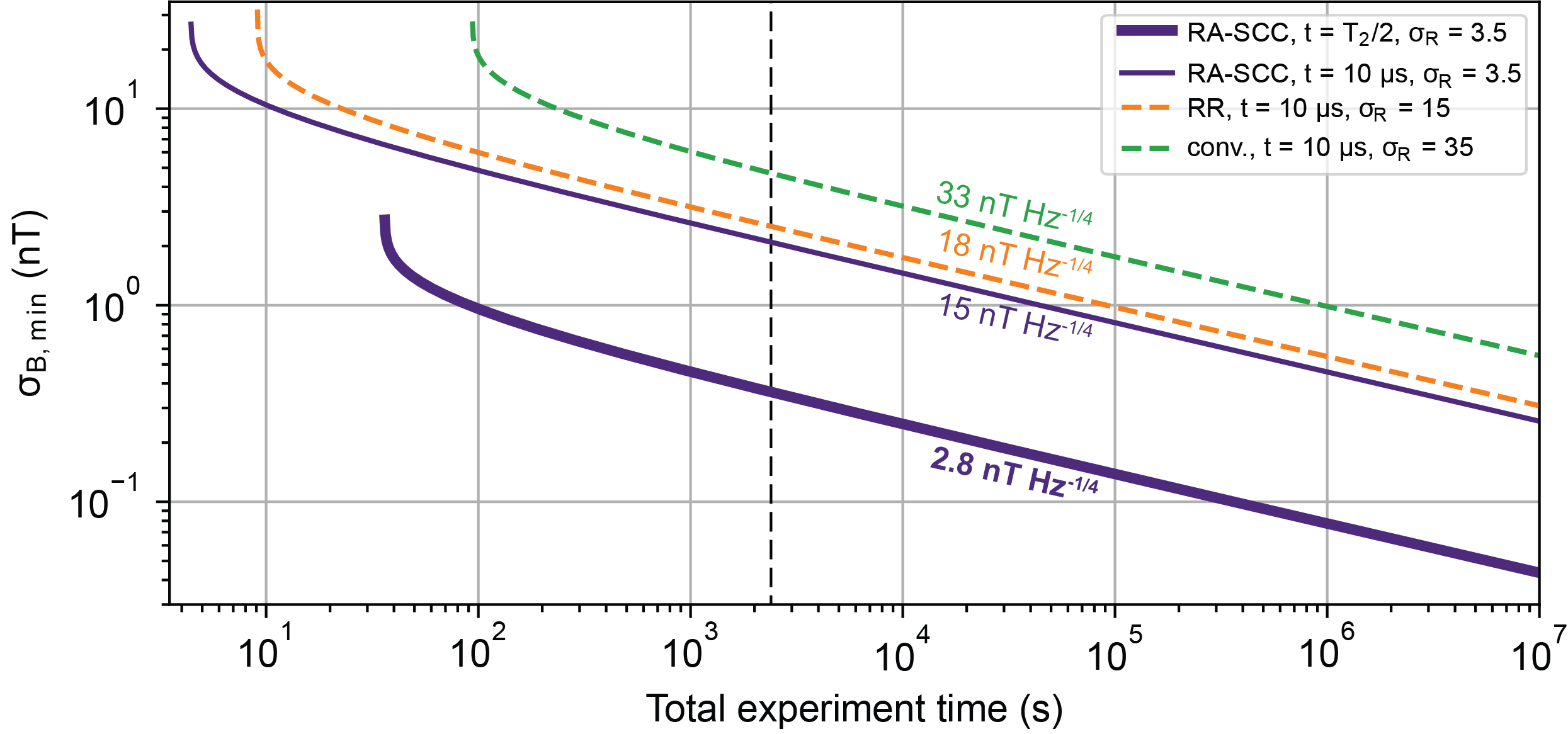}
    \caption{Minimum detectable correlated magnetic field as calculated for different readout methods. For conventional readout (``conv.'', dashed green), we assume the initialization time $t_{\rm init} = 5\,$\textmu s, phase integration time $t = 10\,$\textmu s, similar to Fig.~3 of the main text, and readout time $t_{\rm R} = 300\,$ns. For resonant readout (``RR'', dashed orange), $t_{\rm init} = 30\,$\textmu s, $t = 10\,$\textmu s, and $t_{\rm R} = 2\times2\,$\textmu s due to sequential reading of each NV. For RA-SCC, we show both cases corresponding to $t = 10\,$\textmu s (thin solid purple) and $t = T_{\rm 2}/2\,$ (thick solid purple) respectively, with $T_{\rm 2} = 2\,$ms based on Fig.~\ref{fig::figS7}(a). For both cases, $t_{\rm init} = 1.5\,$ms and $t_{\rm R} = 2\times3\,$ms. The annotations adjacent to each curve indicate magnetic sensitivity of the corresponding readout method. The vertical dashed line denotes the experiment time per datapoint in each curve in Fig.~3(b) of the main text.}
    \label{fig::figS5}
\end{figure*}

Achieving simultaneously low $\sigma_{r}$ for both NVs allows sensitive detection of magnetic correlations. In Fig.~\ref{fig::figS5}, we plot the minimum detectable correlated magnetic field $\sigma_{\rm B,min}$ as a function of total experimental time for different readout schemes. The dashed vertical line indicates the measurement time of 40 minutes for each data point in Fig.~3(b) of the main text. We observe agreement between the error bars of that measurement (4\,nT$^2$/Hz when detecting $f=2.5$\,MHz) and the calculated $\sigma_{\rm B,min}$ for this measurement time (2\,nT, intersection of the vertical line and the thin purple line in Fig.~\ref{fig::figS5}), assuming that the spectrum is locally flat near $f=2.5$\,MHz. Additionally, we plot the projected curve with RA-SCC and a sensing time of 1\,ms (thick purple line in Fig.~\ref{fig::figS5}), which is feasible due to the long coherence time shown in Fig.~\ref{fig::figS7}(a). This could enable the detection of 1\,nT correlated field with signal-to-noise ratio of 2.5 with the same 40 minutes total experiment time.

\section{Driven correlation}
In this section, we derive the expected correlation for the driven correlation experiment in Fig.~2(e) of the main text. The Pearson correlation $r$ is defined as
\begin{align}
    r = \frac{\langle S_1  S_2 \rangle - \langle S_1 \rangle \langle S_2 \rangle}{\sigma_{\rm S1}\sigma_{\rm S2}},
    \label{eqn::r_driven1}
\end{align}
where $S_i \in \{0,1\}$ represents the projected quantum state of the $i$-th NV in each measurement, $\langle\,\cdots\,\rangle$ denotes averaging over all $2M$ experimental iterations, and $\sigma_{{\rm S}i}=\sqrt{\langle S_i^2\rangle - \langle S_i\rangle^2}$ is the standard deviation of $S_i$ across all iterations. For simplicity, we start by assuming no readout noise and only consider the ``correlation'' configuration, i.e., both NVs are always driven to the same state. By alternatively driving to $\ket{\theta}= \begin{pmatrix}
    \cos(\theta/2) \\ \sin(\theta/2) 
\end{pmatrix}
$ and $\ket{\theta+\pi}$ state, the mean value $\langle S_i\rangle$ is
\begin{align}
    \langle S_i\rangle = \frac{M\langle S\rangle_{\theta} + M\langle S\rangle_{\theta + \pi}}{2M},
    \label{eqn::s1s2_driven1}
\end{align}
where $\langle S\rangle_{\theta}$ denotes the expectation of the pseudospin operator \( S = \begin{pmatrix}
    0 & 0 \\ 0 & 1
\end{pmatrix}  \) over the state $\ket{\theta}$. It can be easily shown that $\langle S\rangle_{\theta}$ = $\langle S^2\rangle_{\theta} = \sin^2(\theta/2)$, thus Eq.~\eqref{eqn::s1s2_driven1} becomes
\begin{align}
    \langle S_i\rangle = \langle S_i^2\rangle = \frac{1}{2},
    \label{eqn::s1s2_driven2}
\end{align}
with all $\theta$-dependence being canceled out. It follows that $\sigma_{{\rm S}i}=1/2$.

What is left is to evaluate the cross term $\langle S_1  S_2 \rangle = \displaystyle\frac{1}{2M} \left( M\langle S\rangle_{\theta, \theta} + M\langle S\rangle_{\theta+\pi, \theta+\pi} \right)$, where the pseudospin operator $S$ and its expectation value are with respect to the two-NV Hilbert space. Since the dipolar interaction between NV1 and NV2 is vanishingly small and the experimental protocol does not induce any quantum correlation, the two NVs are always in a product state. Therefore, $\langle S\rangle_{\theta, \theta} = \left( \langle S\rangle_{\theta} \right)^2 = \sin^4(\theta/2)$, and the cross term becomes
\begin{align}
    \langle S_1S_2\rangle = \frac{ \sin^4\left(\frac{\theta}{2}\right) +  \sin^4\left(\frac{\theta+\pi}{2}\right)}{2} = \frac{1+\cos^2(\theta)}{4},
    \label{eqn::s1s2_driven3}
\end{align}
which causes $r$ to depend on $\theta$. Plugging Eq.~\eqref{eqn::s1s2_driven3} and Eq.~\eqref{eqn::s1s2_driven2} into Eq.~\eqref{eqn::r_driven1}, and including the effect of readout noise $\sigma_{{\rm R}i}$ as derived in~\cite{rovny2022supp}, we obtain the simple expression for the theory curves in Fig.~2(e) of the main text:
\begin{align}
    r=\frac{\cos^2(\theta)}{\sigma_{\rm R1}\sigma_{\rm R2}}.
    \label{eqn::r_driven2}
\end{align}

\begin{figure*}[hbt!]  
    \centering
    \includegraphics[width=0.8\textwidth]{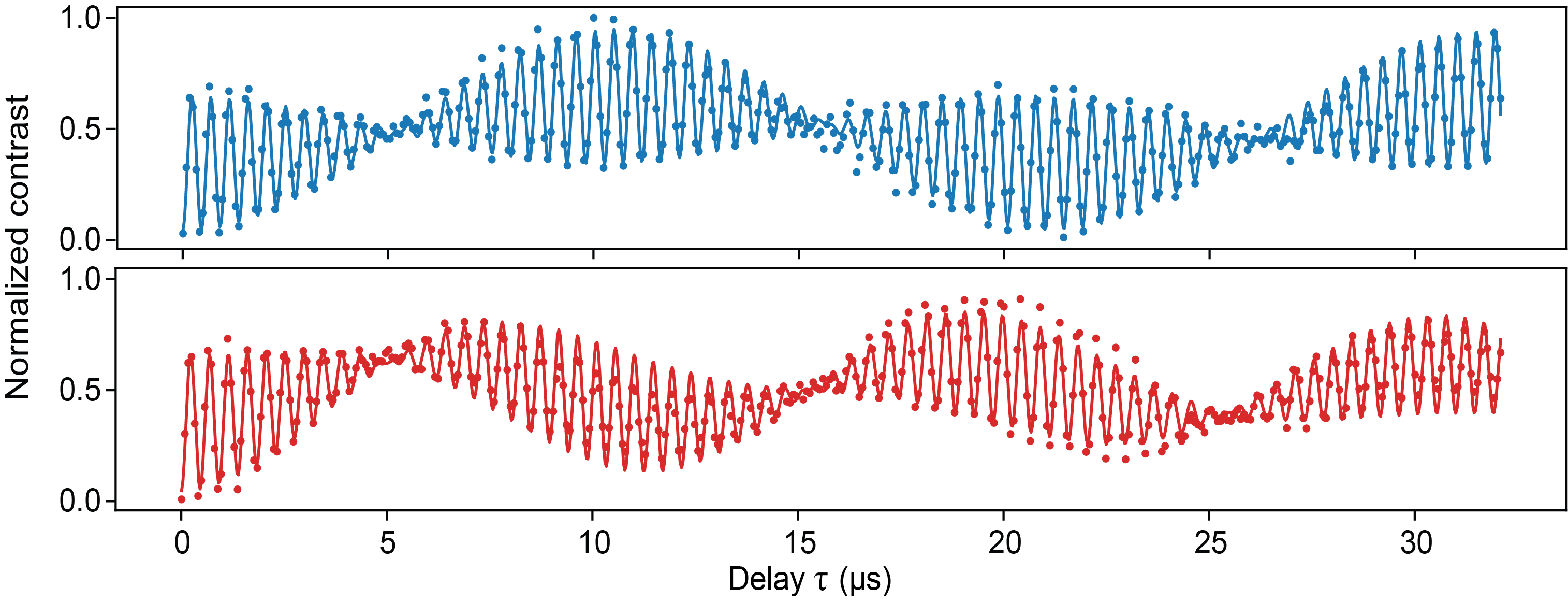}
    \caption{Ramsey experiment for each NV with a fit to three hyperfine oscillations. Data taken using resonant readout and without test noise signals.}
    \label{fig::figS6}
\end{figure*}

\begin{figure*}[hbt!]  
    \centering
    \includegraphics[width=0.9\textwidth]{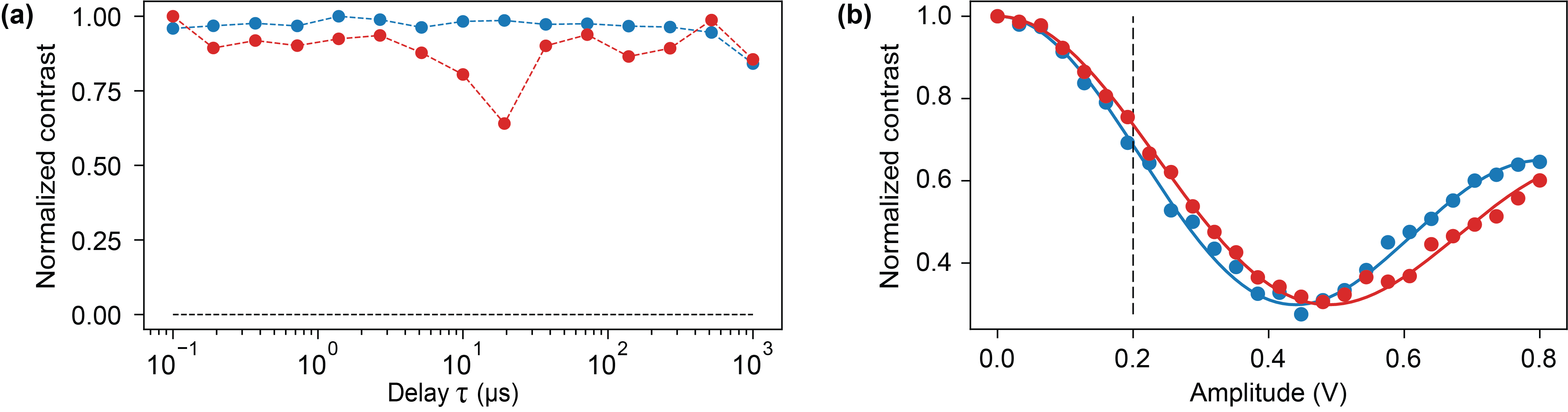}
    \caption{(a) Hahn-echo experiment for each NV with no test signals, showing $T_{\rm 2}$ well exceeding 1\,ms for both NVs. Dashed curves are guides to the eye. (b) Calibration of the magnetic field amplitude from the voltage amplitude of the AC test signal, measured by XY8-5 sequence for NV1 and XY8-8 for NV2. Solid curves are fits from Eq.~\eqref{eqn::fit_single_NV_T2}. Vertical dashed line denotes the amplitude used for Fig.~3(b) of the main text. Data in both panels are measured with the use of resonant readout.}
    \label{fig::figS7}
\end{figure*}

\section{Single-NV $T_2$ spectroscopy} \label{sec:singleNVT2}
In addition to correlation measurements, we also performed single-NV $T_2$ characterization. Figure~\ref{fig::figS6} shows the Ramsey measurements for both NVs, revealing hyperfine oscillations due to coupling to the $^{14}$N nuclear spins within each NV center. Figure~\ref{fig::figS7}(a) presents the spin-echo measurements for both NVs, highlighting long coherence times ($T_{2}>1$\,ms). 

For the sensing experiments, we apply a continuous-wave, constant-amplitude, phase-noisy AC magnetic field $B_i\cos(2\pi f t + \alpha(t))$ as a test signal, where $B_i$ is the field amplitude at the location of NV $i$, $\alpha(t)$ denotes a slowly-varying random phase. $B_i$ is calibrated by measuring the spin contrast as a function of the AWG's output voltage amplitude $V$ with an XY8-$N_i$ sensing sequence. The calibration for the $f=2.5$\,MHz test signal used in Fig.~3 of the main text is shown in Fig.~\ref{fig::figS7}(b), fitted to
\begin{align}
    \frac{1}{2}\left[ 1+ J_0\left( \frac{\gamma \kappa_i VN_i}{\pi f}\right) \right], \label{eqn::fit_single_NV_T2}
\end{align}
where 
\(J_0\) is the Bessel function of the first kind resulting from averaging over the random phase \(\alpha\) (which varies slowly, so the averaging can be approximated as uniform over the interval \([0,2\pi)\)), $\gamma/2\pi$ = 28.025\,GHz/T is the electron gyromagnetic ratio, and $\kappa_i=B_i/V$ is the fit parameter that relates the voltage amplitude to the magnetic field amplitude that the $i$-th NV experiences~\cite{degen2017}. 
From the fit, the 0.2\,V amplitude used for Fig.~3 of the main text corresponds to $B_1 = 1.944(11)$\,\textmu T, and $B_2=0.56B_1$, consistent with the fact that the two NVs  sense different field strength due to different spatial orientations.

\section{Correlated $T_2$ spectroscopy}
\subsection{Expected correlation}
In this section, we derive an analytical expression for the measured Pearson correlation \(r\) between NV1 and NV2, when each NV accumulates a different phase, \(\phi_{\rm C1}(t_1) \neq \phi_{\rm C2}(t_2)\), due to the correlated field.
This is relevant when each NV experiences a different field strength and/or undergoes a different sensing sequence.

As derived in Ref.~\cite{rovny2022supp}, for a sine magnetometry protocol, we have:
\begin{align}
    r = \frac{e^{-[\tilde\chi_1(t_1) + \tilde\chi_2(t_2)]}}{\sigma_{\rm R1}\sigma_{\rm R2}}\langle \sin[\phi_{\rm C1}(t_1)] \sin[\phi_{\rm C2}(t_2)] \rangle,
    \label{eqn::r}
\end{align}
where $i$ is the NV index, $t_i = N_i\tau_i$ is the total sensing time, $N_i$ is the number of $\pi$-pulses in the sensing sequence, $\tau_i$ is the interpulse spacing, $\tilde\chi_i(t_i)$ is the decoherence function at time $t_i$ due to uncorrelated mechanisms (typically probed by variance magnetometry), and $\sigma_{{\rm R}i}$ is the readout noise. 

Given the test signal described in Sec.~\ref{sec:singleNVT2}, and assuming that the random phase $\alpha(t)$ fluctuates on a timescale appreciably slower than $t_i$, we can write the accumulated phase in one experimental repetition as 
\begin{align}
    \phi_{{\rm C}i}(t_i) = \gamma B_{i}t_iW_i(f,\alpha),
    \label{eqn::phi_c}
\end{align}
where $W_i(f,\alpha)$ is the filter function defined by the MW sensing sequence of the $i$-th NV and the properties of the AC field. For an XY8-$N_i$ sequence,:
\begin{align}
    W_i(f,\alpha) = w_i\cos(\alpha + \varphi_i) =\frac{\sin(\varphi_i)}{\varphi_i}\big[1-\sec(\varphi_i/N_i)\big]\cos(\alpha + \varphi_i),
    \label{eqn::W}
\end{align}
 where $\varphi_i = \pi fN_i\tau_i$~\cite{degen2017}. By rewriting $\sin[\phi_{\rm C1}(t_1)] \sin[\phi_{\rm C2}(t_2)]$ = $\frac{1}{2}\big(\cos[\phi_{\rm C1}(t_1) - \phi_{\rm C2}(t_2)]-\cos[\phi_{\rm C1}(t_1) + \phi_{\rm C2}(t_2)]\big)$, substituting $\phi_{{\rm C}i}(t_i)$ from Eq.~\eqref{eqn::phi_c} and $W_i(f,\alpha)$ from Eq.~\eqref{eqn::W} into Eq.~\eqref{eqn::r}, and regrouping the terms appropriately, we arrive at:
\begin{align}
    r = \frac{e^{-[\tilde\chi_1(t_1) + \tilde\chi_2(t_2)]}}{\sigma_{\rm R1}\sigma_{\rm R2}} \frac{
    \big\langle \cos\big[\xi_-\cos(\alpha+\beta_{-})\big]\big\rangle - 
    \big\langle \cos\big[\xi_{+}\cos(\alpha+\beta_{+})\big] \big\rangle}{2},
    \label{eqn::r2}
\end{align}
with dimensionless variables $\xi_{\pm} = \sqrt{\eta_1^2 + \eta_2^2 \pm 2\eta_1\eta_2\cos(\varphi_1-\varphi_2)}$, 
$\eta_i=\gamma B_it_iw_i$, 
and $\beta_{\pm} = \tan^{-1}\Big(\frac{\eta_1\sin(\varphi_1) \pm \eta_2 \sin(\varphi_2)}{\eta_1\cos(\varphi_1) \pm \eta_2 \cos(\varphi_2)}\Big)$. Averaging over the random phase $\alpha$ gives 
\begin{align}
    r = \frac{e^{-[\tilde\chi_1(t_1) + \tilde\chi_2(t_2)]}}{\sigma_{\rm R1}\sigma_{\rm R2}} \frac{
    J_0 (\xi_-) - 
    J_0 (\xi_{+})}{2}.
    \label{eqn::r3}
\end{align}
When both NVs detect the same field amplitude with the same sensing sequence, we have $\varphi_1=\varphi_2=\varphi$, $w_1=w_2=w$, $\eta_1=\eta_2=\eta$, $\xi_+=2\eta$, and $\xi_-=0$. In this case, the first term in Eq.~\eqref{eqn::r3} therefore becomes 1, and the second term becomes $J_0(2\gamma Btw)$, recovering the results in Ref.~\cite{rovny2022supp}. 
With calibrated single-NV decoherence, readout noise, magnetic field amplitude at each NV, and an overall free scaling factor of 0.37 to account for other experimental imperfections (such as imperfect initialization and ionization, or mechanical noise during long sensing experiments), we can use Eq.~\eqref{eqn::r3} to predict the expected correlation as shown in Fig.~3(b) of the main text.


\subsection{Power spectral density of correlated noise}

Given a correlated magnetic noise detected by NV1 and NV2, its power spectral density (PSD) has different components
\begin{align}
    S_{ij}(\omega) = \int_{-\infty}^{\infty} e^{-i\omega t'}\gamma^2 \langle B_i(t + t')B_j(t)\rangle dt',
    \label{eqn::Sij}
\end{align}
where $i$ and $j$ are the NV indices~\cite{szankowski2016, rovny2022supp, degen2017}. The matrix elements $S_{11}(\omega)$ and $S_{22}(\omega)$ contribute to the single-NV decoherence of NV1 and NV2, respectively, while $S_{12}(\omega)=S_{21}(\omega)$ (due to the classical, commutative nature of the test noise) only affect the correlation $r$. Therefore, measuring $r$ provides a unique probe of this cross-correlation component.

Our goal is to derive an expression for $S_{12}(\omega)$ based on the measured $r$ and single-NV coherence functions. First, we note that $S_{12}(\omega)$ relates to the accumulated phases $\phi_{\rm C1}$, $\phi_{\rm C2}$ via
\begin{align}
    \langle \phi_{\rm C1} \phi_{\rm C2} \rangle = \frac{4}{\pi} \int_0^{\infty} S_{12}(\omega) |W(\omega)|^2 d\omega,
    \label{eqn::phiC1phiC2}
\end{align}
where $|W(\omega)|^2$ is the filter function in frequency domain defined by the sensing sequence. For an XY8-$N$ sequence with many pulses and pulse spacing $\tau$, $|W(\omega)|^2 \approx \frac{2}{\pi}t\delta(\omega - \pi/\tau)$~\cite{degen2017}, which simplifies Eq.~\eqref{eqn::phiC1phiC2} to
\begin{align}
    \langle \phi_{\rm C1} \phi_{\rm C2} \rangle = \frac{8t}{\pi^2}S_{12}\Big(\frac{\pi}{\tau}\Big).
    \label{eqn::phiC1phiC2_2}
\end{align}
Next, for a Gaussian noise source, or assuming that the noise is weak and the accumulated phase is small, Eq.~\eqref{eqn::r} can be rewritten as
\begin{align}
    r = 
    \frac{e^{-[\tilde\chi_1(t_1) + \tilde\chi_2(t_2)]}}{2 \sigma_{\rm R1}\sigma_{\rm R2}} \Big[ 
     e^{-\frac{\left\langle \left(\phi_{\rm C1}-\phi_{\rm C2}\right)^2 \right\rangle}{2}} - e^{-\frac{\left\langle \left(\phi_{\rm C1}+\phi_{\rm C2}\right)^2 \right\rangle}{2}}\Big].
    \label{eqn::r4}
\end{align}
Combining this with Eq.~\eqref{eqn::phiC1phiC2_2}, we obtain
\begin{align}
    S_{12}\Big(\frac{\pi}{\tau}\Big) = \frac{\pi^2}{8t}\sinh^{-1}{\left( \frac{r\sigma_{\rm R1}\sigma_{\rm R2}}{C_1(t)C_2(t)} \right)},
    \label{eqn::s12}
\end{align}
where $C_i(t) = e^{-\tilde{\chi}_i(t) - \left\langle \phi_{{\rm C}i}^2 \right\rangle/2}$ describes a single-NV decoherence due to both uncorrelated and correlated noise.
\begin{figure*}[hbt!]  
    \centering
    \includegraphics[width=0.95\textwidth]{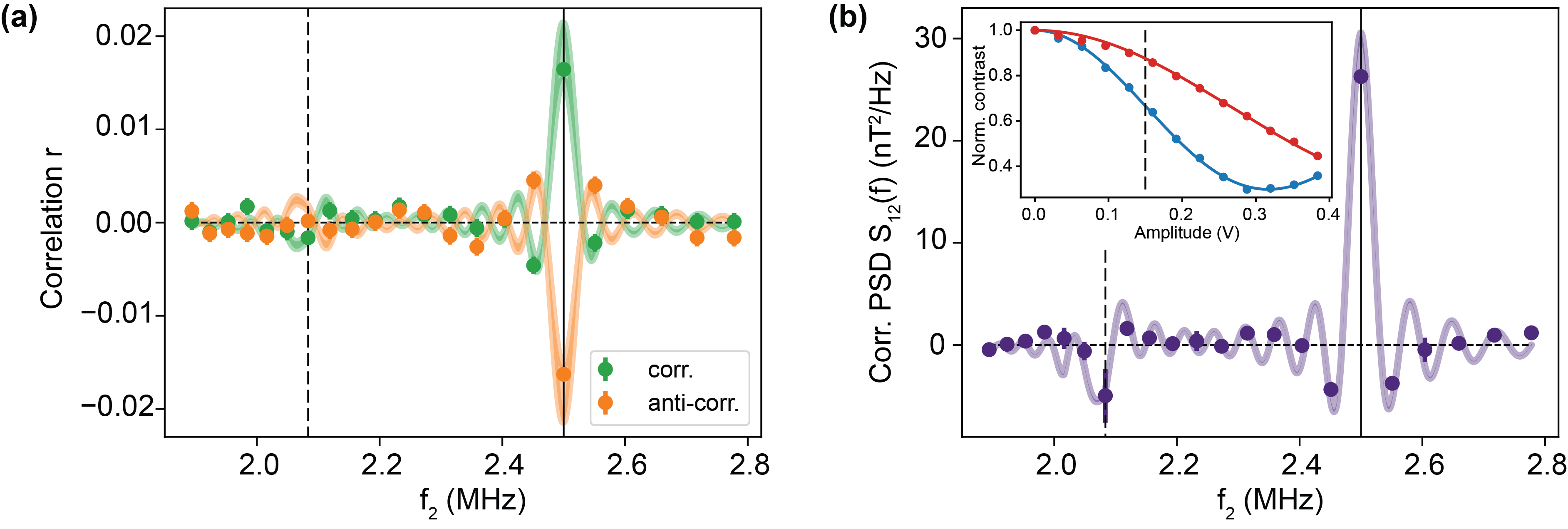}
    \caption{Correlated $T_2$ spectroscopy with each NV probing a different frequency, sweeping $\tau_2=1/(2f_2)$ while fixing $\tau_1=1/(2 \cdot 2.5 \text{ MHz})$. $N_1=N_2=10$. The two test signals at $f_{\rm A} = 2.5\,$MHz and $f_{\rm B} = 2.0833\,$MHz are not phase-locked. PSD of the correlated noise (b) are calculated from the correlations (a). 
    Filled curves in (a) and (b) are theory predictions based on Eq.~\eqref{eqn::r3} and Eq.~\eqref{eqn::s12}, respectively; the curve widths denote uncertainty of measured $\sigma_{\rm R}$, $B_{1}$, and $B_{2}$. 
    Vertical solid (dashed) line in (a) and main panel of (b) indicates $f_{\rm A}$ ($f_{\rm B}$). 
    The inset of (b) shows calibration of magnetic field amplitude $B_{1,{\rm A}}$ of the $f_{A}$ test signal while $f_{\rm B}$ is also applied, using XY8-10 for both NVs. 
    This recalibration (compared to the calibration in Fig.~\ref{fig::figS7}(b) for the main text data) is necessary because, with both channels of the AWG now outputting test signals that are subsequently combined, spurious RF reflections reduce the effective signal strength.
    The fit function (solid curves) is similar to Fig.~\ref{fig::figS7}(b), yielding $B_{1,{\rm A}}=1.010(6)$\,\textmu T at 0.15\,V amplitude (dashed line). The voltage amplitude of $f_{\rm B}$ is 0.3\,V, implying $B_{1,{\rm B}}=2.020(12)$\,\textmu T.}
    \label{fig::figS8}
\end{figure*}

\subsection{Correlation measurements with different sensing sequences for each NV}

We perform an additional correlated $T_2$ spectroscopy experiment, in which each NV probes a different frequency by having different pulse spacings $\tau_1\neq\tau_2$ in the sensing sequence. In this experiment, we apply two phase-random test signals at $f_{\rm A} = 2.5\,$MHz and $f_{\rm B} = 2.0833\,$MHz that are not phase-locked, use NV1 to detect $f_{\rm A}$ by fixing $\tau_1=1/(2f_{\rm A})$, and sweep the probing frequency of NV2 in a range that covers both $f_{\rm A}$ and $f_{\rm B}$.

Although the naive expectation is that meaningful correlation only occurs when both NVs probe $f_{\rm A}$, Fig.~\ref{fig::figS8} shows that both theory and experimental data indicate a small feature in correlation when NV2 probes $f_{\rm B}$. This is due to NV1's filter function being not perfectly zero at $f_{\rm B}$. 
Overall, this experiment confirms the robustness of Eq.~\eqref{eqn::r3} (filled curves in Fig.~\ref{fig::figS8}(a)) in describing the expected spectrum of correlation, capturing well the correlation shape arising from asymmetries between each NV's effective field strengths and sensing sequences. The theory curves are obtained with the same scaling factor of 0.37 as in Fig.~3(b) of the main text, suggesting that this factor originates from the same systematics that are present in both experiments.

\section{Correlated $T_1$ spectroscopy}

\subsection{Theoretical analysis}

The dynamics of the pair of NV centers, effectively two two-level systems, are governed by the Hamiltonian (throughout the theoretical analysis that follows, we set $\hbar = 1$ and use the Pauli matrices $\sigma^{x,y,z}_{i = 1,2}$):
\begin{align}
    {H} = \sum_{i = 1,2} \Big\{ \frac{\Delta_i}{2} \hat{n}_{i} \cdot \bm \sigma_i + \frac{\gamma}{2} \bm B(\bm r_i,t) \cdot \bm \sigma \Big\},
\end{align}
where $\Delta_i$ is the energy difference between $\ket{0}$ and $\ket{-1}$ states, $\hat{n}_i$ represents the quantization axis, $\bm{B}(\bm{r}_i, t)$ denotes the noisy magnetic field at the location $\bm{r}_i$ of the $i$-th NV center, and $\gamma$ is the electron gyromagnetic ratio. 
In our experiment, we generate this MW magnetic field via the stripline:
\begin{align}
    \bm B(\bm r_i, t) = f(t) \hat{B}_i,\qquad f(t) = \big[ B_1 \sin(\omega_1 t + \varphi_1) + B_2 \sin(\omega_2 t + \varphi_2)\big] G(t).
\end{align}
Note that the magnetic field $\bm{B}(\bm{r}_i, t)$ oscillates in time but has a fixed direction.
The frequencies $\omega_1$ and $\omega_2$ are close to but not necessarily equal to $\Delta_1$ and $\Delta_2$, respectively.
Here, $G(t)$ is a random Gaussian noise characterized by:
\begin{align}
    \langle G(t)\rangle  = 0, \qquad \langle G(t) G(t')\rangle = \exp\Big( - \frac{(t - t')^2}{2\tau_c^2} \Big),
\end{align}
where the correlation time $\tau_c$ encodes the noise bandwidth $\propto 1/\tau_c$. 
Within the rotating frame of each NV center, and under the rotating wave approximation, by appropriately selecting local coordinate systems for each NV, the effective Hamiltonian can be written as:
\begin{align}
    H_i \approx \frac{\delta_i}{2} \sigma_i^z +\Big[ {\cal A}_i G(t)\sigma_i^+ +  \text{h.c.} \Big], \qquad {\cal A}_i =  \frac{i e^{-i \varphi_i}}{4} \gamma B_i   \sin\vartheta_i  , \label{eqn::NV_setup_v0}
\end{align}
where $\vartheta_i$ is the angle between $\hat{B}_i$ and $\hat{n}_i$ and $\delta_i = \Delta_i - \omega_i$ is the detuning.

\emph{\textbf{ Stochastic evolution}}---Equation~\eqref{eqn::NV_setup_v0} implies that the two NVs are decoupled from each other and each evolves under its own noisy magnetic field (but the two magnetic fields are correlated by design). 
This allows us to simulate the dynamics of the two NV centers by stochastically sampling the random magnetic fields and evolving the spin dynamics accordingly.
Specifically, by writing a spin-1/2 Hamiltonian as $H(t) = \frac{1}{2} \bm h(t) \cdot \bm \sigma$, its density matrix $\rho = \frac{1}{2} + \frac{1}{2} \bm s(t) \cdot \bm \sigma$, where $s_\alpha(t) \equiv \langle \sigma_\alpha \rangle \equiv \tr\big( \rho(t) \sigma_\alpha \big)$, evolves as: 
\begin{align}
\partial_t \bm s(t) = \bm h(t) \times \bm s(t)\label{eqn:spin_rand_mf}.
\end{align}
Equation~\eqref{eqn::NV_setup_v0} implies that:
\begin{align}
    h_x = ({\cal A}_i + {\cal A}^*_i) G(t),\quad h_y = i ({\cal A}_i - {\cal A}^*_i) G(t),\quad h_z = \delta_i.
\end{align}
In modeling such spin-1/2 time evolution, we note that if the magnetic field were instead static, the corresponding time evolution could be computed analytically~\cite{thibaudeau2012thermostatting}:
\begin{align}
\bm s(t) = {\cal A}(t) \bm s_0 + {\cal B}(t) \bm h + {\cal C}(t) \bm h \times \bm s_0,
\label{eqn:analytical_smf}
\end{align}
where
\begin{gather}
{\cal A}(t) = \cos(h t),\quad {\cal B}(t) = \chi (1-\cos(h t)), \quad {\cal C}(t) = \sin(h t),\qquad \chi = \hat{h}\cdot \bm s_0.
\end{gather}
It is worth noting that the vectors in Eq.~\eqref{eqn:analytical_smf} are generally not orthogonal to each other.
In simulating the stochastic dynamics in Eq.~\eqref{eqn:spin_rand_mf} numerically, we discretize time into steps of size $\delta t$ and update $\bm s_{j+1}$ by evolving it for a time $\delta t$ under the static field $\bm h_{j}$ starting from $\bm s_{j}$ -- this scheme ensures preservation of the norm $|\bm s(t)|$ during the stochastic dynamics. 

Such stochastic simulations have been instrumental in studying the effects of noise bandwidth and in benchmarking the effective Markovian dynamics discussed next, which prevail when the inverse of the correlation time, \( 1/\tau_c \), is large compared to other relevant dynamical scales.

\emph{\textbf{Markovian limit and Lindblad master equation}}---We now assume that the relevant NV processes occur on time scales significantly longer than $\tau_c$. In this case, one can approximate the noise autocorrelation function $\langle G(t) G(t')\rangle$ as Markovian:
\begin{align}
    \langle G(t) G(t')\rangle \to \sqrt{2 \pi} \tau_c \delta(t - t').
\end{align}
In this Markovian limit, the NV dynamics can be described using the quantum master equation:
\begin{align}
    \partial_t \rho(t) & = -i \sum_i \frac{\delta_i}{2} \big[\sigma_i^z, \rho \big] + \sum_{ij}\Big(  {\cal L}_i  \rho {\cal L}_j   - \frac{1}{2} \big\{ {\cal L}_j {\cal L}_i, \rho\big\} \Big)  =-i \sum_i \frac{\delta_i}{2} \big[\sigma_i^z, \rho \big] + \frac{1}{2} \sum_{ij} \big[  {\cal L}_i ,\big[  \rho,  {\cal L}_j \big] \big] ,
    \label{eqn::SI::QME}
\end{align}
where $i, j$ are the NV indices, ${\cal L}_i = (2\pi)^{1/4} \sqrt{\tau_c} ({\cal A}_i \sigma_i^+ + {\cal A}_i^* \sigma_i^-)  = h_{i,x}\sigma^x_i + h_{i,y}\sigma^y_i$, with $h_{i,x} = (2\pi)^{1/4} \sqrt{\tau_c}  \,\text{Re}[ {\cal A}_i]$ and $h_{i,y} =- (2\pi)^{1/4} \sqrt{\tau_c}  \,\text{Im}[ {\cal A}_i]$, is the jump operator arising from the amplitude-noisy drive.

\emph{\textbf{Dephasing effects}}---For future reference and pivotal to capturing the long-time decay of correlation observed in the experiment, we also (phenomenologically) include a dephasing channel:
\begin{align}
    {\cal D}_d[\rho] = \sum_{i j} \gamma^d_{ij} \Big( \sigma_i^z \rho \sigma_j^ z - \frac{1}{2} \big\{ \sigma_j^z \sigma_i^z , \rho \big\} \Big),
\end{align}
where $\gamma_{ij}^d$ is a real symmetric matrix. The matrix elements $\gamma_{11}^d$ and $ \gamma_{22}^d$ represent dephasing of each NV due to its local noise (including both correlated and uncorrelated mechanisms), while $\gamma_{12}^d=\gamma_{21}^d$ encode correlated effects only.

\begin{figure*}[hbt!]  
    \centering
    \includegraphics[width=1\textwidth]{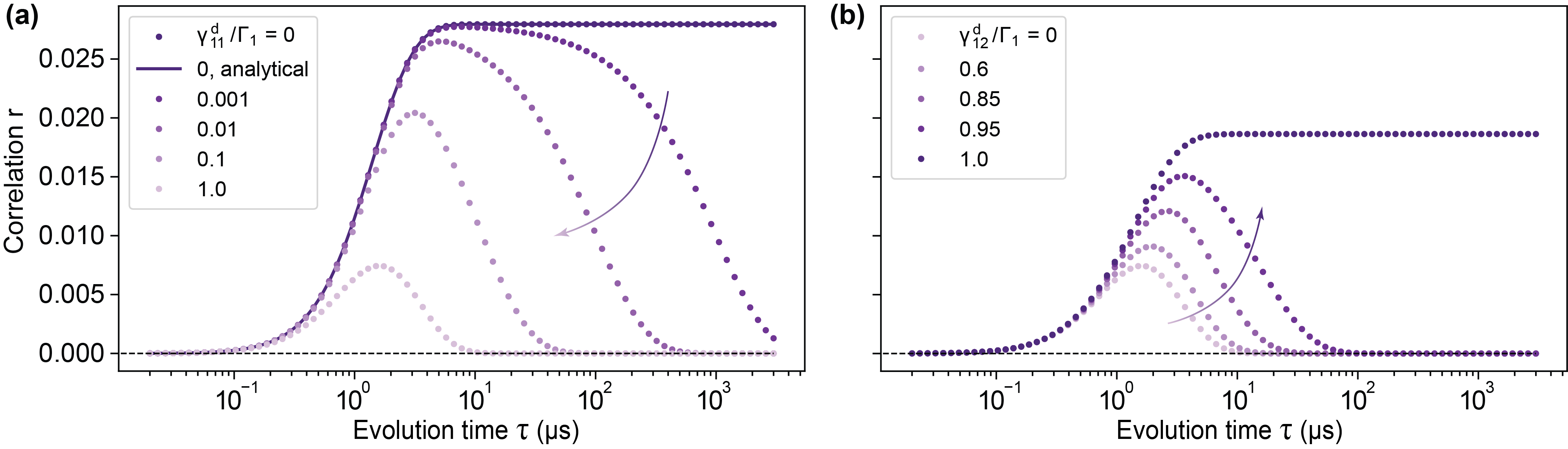}
    \caption{ Simulated effects of dephasing on correlated $T_1$ measurements. 
    (a) The presence of local dephasing, with \( \gamma^d_{12} = 0 \) and \( \gamma^d_{11} = \gamma^d_{22} \), suppresses the build-up of the correlation \( r \), an effect that becomes particularly evident at long times where \( r \) decays rather than plateaus.
    (b) 
    As dephasing becomes progressively more global, i.e., as ${\gamma}^d_{12}$ approaches ${\gamma}^d_{11} = \gamma^d_{22}$ (here, we fixed the latter to be $\Gamma_1$), its effects become less detrimental. The arrow in the left (right) panel indicates the direction of increasing ${\gamma}^d_{11} = {\gamma}^d_{22}$ (${\gamma}^d_{12}$).
    The solid line in panel (a) represents the analytical result in Eq.~\eqref{eqn::analytic}.
}
    \label{fig::figS9}
\end{figure*}

Figure~\ref{fig::figS9} shows the effects of this dephasing channel on correlated $T_1$ measurements.
Local dephasing (we consider \(\gamma_{11}^d = \gamma_{22}^d\) and set \(\gamma_{12}^d = \gamma_{21}^d = 0\)) suppresses the correlated-\(T_1\) signal \(r\), particularly at long times, where \(r\) decays to zero instead of plateauing -- see Fig.~\ref{fig::figS9}(a). This decay occurs because the two NV centers experience distinct random rotations around the \(z\)-axis during their evolution under a given realization of the random magnetic field.
As dephasing becomes more global, i.e., as ${\gamma}^d_{12}$ approaches $\gamma^d_{11} = \gamma^d_{22}$, the random $z$-axis rotations become highly correlated, mitigating the detrimental effects of dephasing on $r$ -- see Fig.~\ref{fig::figS9}(b).  
Global dephasing is thus unimportant in this limit, reminiscent of single-qubit dephasing being unimportant for single-qubit $T_1$-processes.

\emph{\textbf{ Single-qubit observables}}---Equation~\eqref{eqn::SI::QME} predicts that ${s}^\alpha_i(t) \equiv \tr\big(\rho(t)\sigma^\alpha_i \big)$ and $\tr\big(\rho(t)\sigma^\alpha_i \sigma^\beta_i \big)$ exhibit simple relaxation dynamics ($\alpha$, $\beta$ denote axes $x,y,z$ in each NV's Bloch sphere):
\begin{align}
    s_i^z(t) = e^{- 2 (h_{i,x}^2 + h_{i,y}^2 )t } = e^{- \Gamma_i t}, \quad s_i^x(t) = s_i^y(t) = 0, \quad \tr\big(\rho(t)\sigma^\alpha_i \sigma^\beta_i \big) = \begin{bmatrix}
        1 & i e^{- \Gamma_i t}  & 0 \\
        -i e^{- \Gamma_i t}  & 1 & 0 \\
         0 &  0  & 1
    \end{bmatrix},
\end{align}
where $\Gamma_i = 2(\gamma B_i)^2\sqrt{2\pi}\tau_c$ is the $T_1$-relaxation rate. This result agrees with our measurements in Fig.~4(c) and allows extracting the field amplitudes as discussed in the main text.

\emph{\textbf{ Cross-correlations}}---While numerically simulating the QME~\eqref{eqn::SI::QME} is straightforward, our goal is to fit the experimental data, in which case it is highly desirable to have more analytical insight.
To this end, we introduce 
\begin{align}
    \Phi_{\alpha \beta}(t) \equiv \tr\big(\rho(t)\sigma^\alpha_1 \sigma^\beta_2 \big), \qquad  \Phi_{\alpha \beta}(t = 0) = \begin{bmatrix}
        0 & 0 & 0\\
        0 & 0 & 0\\
        0 & 0 & 1
    \end{bmatrix},
\end{align}
which is a real $3\times 3$-matrix that represents the correlation tensor. Equation~\eqref{eqn::SI::QME} yields:
\begin{align}
    \partial_t \Phi_{\alpha \beta} = & \, \delta_1 \varepsilon_{\alpha z \gamma} \Phi_{\gamma \beta} - \delta_2 \Phi_{\alpha \gamma} \varepsilon_{\gamma z \beta} - (\Gamma_1 + \Gamma_2) \Phi_{\alpha\beta} + 2 \Phi_{\alpha\gamma} h_{2,\gamma} h_{2,\beta} + 2 h_{1,\alpha} h_{1,\gamma}\Phi_{\gamma\beta}    + 4 \langle ( \bm h_1  \times  \bm \sigma_1)_\alpha ( \bm h_2  \times  \bm \sigma_2)_\beta  \rangle \notag\\
    &\, 
    + 2 (\gamma^d_{12} + \gamma^d_{21}) \langle ( \hat{z}  \times  \bm \sigma_1)_\alpha ( \hat{z} \times  \bm \sigma_2)_\beta  \rangle - 2 \big(\gamma_{11}^d( 1 - \delta_{\alpha z}) + \gamma_{22}^d( 1 - \delta_{\beta z})\big) \Phi_{\alpha \beta},
\end{align}
where
\begin{align}
    E_z \equiv \varepsilon_{\alpha z \beta} = \begin{bmatrix}
        0 & -1 & 0\\
        1 & 0 & 0\\
        0 & 0 & 0
    \end{bmatrix},\qquad 
    E_x \equiv \varepsilon_{\alpha x \beta}= \begin{bmatrix}
        0 & 0 & 0\\
        0 & 0 & -1\\
        0 & 1 & 0
    \end{bmatrix}.
\end{align}
Since we are primarily interested in the $z$-correlation, i.e., $\alpha = \beta = z$, we are allowed to choose a local coordinate system for each NV such that $\bm h_1 = \hat{x}\sqrt{\Gamma_1/2} $ and $\bm h_2 = \hat{x}\sqrt{\Gamma_2/2}$:
\begin{align}
    \partial_t \Phi_{\alpha \beta} = & \, \delta_1 \varepsilon_{\alpha z \gamma} \Phi_{\gamma \beta} - \delta_2 \Phi_{\alpha \gamma} \varepsilon_{\gamma z \beta} - \gamma_{\alpha \beta} \Phi_{\alpha\beta}   
    + 2 \sqrt{\Gamma_1 \Gamma_2} \langle ( \hat{x}  \times  \bm \sigma_1)_\alpha ( \hat{x}  \times  \bm \sigma_2)_\beta  \rangle 
    + 2 (\gamma^d_{12} + \gamma^d_{21}) \langle ( \hat{z}  \times  \bm \sigma_1)_\alpha ( \hat{z} \times  \bm \sigma_2)_\beta  \rangle, \notag\\
    = &\, \delta_1 E_z \Phi - \delta_2 \Phi E_z - \gamma_{\alpha \beta} \Phi_{\alpha\beta} 
    - 2 \sqrt{\Gamma_1 \Gamma_2} E_x \Phi E_x
    - 4 \gamma^d_{12}   E_z \Phi E_z,
\end{align}
where  $\gamma_{\alpha \beta} \equiv \Gamma_1 (1 - \delta_{\alpha x}) + \Gamma_2(1 - \delta_{\beta x}) + 2\gamma_{11}^d( 1 - \delta_{\alpha z}) + 2\gamma_{22}^d( 1 - \delta_{\beta z})$. 
It follows that $\Phi_{xz} = \Phi_{zx} = \Phi_{yz} = \Phi_{zy} = 0$, and the remaining variables evolve according to:
\begin{align}
    \begin{cases}
        \partial_t \Phi_{zz} = - (\Gamma_1 + \Gamma_2)\Phi_{zz} +  2\sqrt{\Gamma_1\Gamma_2} \Phi_{yy}\\
        \partial_t \Phi_{yy} = - (\Gamma_1 + \Gamma_2 + 2\gamma_{11}^d + 2\gamma_{22}^d)\Phi_{yy} +  2\sqrt{\Gamma_1\Gamma_2} \Phi_{zz} + 4\gamma^d_{12} \Phi_{xx} + \delta_1 \Phi_{xy} + \delta_2 \Phi_{yx} \\
        \partial_t \Phi_{xx} = - 2(\gamma_{11}^d + \gamma_{22}^d)\Phi_{xx} + 4\gamma^d_{12} \Phi_{yy} - \delta_1 \Phi_{yx} - \delta_2 \Phi_{xy}
        \\
        \partial_t \Phi_{xy} = - (\Gamma_2 + 2\gamma_{11}^d + 2\gamma_{22}^d) \Phi_{xy} - \delta_1 \Phi_{yy} + \delta_2 \Phi_{xx} - 4 \gamma_{12}^d \Phi_{yx}
        \\
        \partial_t \Phi_{yx} = - (\Gamma_1 + 2\gamma_{11}^d + 2\gamma_{22}^d) \Phi_{yx} + \delta_1 \Phi_{xx} - \delta_2 \Phi_{yy} - 4 \gamma_{12}^d \Phi_{xy}
    \end{cases}. \label{eqn::QME::main}
\end{align}
The set of equations~\eqref{eqn::QME::main} implies that \(\Phi_{zz}(t)\) is generally expressed as a sum of five exponentials:
\begin{align}
    \Phi_{zz}(t) = \sum_{\lambda = 1}^5 C_\lambda e^{\nu_\lambda t},
\end{align}
where the eigenvalues \(\nu_\lambda\) and coefficients \(C_\lambda\) are determined by numerically diagonalizing the linear matrix on the right-hand-side of Eq.~\eqref{eqn::QME::main}. This insight enables us to efficiently fit the experimental data.

Let us also note that in the absence of both dephasing and detuning, Eq.~\eqref{eqn::QME::main} simplifies to:
\begin{align}
    \begin{cases}
        \partial_t \Phi_{zz} = - (\Gamma_1 + \Gamma_2) \Phi_{zz} +   2\sqrt{\Gamma_1\Gamma_2} \Phi_{yy}\\
        \partial_t \Phi_{yy} = - (\Gamma_1 + \Gamma_2) \Phi_{yy} +  2\sqrt{\Gamma_1\Gamma_2} \Phi_{zz}
    \end{cases}\Rightarrow 
    \langle \sigma^z_1 \sigma^z_2 \rangle -\langle \sigma^z_1 \rangle \langle \sigma^z_2 \rangle = \Big[\cosh\big(2 t\sqrt{\Gamma_1\Gamma_2} \big) - 1 \Big] e^{- (\Gamma_1 + \Gamma_2) t}.
    \label{eqn::analytic}
\end{align}
This analytic form is depicted as a solid line in Fig.~\ref{fig::figS9}(a).

\subsection{Fitting the experimental data}
To fit the experimental dataset in Fig.~4(d) of the main text, which has $\delta_2=0$ and $\delta_1 \in \{0,-250,-500\}\,$kHz, we use the set of equations~\eqref{eqn::QME::main} with 5 free parameters: a prefactor $r_0$ and correlated dephasing rate $\gamma_{12}^d$ kept the same for all three detunings $\delta_1$, and three local dephasing rates $\gamma^d_{\rm tot} = \gamma_{11}^d + \gamma_{22}^d$ for each sub-dataset for a given $\delta_1$ (note that Eq.~\eqref{eqn::QME::main} only gives access to $\gamma^d_{\rm tot}$ as opposed to individual rates $\gamma_{11}^d$, $\gamma_{22}^d$). Detunings $\delta_1$ and $\delta_2$ are known parameters, while single-NV relaxation rates $\Gamma_1$, $\Gamma_2$ are obtained from the fits in Fig.~4(c) of the main text.
\begin{figure*}[hbt!]  
    \centering
    \includegraphics[width=0.4\textwidth]{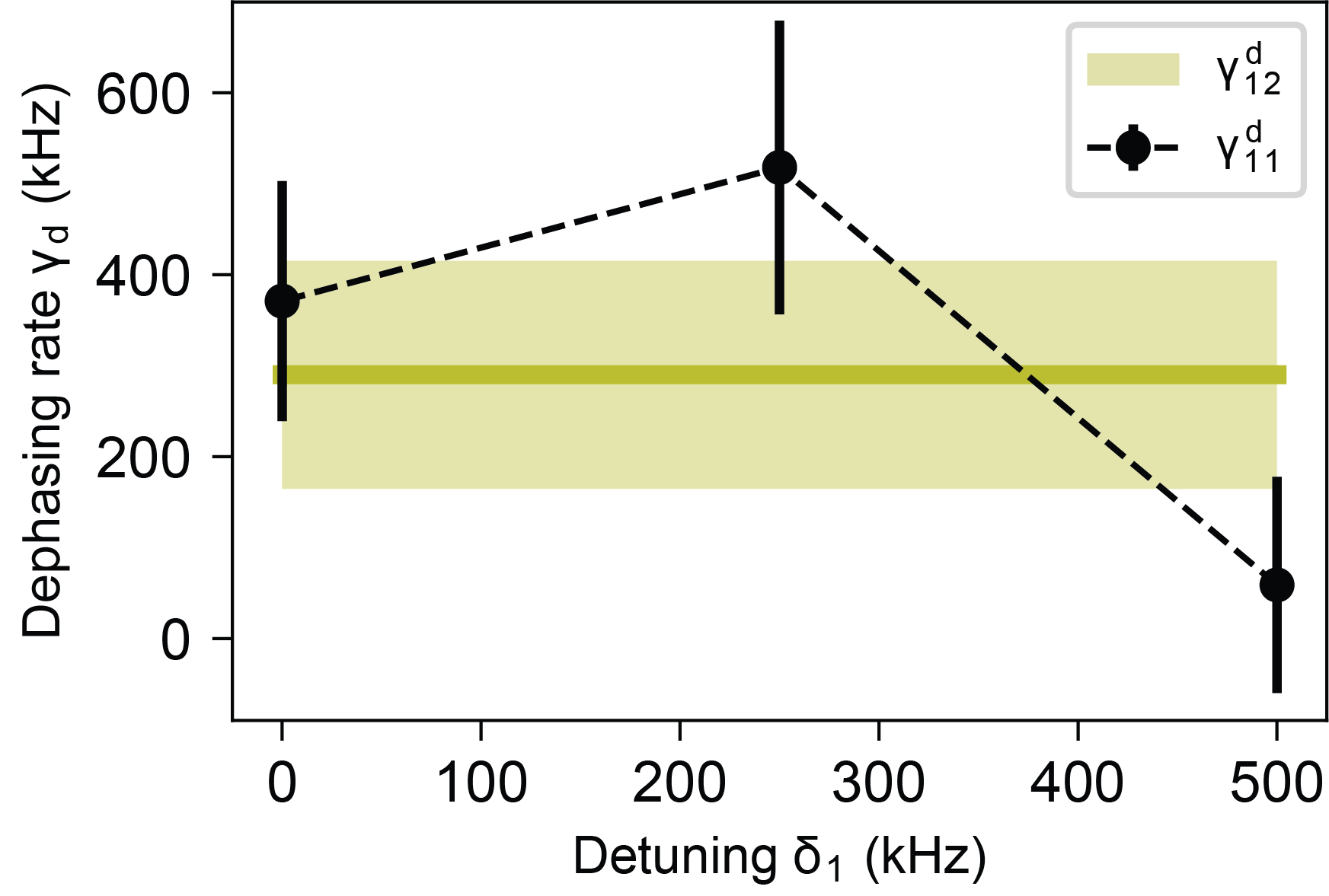}
    \caption{Fitted dephasing rates from correlated $T_1$ experiments. Black denotes the local dephasing rate of NV1 $\gamma_{11}^d$, assuming $\gamma_{22}^d = \left. \gamma_{11}^d \right|_{\delta_1=0}$. Olive green represents the correlated dephasing rate $\gamma^d_{12}$ for all detunings and its uncertainty (shaded region).}
    \label{fig::figS10}
\end{figure*}

Since finite $\gamma_{12}^d \neq 0$ indicates a common dephasing channel to both NV centers, we hypothesize that $\gamma_{12}^d$ could arise from correlated fluctuations of the MW frequency of each NV due to sample drift or thermal fluctuations of the rare-earth magnet's magnetization.

Meanwhile, $\gamma_{11}^d$ and $\gamma_{22}^d$ result from the interaction of each NV with its local magnetic environment, including both the correlated mechanisms such as the noisy drives, and the uncorrelated mechanisms such as coupling to $^{14}$N nuclear spin, similar to the Ramsey experiment. 
For different detunings $\delta_1$, in the rotating frame, the interaction rates of NV1 and its neighboring spins are different, resulting in different dephasing rates~\cite{childress2006}. Therefore, we expect different values of $\gamma_{11}^d$ for each constant-detuning sub-dataset, but the same value of $\gamma_{22}^d$. We thus assume that $\gamma_{22}^d$ = $\left. \gamma_{11}^d \right|_{\delta_1=0} = \frac{1}{2} \left. \gamma_{\rm tot}^d \right|_{\delta_1=0}$ for simplicity, and extract $\left. \gamma_{11}^d \right|_{\delta_1\neq0} = \left. \gamma_{\rm tot}^d \right|_{\delta_1\neq0} - \gamma_{22}^d$. 
The fitted values of \(\gamma_{12}^d\) and \(\gamma_{11}^d\), shown in Fig.~\ref{fig::figS10},  are comparable to the beating frequency observed in the noise-free Ramsey experiment in Fig.~\ref{fig::figS6}, suggesting that uncorrelated NV-nuclear dynamics is indeed a plausible source of local dephasing.

\bibliography{Bib_supp}